\documentclass{aa}
\usepackage{graphicx}
\usepackage{natbib}
\usepackage{enumerate}
\usepackage{enumitem}
\usepackage{txfonts}
\begin{document}
%
   \title{Oscillation mode frequencies of 61 main sequence and subgiant stars observed by {\it Kepler}}

   \author{T.~Appourchaux
          \inst{1,2}
          \and
         W.~J.~Chaplin\inst{3}
           \and
          R.~A.~Garc\'\i a\inst{4}
          \and
          M.~Gruberbauer\inst{5}
          \and
          G.~A.~Verner\inst{3}
          \and
          H.~M.~Antia\inst{6}
          \and
          O.~Benomar\inst{7}
          \and
          T.~L.~Campante\inst{8,9}
          \and
          G.~R.~Davies\inst{4}
          \and
          S.~Deheuvels\inst{10}
          \and
          R.~Handberg\inst{8}
          \and
          S.~Hekker\inst{11,3}
          \and
          R.~Howe\inst{3}
          \and
          C.~R\'egulo\inst{12,13}
          \and
          D.~Salabert\inst{14}
          \and
          T.~R.~Bedding\inst{7}
          \and
          T.~R.~White\inst{7}
          \and
          J.~Ballot\inst{15,16}
          \and
          S.~Mathur\inst{17}
          \and
          V.~Silva Aguirre\inst{18}
          \and
          Y.~P.~Elsworth\inst{3}
          \and
          S.~Basu\inst{10}
          \and
          R.L~Gilliland\inst{19}
          \and
          J.~Christensen-Dalsgaard\inst{8}
          \and
          H.~Kjeldsen\inst{8}
          \and
          K.~Uddin\inst{20}
          \and
          M.~C.~Stumpe\inst{21}
          \and
          T.~Barclay\inst{22}
          }

   \institute{Univ Paris-Sud, Institut d'Astrophysique Spatiale, UMR8617, CNRS, B\^atiment 121, 91405 Orsay Cedex, France
         \and
         Kavli Institute for Theoretical Physics, University of California, Santa Barbara, CA 93106-4030, USA
         \and
          School of Physics and Astronomy, University of Birmingham, Edgbaston, Birmingham B15 2TT, United Kingdom
         \and
          Laboratoire AIM, CEA/DSM-CNRS-Universit\'e Paris Diderot, IRFU/SAp, Centre de Saclay, 91191 Gif-sur-Yvette Cedex, France
          \and
          Institute for Computational Astrophysics, Department of Astronomy \& Physics, Saint Mary's University, Halifax, NS B3H 3C3, Canada
          \and
          Tata Institute of Fundamental Research, Homi Bhabha Road, Mumbai 400005, India
          \and
          Sydney Institute for Astronomy (SIfA), School of Physics, University of Sydney, New South Wales 2006, Australia
          \and
          Stellar Astrophysics Centre, Department of Physics and Astronomy, Aarhus University, Ny Munkegade 120, DK-8000 Aarhus C, Denmark
          \and
          Centro de Astrof\'isica and Faculdade de Ci\^encias, Universidade do Porto, Rua das Estrelas, 4150-762 Porto, Portugal
          \and
         Department of Astronomy, Yale University, P.O. Box 208101, New Haven CT 06520-8101, USA
          \and
          Astronomical Institute "Anton Pannekoek", University of Amsterdam, Science Park 904, 1098 XH Amsterdam, The Netherlands%
          \and
          Instituto de Astrof\'isica de Canarias, 38205 La Laguna, Tenerife, Spain
          \and
          Universidad de La Laguna, Dpto de Astrof\'isica, 38206 La Laguna, Tenerife, Spain
          \and
          Universit\'e de Nice Sophia-Antipolis, CNRS UMR 6202, Observatoire de la C\^ote d'Azur, BP 4229, 06304 Nice Cedex 4, France
          \and
          Institut de Recherche en Astrophysique et Plan{\'e}tologie, CNRS, 14 avenue E. Belin, 31400 Toulouse, France
          \and
          Universit{\'e} de Toulouse, UPS-OMP, IRAP, 31400 Toulouse, France
          \and
          High Altitude Observatory, NCAR, P.O. Box 3000, Boulder, CO 80307, USA
          \and
          Max Planck Institute for Astrophysics, Karl-Schwarzschild-Str. 1, 85748, Garching bei M\"{u}nchen, Germany
          \and
          Center for Exoplanets and Habitable Worlds, The Pennsylvania State University, University Park, PA 16802, USA
          \and
          Orbital Sciences Corporation/NASA Ames Research Center, Moffett Field, CA 94035, USA
          \and
          SETI Institute/NASA Ames Research Center, Moffett Field, CA 94035, USA
          \and
          Bay Area Environmental Research Institute/NASA Ames Research Center, Moffett Field, CA 94035, USA
          }

   \date{Received 2 February 2012; accepted 13 April 2012}

 
  \abstract
  {Solar-like oscillations have been observed by {{\it Kepler}} and CoRoT in several solar-type stars, thereby providing a way to probe the stars using asteroseismology}
   {We provide the mode frequencies of the oscillations of various stars required to perform a comparison with those obtained from stellar modelling.}
   {We used a time series of nine months of data for each star.  The 61 stars observed were categorised in three groups: simple, F-like and mixed-mode.  The simple group includes stars for which the identification of the mode degree is obvious.  The F-like group includes stars for which the identification of the degree is ambiguous.  The mixed-mode group includes evolved stars for which the modes do not follow the asymptotic relation of low-degree frequencies.  Following this categorisation,  the power spectra of the 61 main sequence and subgiant stars were analysed using both maximum likelihood estimators and Bayesian estimators, providing individual mode characteristics such as frequencies, linewidths, and mode heights. We developed and describe a methodology for extracting a single set of mode frequencies from multiple sets derived by different methods and individual scientists. We report on how one can assess the quality of the fitted parameters using the likelihood ratio test and the posterior probabilities.}
   {We provide the mode frequencies of 61 stars (with their 1-$\sigma$ error bars), as well as their associated \'echelle diagrams.}
   {}

   \maketitle
%

\section{Introduction}
\label{sec:intro}

Stellar physics is undergoing a revolution thanks to the great wealth of asteroseismic data that have been made available by space missions such as CoRoT \citep{Baglin2006} and {\it Kepler} \citep{Borucki2009}.
With the seismic analyses of these stars providing the frequencies of the stellar eigenmodes and the large number of high quality observations, asteroseismology is rapidly becoming a valuable tool for understanding stellar physics.  
 
Solar-type stars have been 
observed over periods exceeding
six months using CoRoT providing significant results from their seismic analysis \citep{Appourchaux2008,Benomar2009,Gaulme2009,Gruberbauer2009,Barban2009,Garcia2009,Mosser2009,Benomar2009b,Gaulme2010,Mathur2010a,Deheuvels2010,Ballot2011a}.
The {\it Kepler} mission now provides a larger sample of stars observed for even longer  durations \citep{Chaplin2011}.  The seismic analyses of several solar-type and subgiant stars observed by {\it Kepler} were reported by \citet{JCD2010}, \citet{Metcalfe2010}, \citet{Campante2011}, \citet{Mathur2011} and \citet{Howell2011}.  

Owing to the ability of {\it Kepler} to perform longer observations of stars, the measurement of mode frequencies on several hundreds of stars becomes a challenge.  The large-scale fitting of many stellar power spectra was anticipated by \citet{TA2003} for the now-defunct Eddington mission.  All of the steps currently used to fit the p-mode power spectra were described in that paper.  \citet{TA2003} also anticipated the difficulties that would be encountered for stars having modes departing from a simple frequency relation, i.e., with mixed modes.  On the other hand,  the problem of the degree tagging for HD49933 due to its large mode linewidth, which was first reported by \citet{Appourchaux2008}, was not anticipated by \citet{Appourchaux2007b}, even though they simulated such large linewidths.

In this paper, we provide mode frequencies for 61 {\it Kepler} main sequence and subgiant stars observed for about nine months by {\it Kepler}. Some of these stars have characteristics that create difficulties cited above when fitting power spectra.

The next section describes how the time series and power spectra were obtained.  Section 3 describes the peak bagging procedure.  Section 4 details how we derive a single data set from the several frequency sets provided by the fitters.  Section 5 provides product-and-assurance-quality tools needed to validate  the mode frequencies.  Finally, we provide a short conclusion.     The paper includes five examples of the table of frequencies and  \'echelle diagrams, while tables of frequencies of 56 stars and \'echelle diagrams for all 61 stars are available online.


   
     \begin{figure*}[!]
   \centering
   \hbox{
   \includegraphics[angle=90,width=9.cm]{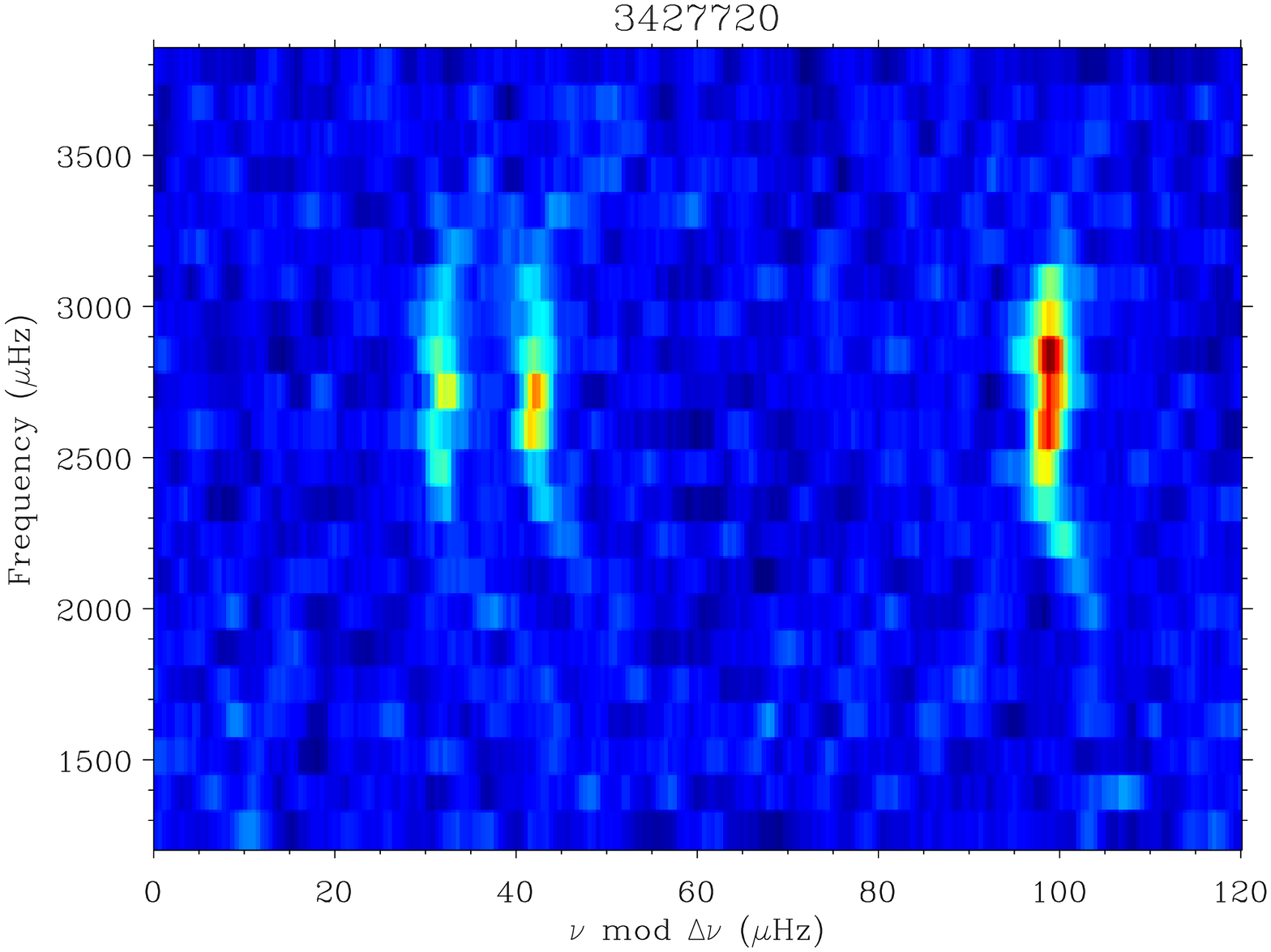}
      \includegraphics[angle=90,width=9.cm]{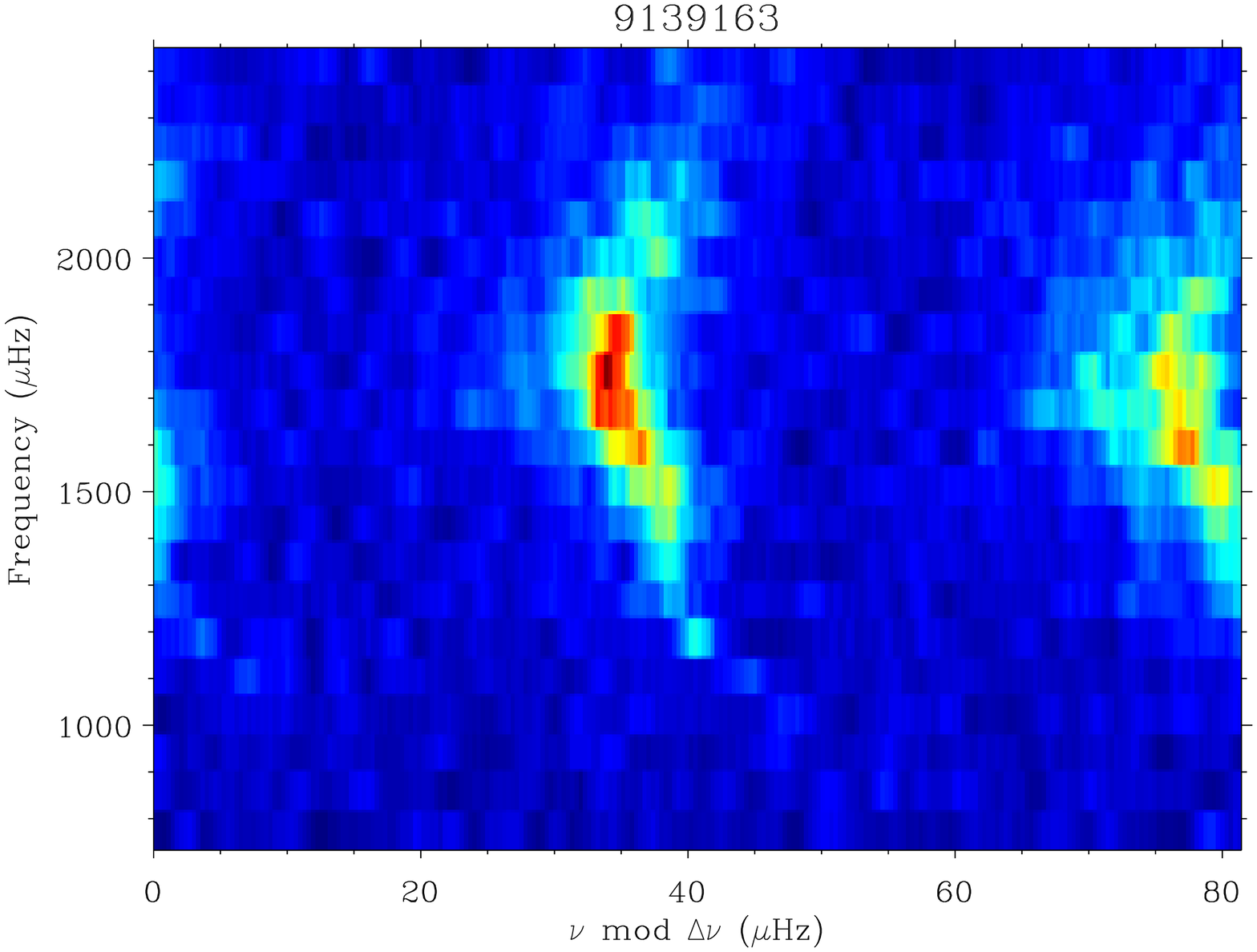}
      }
      \caption{Echelle diagrams of the power spectra of two {\it simple} stars (KIC 3427720 and KIC 9139163). The power spectra are normalised by the background and smoothed over 3~$\mu$Hz.}
         \label{fig1}
   \end{figure*}
   
        \begin{figure*}[!]
   \centering
         \hbox{
   \includegraphics[angle=90,width=9.cm]{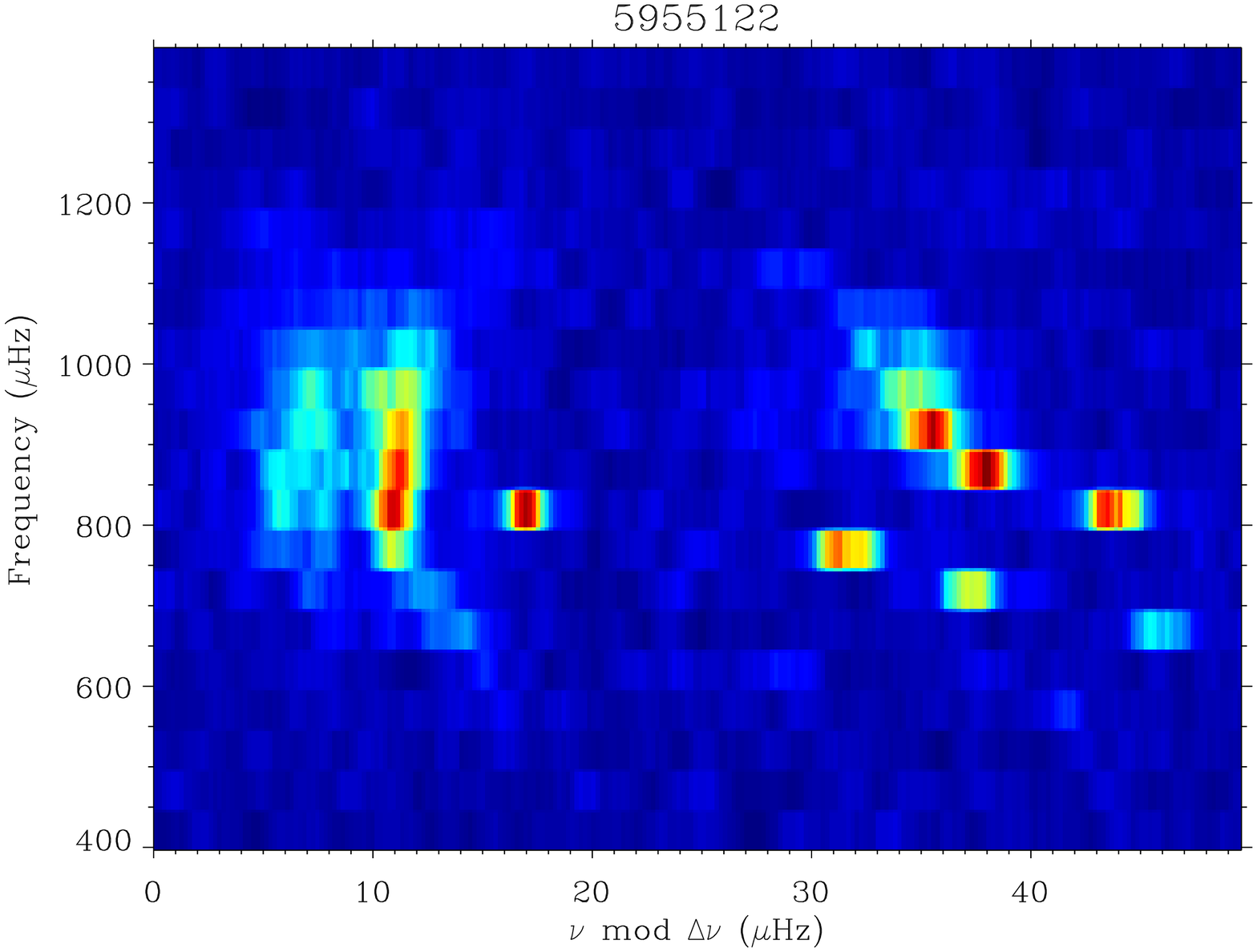}
   \includegraphics[angle=90,width=9.cm]{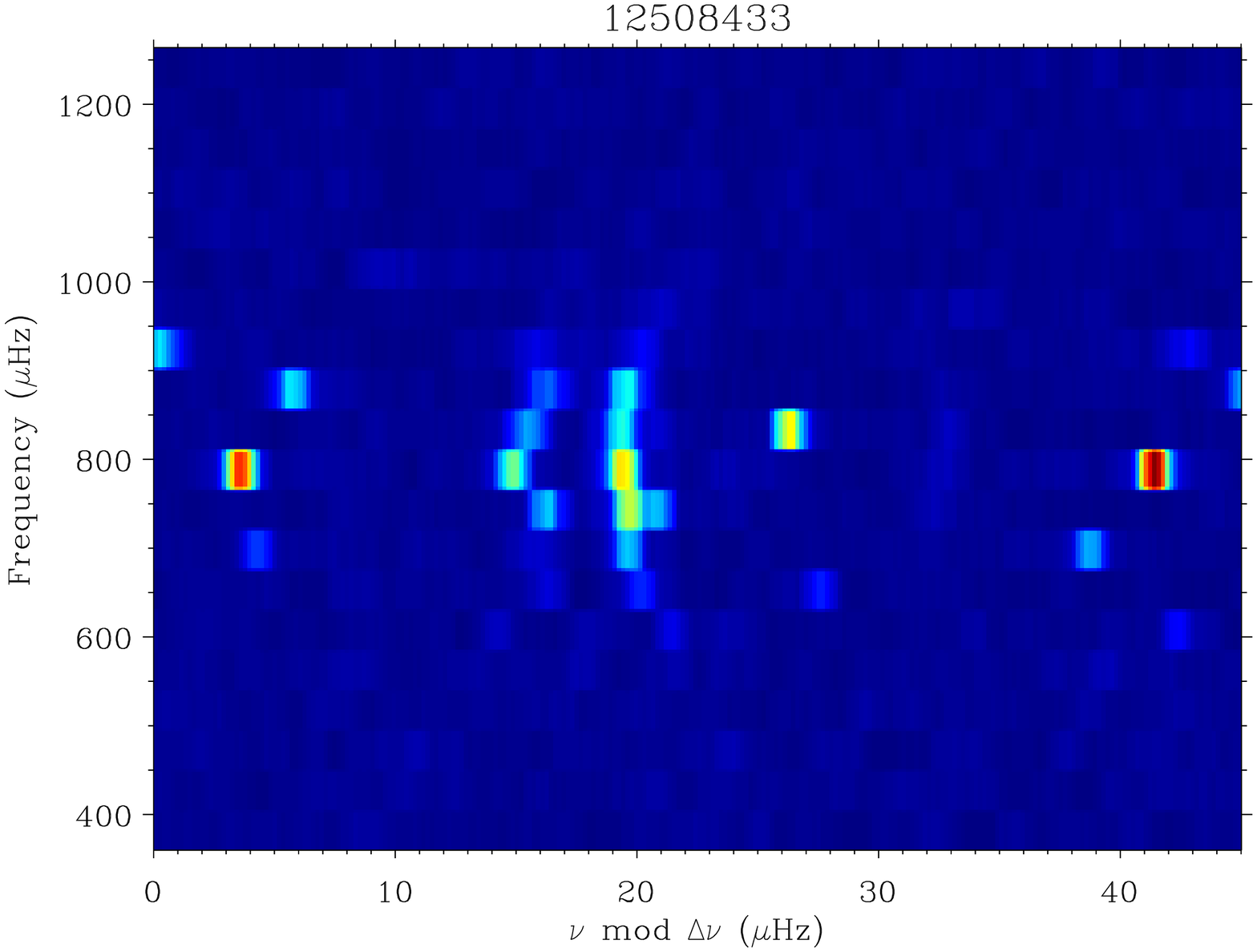}
      }
      \caption{Echelle diagrams of the power spectra of two {\it mixed-mode} stars (KIC 5955122 and KIC 12508433). The power spectra are normalised by the background and smoothed over 3~$\mu$Hz.  The $l=1$ ridges are broken and pass through the $l=0-2$ ridges;  the plot on the left hand side shows an example of a {\it mixed-mode} star with the $l=1$ modes not aligned along a ridge but still passing through the $l=0-2$ ridges. The plot on the right hand side shows a faint $l=3$ ridge at 32~$\mu$Hz.}
         \label{fig2}
   \end{figure*}
   
        \begin{figure*}[!]
   \centering
    \hbox{
       \includegraphics[angle=90,width=9.cm]{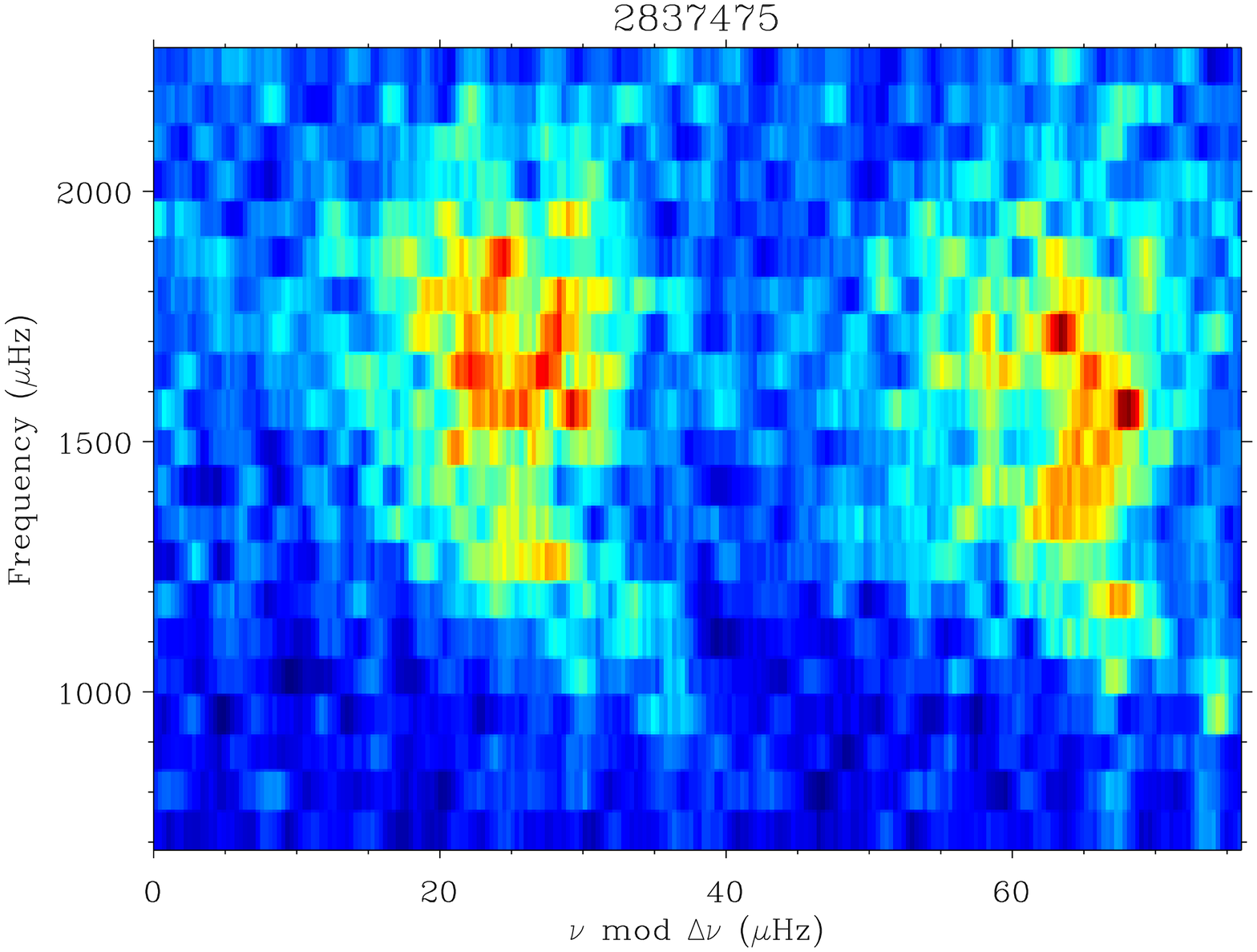}
       \includegraphics[angle=90,width=9.cm]{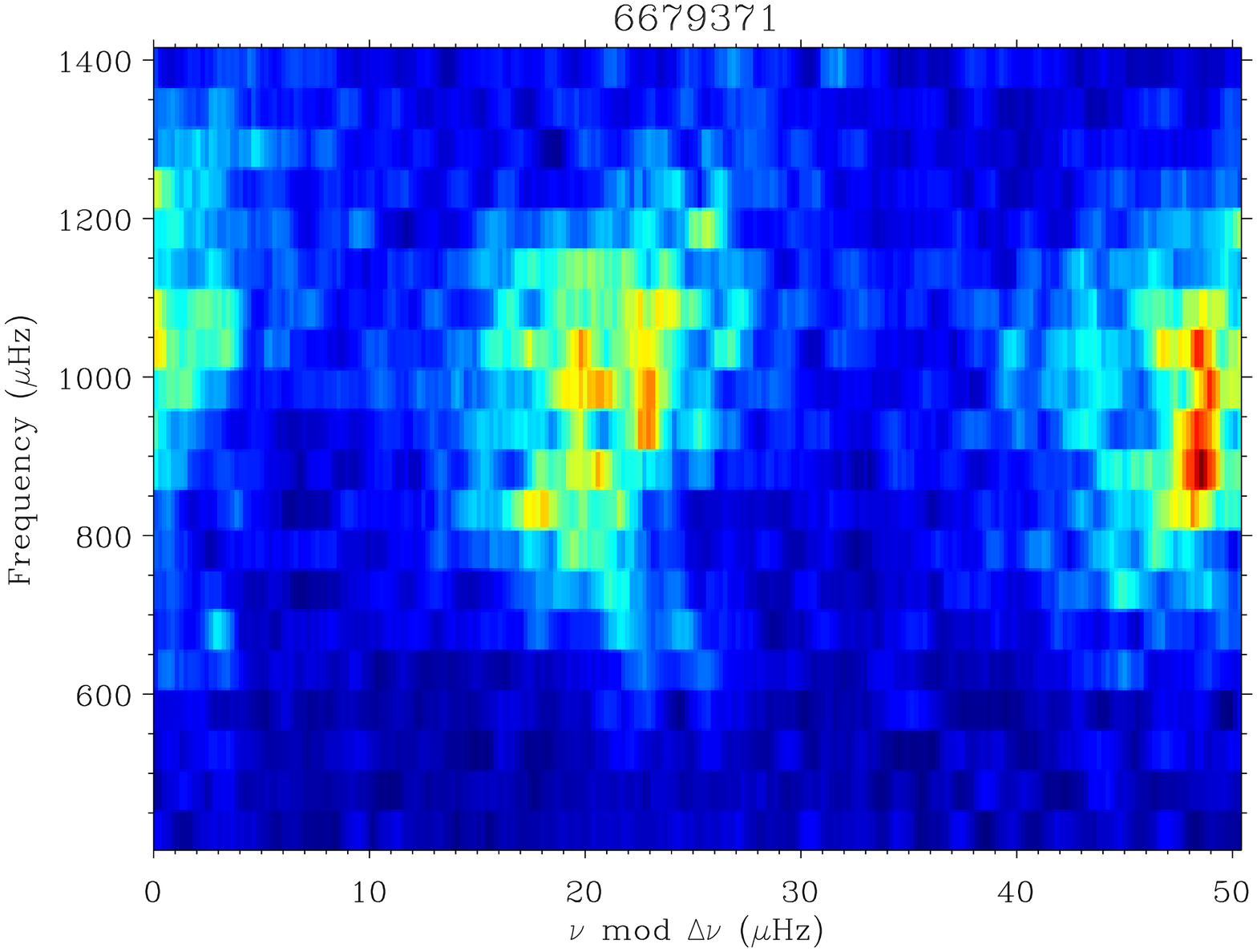}
}
      \caption{Echelle diagrams of the power spectra of two {\it F-like} stars (KIC 2837475 and KIC 6679371). The power spectra are normalised by the background and smoothed over 3~$\mu$Hz.}
         \label{fig3}
   \end{figure*}


\section{Time series and power spectra}
{\it Kepler} observations are obtained in two different operating modes: long cadence (LC) and short cadence (SC) \citep{Gilliland2010,Jenkins2010}.  This work is based on SC data. For the brightest stars (down to {\it Kepler} magnitude, $Kp \approx12$), SC observations can be obtained for a limited number of stars  (up to 512 at any given time)  with a faster sampling cadence of 58.84876~s (Nyquist frequency of $\sim$ 8.5~mHz), which permits a more precise transit timing and the performance of asteroseismology.  {\it Kepler} observations are divided into three-month-long {\it quarters} (Q).  A subset of 61 solar-type stars observed during quarters Q5, Q6, and Q7 (March  22, 2010 to December 22, 2010) were chosen because they have oscillation modes with high signal-to-noise ratios. This length of data gives a frequency resolution of about 0.04~$\mu$Hz.

To maximise the signal-to-noise ratio for asteroseismology, the time series were corrected for outliers, occasional jumps, and drifts \citep[see][]{RAG2011}, and the mean levels between the quarters were normalised.  Finally, the resulting light curves were high-pass filtered using a triangular smoothing of width of one day, to minimise the effects of the long-period instrumental drifts.   The typical amount of data missing from the time series ranges from 3\% to 7\%, depending on the star.  All the power spectra were produced by one of the co-authors using the Lomb-Scargle periodogram \citep{Scargle82}, properly calibrated to comply with
Parseval's theorem \citep[see][]{Appourchaux2011}.


\section{Star categories}

To simplify the extraction of mode parameters, three categories of star were identified: simple (sun-like), F-like (also known as the HD49933 syndrome), and mixed modes.  The categorisation was performed using the \'echelle diagram that was first introduced by \citet{GG81}.  The construction of the diagram is based on the low-degree modes being essentially equidistant in frequency for a given $l$, with a typical spacing of the large separation ($\Delta\nu$).  The equidistance of the mode frequency $\nu_{nl}$ is the result of an approximation derived by \citet{Tassoul80} as
\begin{equation}
\nu_{nl} \approx \Delta\nu \left(n+\frac{l}{2}+\epsilon\right)-\delta_{nl},
\label{asymp}
\end{equation}
where $l$ is the degree of the mode, $n$ is the radial order, $\epsilon$ is a parameter related to stellar surface properties, and $\delta_{nl}$ is the small separation.  The spectrum is cut into pieces of length $\Delta\nu$, which are stacked on top of each other.  Since the modes are not exactly equidistant in frequency, the \'echelle diagram shows up power due to the modes as curved ridges.  Examples of these \'echelle diagrams are given in Figs.~\ref{fig1} to \ref{fig3}, which represents the three main categories used in this paper.  Fig~\ref{fig1} shows examples for {\it simple} stars.  Fig~\ref{fig2} shows examples for {\it mixed-mode} stars where an avoided crossing\footnote{ Mixed modes occur in evolved stars and their frequencies are shifted from the usual regular pattern by avoided crossings \citep{Osaki1975,Aizenman1977}.} is present.  Fig~\ref{fig3} shows examples for {\it F-like} stars, which are hotter stars (spectral type F, having large linewidths).

Figure~\ref{fig-hr} shows the measured median large-frequency-separation of the 61 stars as a function of their effective temperature, together with their categories resulting from the visual assessment of the \'echelle diagram.    Out of the 61 stars, we have 28 simple stars, 15 F-like stars, and 18 mixed-mode stars.  Figure~\ref{fig-hr} shows that the boundary between simple stars and F-like stars is about 6400~K, which roughly corresponds to a linewidth at maximum mode height of about 4~$\mu$Hz \citep{Appourchaux2012}.  For these F-like stars, the frequency separation between the $l=0$ and $l=2$ modes (=small separation) ranges from 4~$\mu$Hz to 12~$\mu$Hz, which, combined with a linewidth of at least 4~$\mu$Hz, explains why the detection of the $l=0$ and 2 ridges is more difficult for these stars.



\section{Mode parameter extraction}
\subsection{Power spectrum model}

\begin{table*}[t]
\caption{Characteristics of the fit performed by each fitting group.  The first column provides the fitter identification; the fitter in italics indicates whether it was a final fitter.  The second column provides methods used by the fitters.  The third column provides the category of stars fitted; the category in italic indicates which stars were finally fitted.  The fourth column provides the number of identifications used.  The fifth column provides the number of parameters used per order.  The sixth column provides the number of additional parameters common to the modes and the background.  The seventh column provides the number of fitted orders.  The eighth column provides the range over which the fit is performed.  The last column provides how the error bars are computed.  The final fitters are indicated in bold.  MLE stands for maximum likelihood estimators.  MCMC stands for Monte Carlo Markov chain.}             
\label{tab_summary}      
\centering                          
\begin{tabular}{c c c c c c c c c}        
\hline                 
\hline     
Fitter&Method&Star&Iden.&Param.& Add.&Orders& Window& Error\\
&&category&&per order& parameters&&size&\\
\hline  
\hline                           
{\it Appourchaux, IAS}&MLE Global$^{a}$&{\it simple} / mixed-mode&1&5&5& $\le$20&$\le 20 \Delta \nu$&Hessian\\
Howe, NSO&MLE Global$^{b}$&simple / mixed-mode&1&5&5&$\le$15&$\le 15 \Delta \nu$&Hessian\\
Salabert, A2Z&MLE Global$^{a}$&simple / mixed-mode&1&5&5& $\le$15&$\le 15 \Delta \nu$&Hessian\\
{\it Chaplin, BIR}&MLE Pseudo-global$^{c}$&simple / {\it mixed-mode}&1&5&4& $\le$20&$\Delta \nu$, $\le 20 \Delta \nu$&Hessian\\
Deheuvels, YAL&MLE Global$^{a}$&simple / mixed-mode&1&5&6& $\le$16&$\le 16 \Delta \nu$&Hessian\\
Antia, TAT&MLE Local$^{d}$&simple / mixed-mode&1&12&None& $\le$15&$\Delta \nu$&Hessian\\
{\it Verner, QML}&MLE Global$^{a}$&{\it simple / mixed-mode}&1&5&5& $\le$14&$\le 14 \Delta \nu$&Hessian\\
Benomar, SYD&MCMC$^{e}$&F-like&2&5&10& $>$10&$> 10 \Delta \nu$&Credible\\
{\it Gruberbauer, MAR}&Nested sampling$^{f}$&{\it F-like}&2&5&5& $\le$15&$\le 15 \Delta \nu$&Credible\\
Handberg, AAU&MCMC$^{g}$&F-like&2&5&5& $>$10&$> 10 \Delta \nu$&Credible\\
\hline  
\hline  
\end{tabular}
\begin{list}{}{}
\item[$^{a}$] \citet{Appourchaux2008}
\item[$^{b}$] derived from \citet{Howe1998}
\item[$^{c}$] \citet{Fletcher2009} 
\item[$^{d}$] \citet{EA90}
\item[$^{e}$] \citet{Benomar2009}
\item[$^{f}$]  \citet{Gruberbauer2009, Feroz2009}
\item[$^{g}$] \citet{Handberg2011}
\end{list}
\label{methods}
\end{table*}


  \begin{figure}[!]
   \centering
   \includegraphics[angle=90,width=8.25cm]{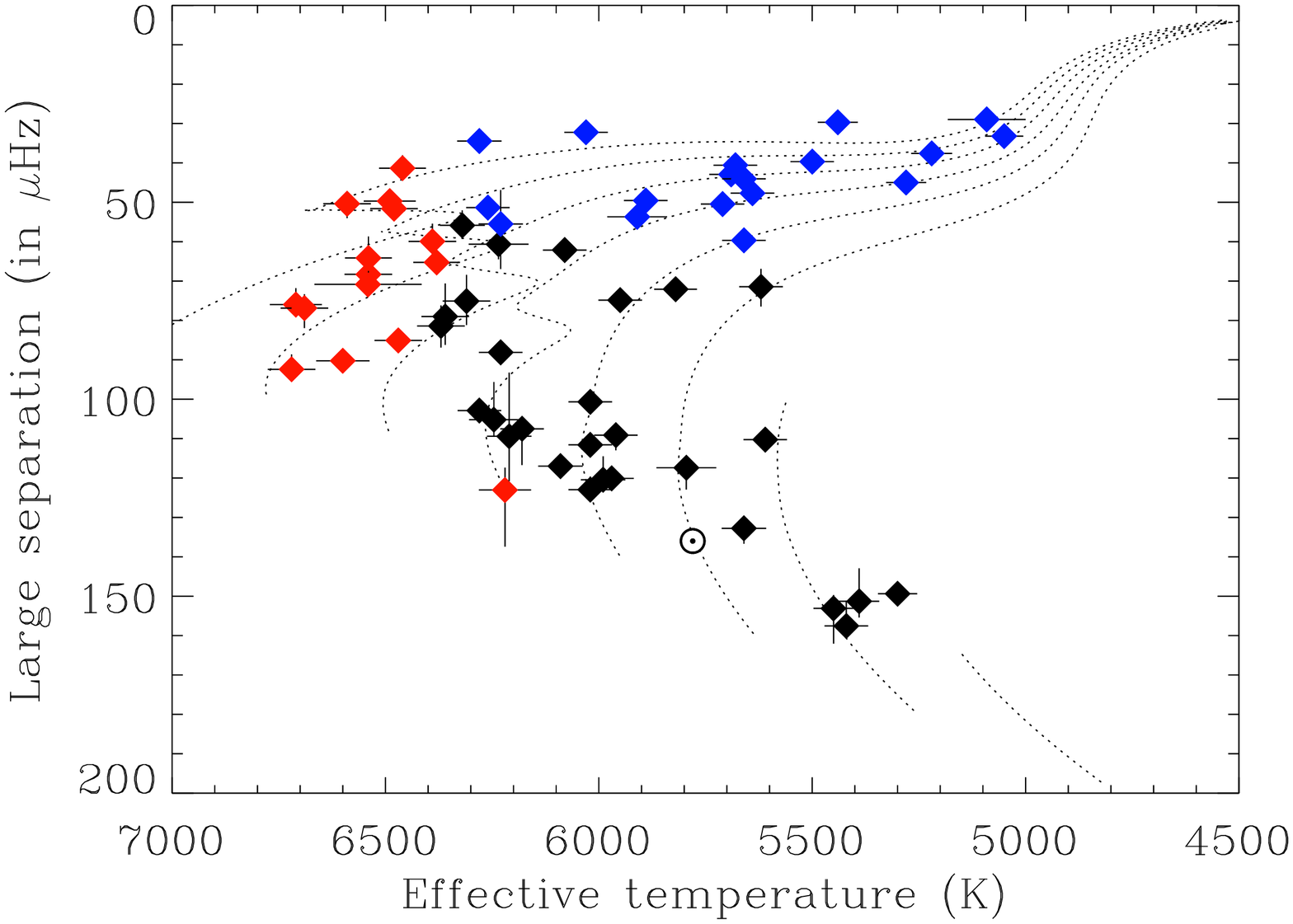}
      \caption{Large separation as a function of effective temperature for the stars in this study; (black) {\it simple} stars, (blue) {\it mixed-mode} stars, (red) {\it F-like} stars, and ($\odot$) the Sun.  The effective temperatures were derived from  \citet{Pins2011} except where noted in Table~2.  The uncertainties in the large separation represent the minimum and maximum variations with respect to the median measured in this study  (see Table 2); some of these uncertainties are within the thicknesses of symbols.  The dotted lines are evolutionary tracks for stars of mass from 0.8 M$_{\odot}$(farthest right) to  1.5 M$_{\odot}$ (farthest left), in steps of 0.1 M$_{\odot}$ derived from \citet{JCDGH2010} for solar metallicity.}
         \label{fig-hr}
   \end{figure}

The mode parameter extraction was performed by ten teams of fitters whose leaders are listed in Table~\ref{methods}.  The power spectra were modelled over a frequency range typically covering 10 to 20 large separations (=$\Delta\nu$).  The background was modelled using a multi-component Harvey model \citep{Harvey85}, each component with two parameters,  and a white noise component.  The background was fitted prior to the extraction of the mode parameters and then held at a fixed value. For each radial order, the model parameters were mode frequencies (one each for $l$=0,1,2), a single mode height (with assumed ratios of $h_1/h_0=1.5$, $h_2/h_0$=0.5 where the subscript refers to the degree), and a single mode linewidth.  The ratios taken in this paper are typical of those expected for {\it Kepler} stars \citep{Ballot2011}.  In the case of  the AAU fitter only, the $l=0$ linewidths were fitted and the linewidths of the other degrees were interpolated in between two $l=0$ mode linewidths.  The relative heights $h_{(l,m)}$ (where $m$ is the azimuthal order) of the rotationally split components of the modes depend on the stellar inclination angle, as given by \citet{Gizon2003}.  For 
each star, the rotational splitting and stellar inclination angle were chosen to be common across all the 
modes.
The mode profile was assumed to be Lorentzian.  We used a single Harvey component for all stars other than the 16 stars fitted by QML for which we used a double component.  In total, the number of free parameters for 15 orders was at least $5 \times 15+2=77$.  

The two models described above were used to fit the parameters of 46 stars (28 simple and 18 mixed-mode) using maximum likelihood estimators (MLE).  Formal uncertainties in each parameter were derived from the inverse of the Hessian matrix \citep[for more details on MLE, significance, and formal errors, see][]{Appourchaux2011}.

The 15 F-like stars were fitted with a Bayesian approach using different sampling methods. Both SYD and AAU employed Monte Carlo Markov chain (MCMC) \citep{Benomar2009,Handberg2011}, while MAR used nested sampling via the code MultiNest \citep{Feroz2009}. For the latter sampling approach, the large number of parameters forced us to use MultiNest's constant-efficiency, mono-modal mode. The priors on the central frequency and inclination angle were uniform. The prior on the splitting was either uniform in the range $0-15~\mu$Hz (MAR) or a combination of a uniform prior in the range $0-2~\mu$Hz and a decaying Gaussian (SYD, AAU). The priors on mode height were modified Jeffreys priors \citep[][]{Jeffreys,Benomar2009,Gruberbauer2009}, and the priors on the linewidth were either uniform (MAR) or modified Jeffreys priors (SYD, AAU). 
The error bars were derived from the marginal posterior distribution of each parameter.  Each Bayesian fitter had seven stars to fit: four stars + three common stars.  The latter were used for comparison with the Bayesian methods.  Priors on the frequencies were set after visual inspection of the power spectrum. Modes of degree $l=2$ were assumed to be on the low-frequency side of the $l=0$ (\emph{i.e.}, the small spacing $d_{02}$ was assumed to be positive).  To avoid spurious results, two of the Bayesian fitters (SYD, MAR) also used a smoothness condition on the frequency for each degree, in a similar way to \citet{Benomar2012}.

Table~\ref{tab_summary} provides a summary of how the different fits were performed.

\subsection{Initial guesses for parameters}
Of all the parameters describing the modes, the frequencies are the most difficult to guess.  Given the number of stars, initial guesses for these mode frequencies were obtained using different techniques:
\begin{itemize}
\item The automatic detection of modes based on the values of $\Delta \nu$ and $\nu_{\rm max}$ \citep[see][and references therein]{Verner2011}, that were then manually tweaked if required.
\item Visual detection using the \'echelle diagram (especially for mixed-mode stars).
\item Derivation from fitted parameters obtained from previous reported observations of the stars (see Table 2).
\end{itemize}

The degree tagging could then be done quite easily for the simple and mixed-mode stars (see Figs.~\ref{fig1} and \ref{fig2}).  The mere visual assessment of the \'echelle diagrams was enough to permit the tagging of the ridges with the proper degree where  the $l=1$ stands alone, while the $l=0-2$ pair appears as a double ridge.  For the mixed-mode stars, the tagging was also done by inspection of the \'echelle diagram, but required the input of model frequencies as the $l=1$ modes go through the $l=0-2$ ridges; some ambiguity could be caused by the avoided crossing.  For the F stars, the fit was performed for both possible identifications ($l=0-2$ and $l=1$, or vice versa), and the model probabilities were calculated to obtain the most likely identification \citep{Benomar2009,Handberg2011}.  For these latter stars, other tools are available that compare the measured value of $\epsilon$ (see Eq.~\ref{asymp}) with that of the theoretical values derived for other stars with similar effective temperatures \citep{White2011}.

\subsection{Fitting procedure}
The steps that we adopted to perform the fit are as follows:
\begin{enumerate}
\item We fit the power spectrum as the sum of a background made of a combination of Lorentzian profiles (one or two) and white noise, as well as a Gaussian oscillation mode envelope with three parameters (the frequency of the maximum mode power, the maximum power, and the linewidth of the mode power).
\item We fit the power spectrum with $n$ orders using the mode profile model described above, with no splitting and the background fixed as determined in step 1
\item We follow step 2 but define the splitting and the stellar inclination angle as free parameters, and then apply a likelihood ratio test to assess the significance of the fitted splitting and angle.
\end{enumerate}
The steps above were sometimes varied slightly depending on the assumptions that were made.  For instance, the mode height ratio could instead be defined as a free parameters to study the impact of its variations on the derivation of the mode parameters such as linewidth and mode height.

\section{Derivation of mode frequency set}
The derivation of a single frequency set from the several sets obtained by different fitters has been a source of considerable concern since we started this work.  The measurements performed by {{\it Kepler}} for a given star are done only by this mission; there is no additional source of photometry.  Hence, this single measurement of oscillation modes in the photometry of a given star must provide a single frequency set.  The main question that arose was how to derive a single frequency set from the different fits.

We now explain the chain of thought that led to the procedure used in this paper.  This procedure had constantly evolved until we reached the {\it final} procedure.    Since anyone having to derive a common data set from several data sets derived from a single observable will face the same challenge as we faced, it is useful to understand how we decided on this procedure 
to avoid repeating
the same chain of thoughts.  

The {\it final} procedure can be applied to any data set to be derived from a single observable, such as mode linewidth, mode amplitude, and so forth.  An example of the application of this procedure for mode linewidth can be found in \citet{Appourchaux2012}.

Hereafter, we describe the final procedure in three steps.  The first step is common to all variants of the procedure, while the second step had several versions.  We present all the second-step versions used, including the step used in the {\it final} procedure.

\subsection{Common first step}
The goal of the first step is to provide an {\it average} frequency set and to quantify how a given frequency set provided by a fitter differs from the {\it average}.

\subsubsection{Rejection of outliers}
For a given ($n,l$) mode, one frequency is derived from several (if not all) fitters and outliers may need to be rejected.  The rejection of outliers can be
done either using the well-known 3-$\sigma$ threshold or using {\it Peirce's criterion} \citep{Peirce1852}.  The main disadvantage of the 3-$\sigma$ threshold is that it is not applicable when the number of measurements is small.  Peirce's criterion explicitly assumes that the number of measurements is small and that the root mean square (rms) deviation can be corrected for the rejection of outliers.  Peirce's criterion is based on rigourous probability calculation and not on any ad hoc assumption.
To cite Peirce's explanation of his criterion: ``The proposed observations should be rejected when the probability 
of the system of errors obtained by retaining them is less than that of the system of
errors obtained by their rejection multiplied by the probability of making so many, and
no more, abnormal observations.''   This logic calls for an iterative assessment of the rejection
when one or more datasets are rejected.  The iteration stops when no improvement is possible.  

Following the work of \cite{Gould1855}, we implemented Peirce's criterion for a sample of $x_i$ as follows:
\begin{enumerate}
	\item We compute the mean value $x_m$ and rms deviation $\sigma$ of a data set $x_i$.
	\item We compute the rejection factor $r$ by solving Eq. (D) of \cite{Gould1855}, assuming one doubtful observation.
	\item We reject data if $|x_i-x_m| > r \sigma$.
	\item If $n$ data have been rejected then we compute the new rejection factor $r$, assuming $n+1$ doubtful observations.
	\item We repeat steps 3 to 4 until no more data are rejected.
\end{enumerate}

The assumption behind Peirce's criterion is that all the observed data have the same statistical distribution, {\em i.e.}, the same mean and standard deviation.  In our case, this assumption is valid because all the methods used (see Table~\ref{methods}) are either exactly or approximately akin to MLE, so that the error bars in our mode frequencies are locally related to the curvature of the MLE at the location of the maximum.

The rejection of outliers following Peirce's criterion was implemented as an option for the common first step.

\subsubsection{Deviation from the average frequency set}
For a given star and for a given ($n,l$) mode, one computes the mean mode frequency $\langle \nu_{n, l} \rangle$ from the frequencies provided by the fitters who detected this mode.  The frequency set consisting of the data $\langle \nu_{n, l} \rangle$ is then designated as the {\it average} frequency set.  For each fitter labelled k, one then computes for each mode the mean normalised rms deviation from the mean mode frequency $\langle \nu_{n, l} \rangle$, and the average deviation over all modes:
\begin{equation}
\delta_{{\rm k}} = \sqrt{\frac{1}{N_{\rm k}}\sum_{n, l}\frac{
\left (
\nu_{n, l} ^{\rm k}- \langle \nu_{n, l} \rangle
\right )^2}
{(\sigma_{n,l}^{\rm k})^{ 2}}},
\end{equation}
where, $\nu_{n, l}^{\rm k}$ and $\sigma_{n,l}^{\rm k}$ are the frequency and its uncertainty
returned by fitter k, and $N_{\rm k}$ is the number of modes fitted by the fitter k.  The normalised rms deviation $\delta_{{\rm k}}$ then provides  a way of assessing how far the fitter k deviates from the average value of the frequencies.

\subsection{Second step}
The goal of the second step is to provide the {\it selected} frequency set using the results of the first step.  The {\it selected} frequency set is then used for either subsequent modelling or fitting.  

\subsubsection{Method 1}
Peirce's criterion is not applied.  The {\it selected} mode set comprises the modes for which the frequencies of {\it all} fitters agree, within their own $1-\sigma$ error bars, with the {\it average} frequency set.  This set is supplemented by additional modes for which the frequencies of only a smaller group of the fitters agree, {within their own $1-\sigma$ error bars, with the {\it average} frequency set}.  The {\it selected} frequency set of the fitter with the smallest $\delta_{{\rm k}}$ is then designated the {\it reference} fit.  This method was used by \citet{Appourchaux2008} for HD49933.

The major drawback of this method is that if a single fitter disagrees, owing to the very small error bars of their measurements, then there is no {\it selected} set.  In addition, the good modes fitted by a single fitter are automatically rejected.

\subsubsection{Method 2}
Peirce's criterion is not applied.  Instead of having only one mode set, we derive {\it minimal} and {\it maximal} mode sets as follows.  The minimal mode set is, as previously, the one for which all fitters agree {within their own $1-\sigma$ error bars, with the {\it average} frequency set}.  The maximal mode set is made up of the frequencies for which {\it at least} two fitters agree {within their own $1-\sigma$ error bars, with the {\it average} frequency set}.  The frequency set of the fitter with the smallest $\delta_{{\rm k}}$ for the minimal mode set is then designated the {\it minimal} frequency set, and the frequency set of the fitter with the smallest $\delta_{{\rm k}}$ for the maximal mode set is designated the {\it maximal} frequency set.  This method was used by \citet{Metcalfe2010} for the {\it Kepler} star KIC 11026764.

There are several drawbacks to this method.  Firstly, the minimal and maximal frequency sets can be produced by different fitters.  This ``feature'' is a great nuisance when one derives different stellar models for the same star, that are the result of two different frequency sets.  In addition, the drawbacks of method 1 are not alleviated at all by this scheme, and the minimal set may not exist at all.

\subsubsection{Method 3}
This is the same as method 2 but with the Peirce's criterion applied.  The drawbacks are the same as method 2.

\subsubsection{Method 4}
Peirce's criterion is applied.  We still derive a minimal and a maximal mode set, but now, for deriving the minimal mode set, we use a voting scheme.  The minimal mode set contains the modes for which at least half the fitters agree {within their own $1-\sigma$ error bars, with the {\it average} frequency set}.  The maximal mode set contains the modes for which {\it at least} two fitters agree {within their own $1-\sigma$ error bars, with the {\it average} frequency set}.    The frequency set of the {\it best} fitter with the smallest $\delta_{{\rm k}}$ for the minimal mode set is then designated as the {\it best} frequency set.  The maximal and minimal frequency sets are then derived from the minimal and maximal mode sets of the {\it best}~fitter.  With such a scheme, one of the drawbacks of method 2 is removed: the two sets come from the same~{\it best}~fitter.  This fourth method was used by \citet{Mathur2011}, \citet{Campante2011}, and \citet{Howell2011}.  
The remaining drawback is that the good modes fitted by only one fitter are still rejected.


\begin{figure*}[!]
\center{
\hbox{
\hspace{-1.7cm}
\includegraphics[width=0.6\textwidth,angle=0]{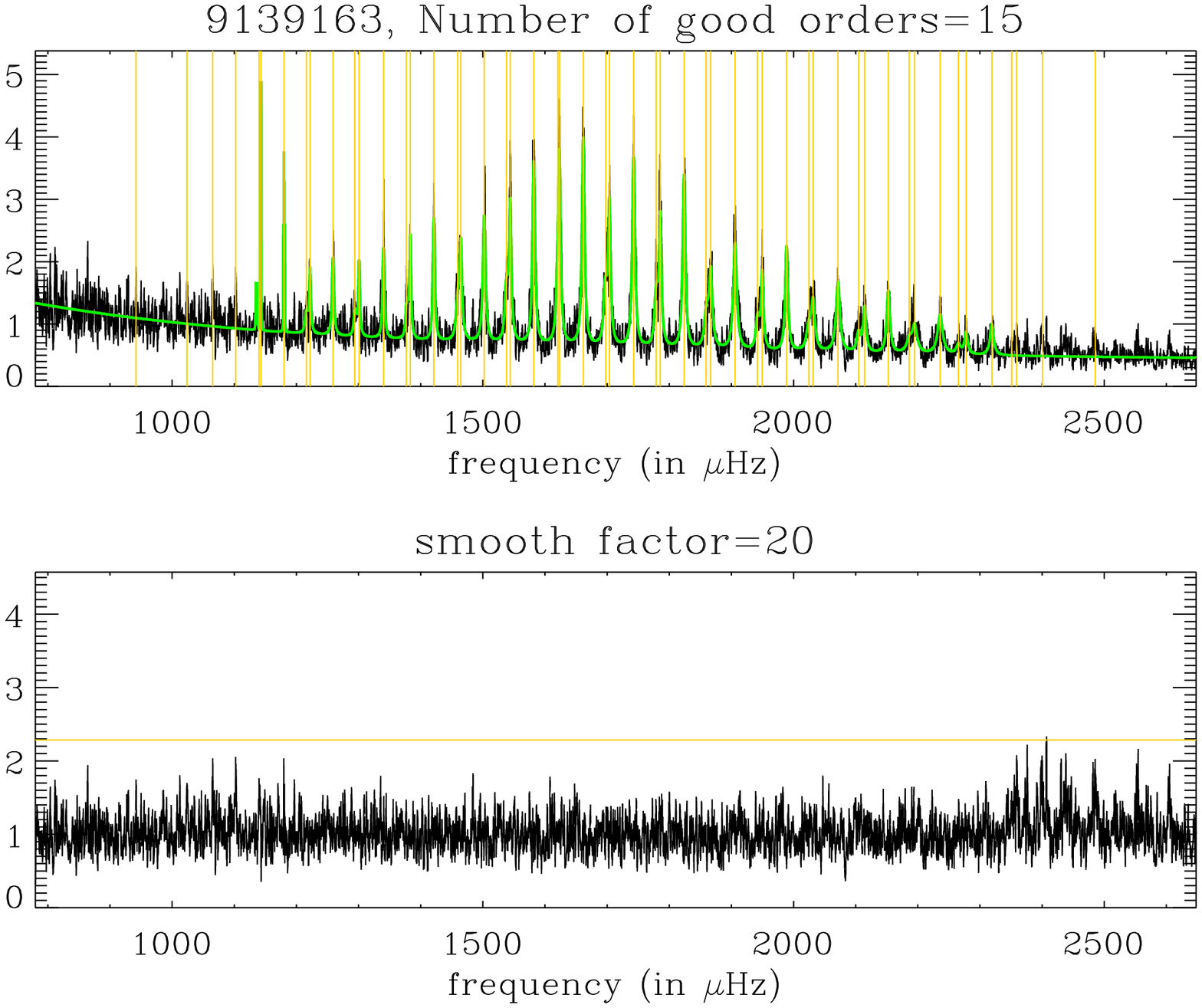}
\hspace{-1.7cm}
\includegraphics[width=0.6\textwidth,angle=0]{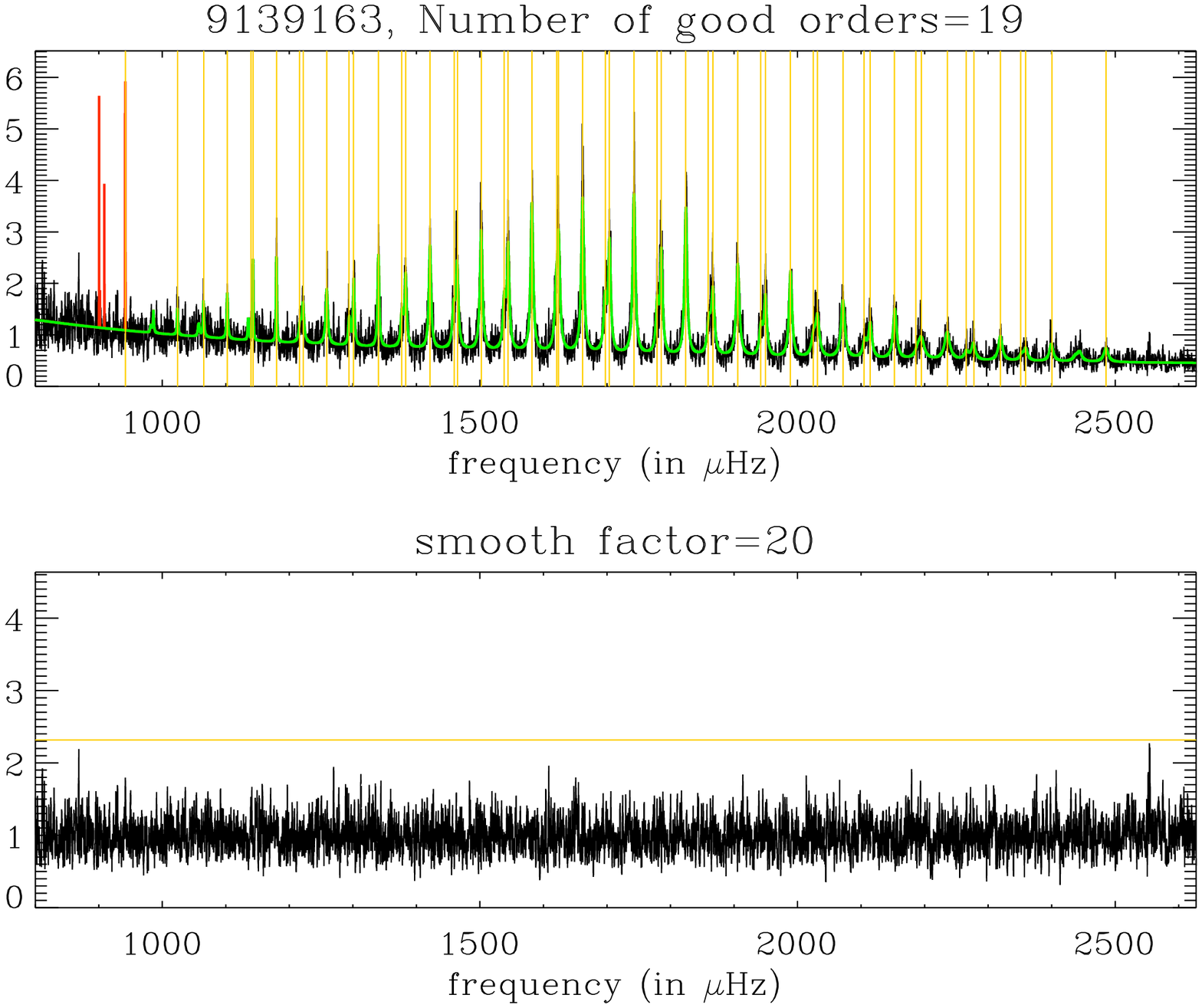}
}
}
 \caption[]{Detection of left-out modes. All power spectra are shown as a function of frequency smoothed over 20 bins.  Before the refit: (Top, left)  Power spectrum of KIC 9139163 with the fitted model with accepted modes (green) and with the guess frequencies provided by the fitters (orange lines).  (Bottom, left) Power spectrum normalised to the fitted model;  the orange horizontal line indicates the level at which the null hypothesis is rejected at the 1\% level.  After the refit: (Top, right) Power spectrum of KIC 9139163 with the fitted model for either accepted modes (green), rejected modes (red), or the guess frequencies provided a posteriori by the fitters (orange lines).  (Bottom, right) Power spectrum normalised to the fitted model;  the orange horizontal line indicates the level at which the null hypothesis is rejected at the 1\% level.}
\label{roc1}
\end{figure*}

\subsection{Third step: the final fitter}
The goal of the third step is to provide a single homogenous frequency set using the {\it maximal} frequency set as guess frequencies.  This third step was used by \citet{Metcalfe2012} for 16 Cyg A/B stars,  by \citet{Mathur2012} for 22 {\it Kepler} stars, and in this paper.

The current solution that we adopt here is to ensure that a {\it final} fitter refits the spectrum using the maximal mode set derived in method 4 of the Second Step, and also using some visual assessment and/or statistical tests where all the modes are included.  In this case, even the modes provided by only one fitter are included.  In addition, the solution of having a {\it final} fitter provide the frequency sets produces a homogenous data sets with systematic errors traceable to a single origin.  This is now the current strategy used to provide seismic parameters of the solar-like stars from the {\it Kepler} data.

After having applied the procedure described above, the 61 stars were fitted by 4 {\it final} fitters: BIR fitted 14 mixed-mode stars, IAS fitted 16 simple stars, QML fitted 13 simple and 3 mixed-mode stars, and MAR fitted the 15 F-like stars (see Table 1).  The division of the work for the MLE-based fit was based upon the availability of both computer and personal time.  As for the F-like stars, the work was performed by a single fitter for consistency.

\section{Product and quality assurance of the frequencies}
After producing the frequency sets for each star, we needed to assess whether the frequencies provided were of significant quality to be used in subsequent stellar modelling.
Several tests are at our disposal for gauging the likelihood of having either a good or bad mode.  Here the term {\it bad} refers to a mode that is obviously either a noise peak mistaken for a mode (for example, at low frequency where the mode linewidth is narrow) or to a broad spurious excess of power at high frequency (where the mode linewidth is large).  Statistical tests fall into two major categories: frequentist and Bayesian.  Both type of tests are addressed in \citet{Appourchaux2011}.

\subsection{Significance of fitted parameters}
To assess the significance of a given mode, one can use the likelihood ratio (LR) test, which is a frequentist test.  The test simply checks the likelihood that the fitted mode is due to spurious noise, {\em i.e.}, it tests whether the {\it null} or H$_{0}$ hypothesis can be accepted or rejected. The test consists of computing $\log(\Lambda)$, where $\Lambda$ is the ratio of the likelihood obtained when fitting $m$ parameters to the likelihood obtained when fitting $n$ parameters ($m>n$).  The test statistic $-2\log(\Lambda)$ is then known to be distributed as $\chi^2$ with ($m-n$) degrees of freedom (d.o.f) under the null hypothesis.

\subsubsection{Splitting and inclination angle}
The LR test can be simply applied to check the significance of the splitting and inclination angle, which are usually assumed to be common amongst the fitted orders.
The LR test is applied as follows:
\begin{enumerate}
\item We fit the whole spectrum with all the parameters apart from the angle and splitting (both set to 0).  We compute the likelihood of the fit.
\item We fit the whole spectrum with all the parameters, including the angle and splitting, and then compute the likelihood of the fit.
\item We compute LR($p$) ($p=2$).
\item We reject the pair (splitting,angle) at the 1\% level.
\end{enumerate}
This test was used to derive the frequencies provided in this paper. However, rotational splitting in these stars is the subject of another study, currently in progress.

\subsubsection{Orders and modes}
The major drawback of the LR test is that it requires a new fit each time we need to test the significance of several parameters.  In our case, since we fitted up to 80 parameters, it became impractical to  ``switch of'' each of the orders in turn, let alone each of the modes.  We devised a simplified LR test based on the assumption that some of the parameters are independent of each other.  For example, if we remove a given order then to a good approximation this will not affect the result of the other orders.  Therefore, we applied the simplified LR test as follows:
\begin{enumerate}
\item We fit the whole spectrum with all the parameters and compute likelihood of the fit.
\item We ``switch of'' orders (one at a time) and compute the new likelihood without making a new fit.
\item We compute LR($n$) ($n=5$, three mode frequency, one linewidth, one amplitude).
\item We reject order at the 1\% level.
\end{enumerate}
If the whole order is rejected, then we apply the same simplified LR test to each mode within it, with the same cut level ($n=1$, 1 frequency).


\begin{figure}[!]
\hspace{-0.8cm}
\includegraphics[width=0.4\textwidth,angle=90]{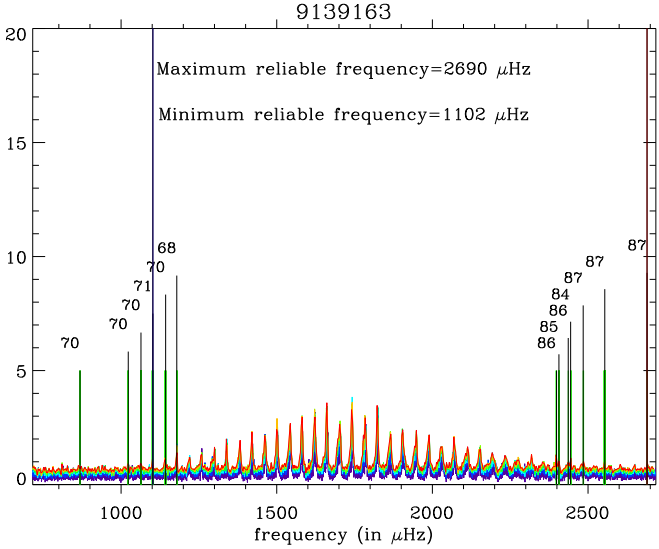}
 \caption[]{Smoothed power spectrum normalised to the detection $x_{\rm det}$ as a function of frequency for several smoothing factors (several colours; the redder the colour the higher the smoothing factor, which varies from 2 to 70 frequency bins).  The short vertical black lines show the minimum and maximum frequency detected with a single smoothing factor and their associated posterior probabilities ($p(x_{\rm det} | {\rm H}_1)$) for H$_{1}$, in \%.  When the frequency is detected several times to within 4~$\mu$Hz, only the highest posterior is shown.  The long vertical black lines show the minimum minimorum and maximum maximorum frequencies.}
\label{roc3}
\end{figure}

\subsection{Are all modes detected?}
When the power spectra were fitted, the frequencies of the modes were guessed using either a simplified detection test that looks for high peaks around the region of maximum mode power,
or a search for clumps of power in several adjacent or close bins.  We were also able to guess the mode frequencies by eye in the \'echelle diagram.  These statistical tests or visual tests
rely either on testing the null hypothesis (frequentist) or explicit knowledge of the mode frequency behaviour (Bayesian).  Hereafter, we detail both types of tests that were used
to ensure that all modes are detected.

\subsubsection{Frequentist test}
After performing the fit, we computed the ratio of the power spectrum to the fitted model.  Since the modes are stochastically excited harmonic oscillators, the resulting ratio was
simply the power spectrum of the forcing functions (mode excitation + noise), which is distributed as a $\chi^2$ distribution with two d.o.f.  The resulting ratio power spectrum could then be smoothed to detect signals that are spread over several bins.  The left-hand side of Fig.~\ref{roc1} shows the result of applying this procedure: power is clearly detected above 2300~$\mu$Hz.  The right-hand side of Figure~\ref{roc1} shows the result of applying the procedure after including the missing modes: nothing is detected in the smoothed ratio power spectrum above a frequency of 2300~$\mu$Hz.    In Fig.~\ref{roc1}, we note that the modes at low frequencies below 1100~$\mu$Hz are not detected by the mere application of a single smoothing factor.  We devised a more sensitive test that allows the detection of these modes at low frequency, but also the detection of modes at higher frequencies that are not provided by the fitters.

\subsubsection{Bayesian approach}
The previous test described in Sect. 6.2.1 is quite effective in rejecting the null hypothesis but fails to achieve what is required: to provide a quantitative likelihood that a mode has been detected.  To achieve this aim, we must instead use some a priori knowledge of the behaviour of the modes at low and high frequencies.  Here we used the work of \citet{Appourchaux2004} to detect short-lived modes.  This work had been applied to two CoRoT stars \citep{Appourchaux2009a, Deheuvels2010}.  We used a variant of the procedure that included a test based on a priori empirical knowledge of the mode linewidth at low and high frequency.   The procedure adopted is as follows:
\begin{enumerate}
\item We fit the power spectrum as the sum of a background made of a combination of one or two Lorentzian profiles centred at the zero frequency and white noise, with a Gaussian oscillation mode envelope.  The combination of the Lorentzian profiles and the white noise provides a model for the observed stellar background noise.
\item We compute the ratio of the power spectrum to the background but put the mode envelope to zero.  The signal-to-noise ratio of the modes of oscillation is then the observed power spectrum divided by the modelled background noise.  This ratio of the power spectrum to the background contains the modes of oscillation multiplied by the $\chi^2$ 2 d.o.f. forcing functions. 
\item We smooth the ratio power spectrum over $n$ bins to maximise the signal in power due to modes of oscillation that are distributed over many bins.  To maintain the scaling of the signal-to-noise ratio, the smoothed power spectrum must be multiplied by $n$.  The smoothing, of course, modifies the statistical distribution, and the modified distribution is known to be $\chi^2$ with  2$n$ d.o.f.
\item We accept or reject the H$_0$ hypothesis with a detection probability of $p_{\rm det}^{\rm win}$ over a window covering half the large separation (=$\Delta\nu_0/2$), taking into account that in each window the number of independent bins is $N_{\rm ind}=\Delta\nu_0 /(2 n {\delta \nu})$ where $\delta \nu$ is the frequency resolution of the original power spectrum.  The detection probability per independent bin is then $p_{\rm det}=p_{\rm det}^{\rm win}/N_{\rm ind}$.  The detection probability we used is $p_{\rm det}^{\rm win}=0.1$ or 10\%.
\item To determine the detection level $x_{\rm det}$, we compute the probability $p_{\rm det}$ of observing $x_{\rm det}$ or greater, which is $p_{\rm det}~=~\frac{1}{\Gamma(n)}\int_{x_{\rm det}}^{+\infty} u^{n-1} {\rm e}^{-u}{\rm d}u$ (where $\Gamma$ is the gamma function, and $u$ is a dummy symbol), and then solve this equation for $x_{\rm det}$ given $p_{\rm det}$.
\item In each window, we then select  the bins in the smoothed ratio power spectrum that are greater than $x_{\rm det}$, {\em i.e.}, we reject the H$_{0}$ hypothesis.
\end{enumerate}
This procedure from step 1 to step 6 is very similar to the one described in the previous section for the detection of left-out modes.  

Additional assumptions are now taken into account in the Bayesian approach.  If the H$_0$ hypothesis is rejected, we can write the detection likelihood, which is given by Eq.~(15) of  \citet{Appourchaux2009a},
\begin{equation}
p(x_{\rm det} | {\rm H}_0)=\frac{x_{\rm det}^{n-1}{\rm e}^{-x_{\rm det}}}{\Gamma(n)}.
\end{equation}
The next step is to derive $p(x_{\rm det} | {\rm H}_1)$ subject to the H$_1$ hypothesis, which is to assume that a true mode has been detected.  With the many stars at our disposal, we preferred to have an educated a priori knowledge of the mode height and mode linewidth, in lieu of their theoretical model parameters.  We can rewrite Eq.~(18) \citep{Appourchaux2009a} assuming uniform priors for the mode height and linewidth
\begin{equation}
p(x_{\rm det} | {\rm H}_1)=\frac{1}{h_{\rm u} W_{\rm u}} \int_0^{h_{\rm u}} \int_0^{W_{\rm u}}\frac{\lambda^{\nu}}{\Gamma(\nu)} x_{\rm det}^{\nu-1}{\rm e}^{-\lambda {x_{\rm det}}} {\rm d}h'{\rm d}W',
\end{equation}
where $\lambda$ and $\nu$ are the parameters defining the Gamma statistical distribution, $\lambda$ and $\nu$ are functions of $h'$ and $W'$ with $h'$ being the mode height in noise units, $W'$ is the mode linewidth (see Appendix A),  $h_{\rm u}$ is the maximum mode height, $W_{\rm u}$ is the maximum mode linewidth.   This formula is obviously more observer-oriented.  Since we wished to detect faint modes at either end of the spectrum, we assumed that the maximum mode height would be no larger than twice the noise, and that the mode linewidth no larger than 1~$\mu$Hz, at low frequency and 10~$\mu$Hz at high frequency.  We then computed the posterior probability of H$_{1}$ as
\begin{equation}
p({\rm H}_{1} | x_{\rm det})= \left(1+\frac{p(x_{\rm det} | {\rm H}_0)}{p(x_{\rm det} | {\rm H}_1)}\right)^{-1}.
\label{posterior}
\end{equation}
The procedure then continues with the steps:
\begin{enumerate}[resume]
\item From all the frequencies for which the null hypothesis was rejected, we keep the highest and lowest detected frequencies
\item We then compute the posterior probability of H$_{1}$ as given by Eq.~(\ref{posterior}) for these two extreme frequencies.
\end{enumerate}
The steps 1 to 8 are repeated for a range of smoothing factors $n$ from 2 to 70, corresponding to the resolution of 0.08~$\mu$Hz to 2.8~$\mu$Hz.  The variable amount of smoothing allows
for the detection probability depending on both the smoothing factor and the mode linewidth.  The maximum detection probability is reached for different values of the smoothing factor, depending on the mode linewidth.  

We then defined the maximum maximorum detectable frequency as the highest detected frequency for which the posterior probability is either greater than 90\% or has the highest value if this is lower than 90\%.  The minimum minimorum frequency was defined as the lowest detected frequency for which the posterior probability is either greater than 90\% or has the highest value if this is lower than 90\%.  Figure~\ref{roc3} provides an example of the application of the procedure.  

\begin{figure*}
\hbox{
\hspace{-0.8cm}
\includegraphics[width=0.4\textwidth,angle=90]{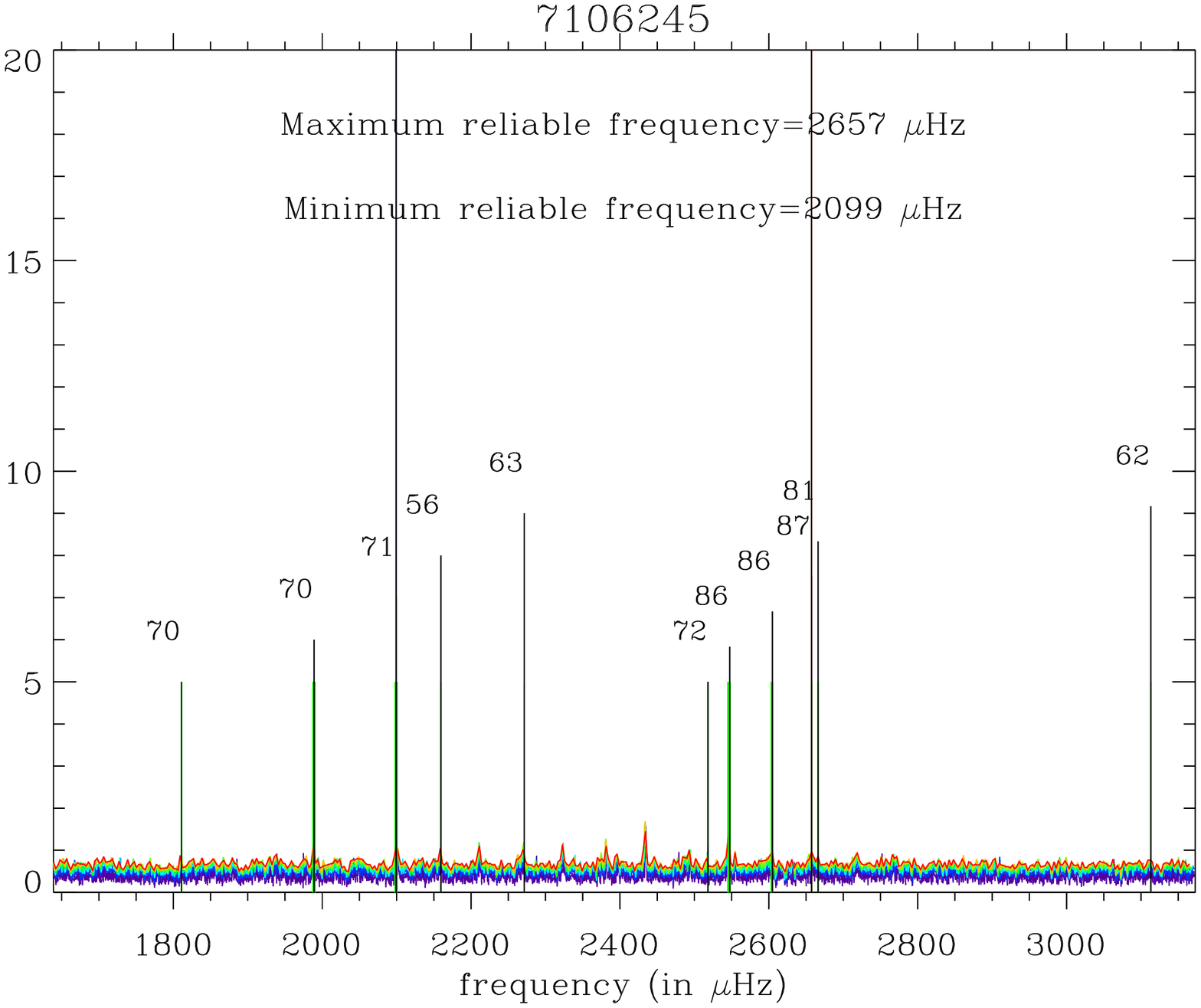}
\hspace{-0.8cm}
\includegraphics[width=0.4\textwidth,angle=90]{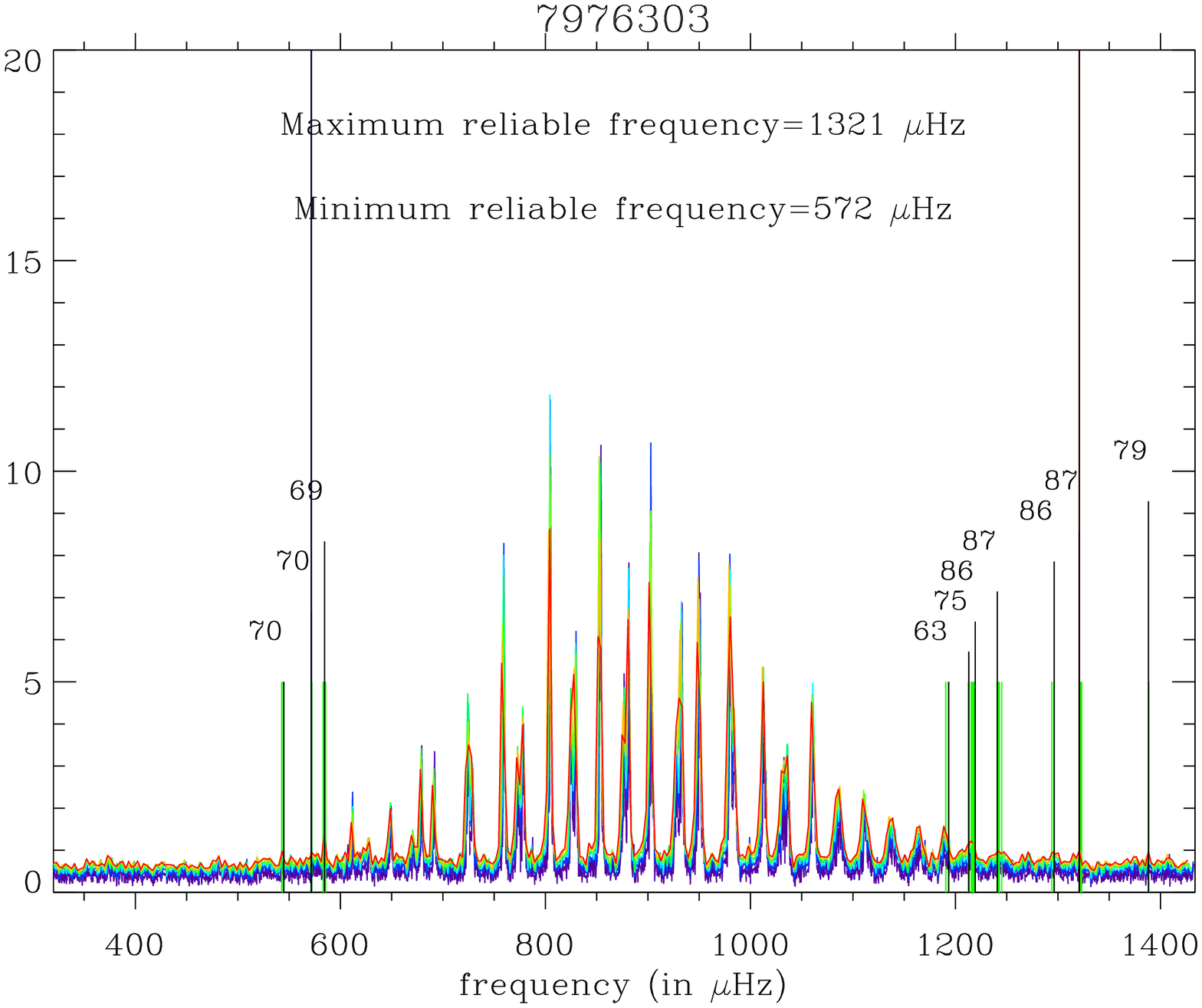}
}
\hbox{
\hspace{-0.8cm}
\includegraphics[width=0.4\textwidth,angle=90]{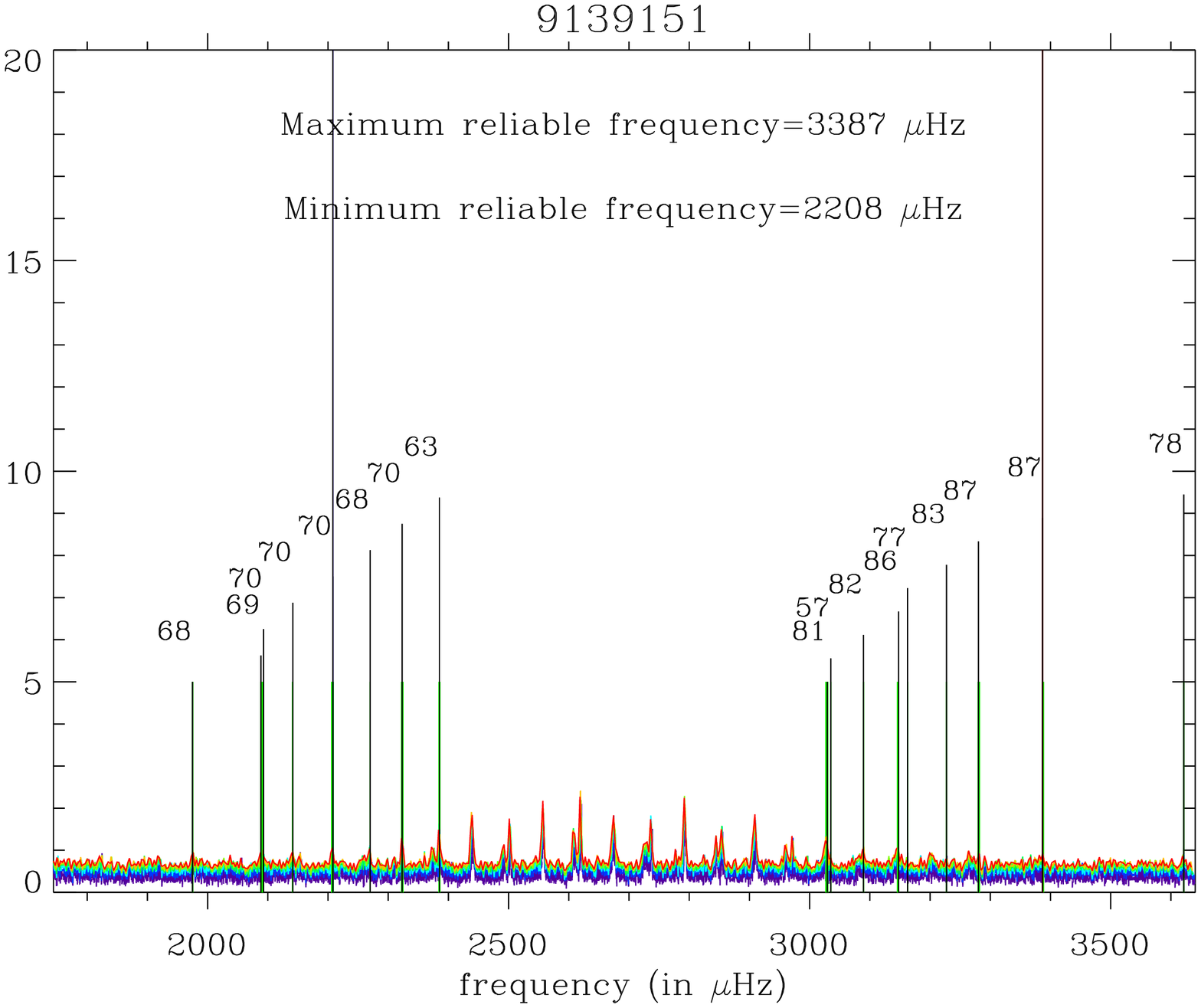}
\hspace{-0.8cm}
\includegraphics[width=0.4\textwidth,angle=90]{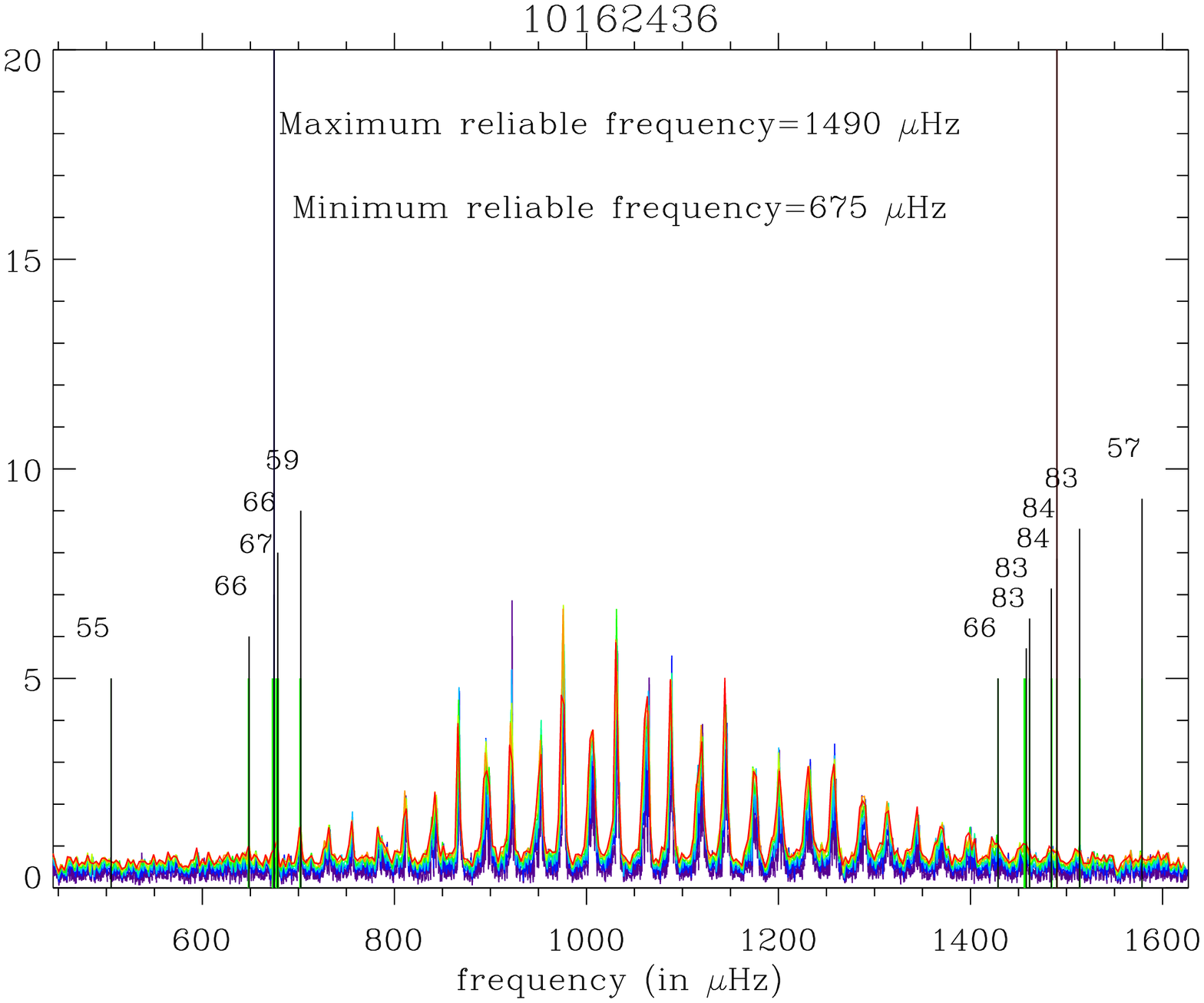}
}
 \caption[]{Smoothed power spectrum normalised to the detection $x_{\rm det}$ as a function of frequency for several smoothing factor values (several colours; the redder and the higher the smoothing factor) for four stars (KIC 7106245, KIC 7976303, KIC 9139151, KIC 10162436).  The short vertical black lines show the minimum and maximum frequency detected with a single smoothing factor with their associated posterior probability ($p(x_{\rm det} | {\rm H}_1)$) for H$_{1}$, in \%.  When the frequency is detected several times within 4~$\mu$Hz, only the highest posterior is shown.  The long vertical black lines show the minimum minimorum and maximum maximorum frequencies.}
\label{quality1}
\end{figure*}

Table~\ref{9139163} shows an example of an application of the quality assurance test to the fitted frequencies, which can be compared with Figure~\ref{roc1}.  Figure~\ref{quality1} shows more examples of the application of the procedure resulting in different cases of {\it no detection}, {\it no fit}, and {\it detection with a posterior probability less than 90\%}.   For the star KIC 7106245, several modes were {\it not detected}, as listed in Table~\ref{7106245}.  For the star KIC 7976303, several modes were {\it not fitted}, as listed in Table~\ref{7976303}.  For the stars KIC 9139151 and KIC 10162436, a few modes were either {\it not detected} or {\it not fitted} or {\it detected with a posterior probability lower than 90\%}, as listed in Table~\ref{9139151} and \ref{10162436}, respectively.

The procedure was applied to all stars in this paper irrespective of the {\it final} fitter for providing a quality assessment at all the frequencies.  If one of the frequencies from the test was not provided by the fitter, we verified whether it could be a mode close to an integer value of the large separation, and then made the most likely degree identification.  

The application of the product assurance resulted in 61 tables of frequencies.  The mode frequencies of 61 stars are provided in Tables 3 to 7 with the paper, and Tables A.1 to A.56 as online material.  All modellers are advised to use all of the mode frequencies labelled {\it OK} and to use with caution all other frequencies.  In addition, the 61 \'echelle diagrams are shown in Figs A.1 to A.10, and also as online material.


\section{Conclusions}
We have analysed the oscillation power spectra of 61 main sequence and subgiant stars for which we fitted the p-mode parameters.  We have divided the stars into three categories related to the visual appearance of their \'echelle diagrams called simple, F-like, and mixed-mode stars.  We have shown that we are now able to perform {\it nearly} automated fits of many stellar power spectra derived from {\it Kepler} light curves.  There are two steps that still require manual intervention:  the identification of the star category provided by the \'echelle diagram and the derivation of first-guess frequencies.  In the future, we plan to reduce the amount of manual intervention by using the areas delimited in the ($\Delta\nu$, $T_{\rm eff}$) diagram of Fig.~\ref{fig-hr} to identify the star category; and by using the asymptotic relation of frequencies used by \citet{Benomar2011} to derive automated guess frequencies.

We devised a procedure to use the mode frequencies from several fitters to choose a single fitter to re-fit all the spectra (within workload constraints).  When the power spectra are fitted, we are now also able to make an automated assessment of the fit quality and the mode frequencies obtained; we give several techniques for this assessment.  We provide the \'echelle diagrams of 61 stars and the associated list of mode frequencies for these stars.



\begin{acknowledgements}
The authors wish to thank the entire {\it Kepler} team, without whom
these results would not have been possible. Funding for this Discovery mission is provided by NASA's
Science Mission Directorate.  We also thank all funding
councils and agencies that have supported the activities
of KASC Working Group~1, as well as the International Space Science
Institute (ISSI).  TA gratefully acknowledges the financial support of the Centre National d'Etudes Spatiales (CNES) under a PLATO grant.  TA acknowledges the KITP staff of UCSB for their hospitality
during the research programme ``Asteroseismology in the Space Age". This research was supported in part by the National Science Foundation under Grant No. NSF PHY05-51164.  A special thanks to my wife for having made this paper possible, needless to say that Kirby Cove is in our minds.  TLC acknowledges financial support from project PTDC/CTE-AST/098754/2008 funded by FCT/MCTES, Portugal.  WJC, GAV and YE acknowledge financial support from the UK Science and Technology Facilities Council (STFC).  Funding for the Stellar Astrophysics Centre is provided by The Danish National Research Foundation. The research is supported by the ASTERISK project (ASTERoseismic Investigations with SONG and Kepler) funded by the European Research Council (Grant agreement no.: 267864).  RAG and GRD has received funding from the European Community's Seventh Framework Programme (FP7/2007-2013) under grant agreement no. 269194.  MG received financial support from an NSERC Vanier scholarship. This work employed computational facilities provided by ACEnet, the regional high performance computing consortium for universities in Atlantic Canada.  SH acknowledges funding from the Nederlandse Organisatie voor Wetenschappelijk Onderzoek (NWO). GH acknowledges support by the Austrian Science Fund (FWF) project P21205-N16.  RH acknowledges computing support from the National Solar Observatory.  DS acknowledges the financial support  from CNES.  NCAR is partially supported by the National Science Foundation.  The authors thanks an anonymous referee for contributing to the clarity of the paper.
\end{acknowledgements}

\bibliographystyle{aa}
\bibliography{thierrya}

\begin{table*}[!]
\caption{Table of key stellar parameters.  The first three columns provide the KIC, HIP, and HD numbers.  The fourth column provides the effective temperature and its error bar taken from \citet{Pins2011}.  The fifth column provides the {\it Kepler} magnitude.  The sixth column gives the median of the large separation as measured using the mode frequencies given in this paper.  The seventh column provides the frequency of the maximum of oscillation power.  The eighth column gives the number of detected modes.  The penultimate column gives the star category.  The last column provides notes on whether the star has been previously observed or has mixed modes.}
\centering
\begin{tabular}{r r r c r r r c c c} 
\hline
\hline
KIC&HIP&HD&$T_{\rm eff}$ (K)&Kp& $\Delta\nu$ ($\mu$Hz)&$\nu_{\rm max}$($\mu$Hz)& Number of modes &Star category&Notes\\
\hline
\hline
1435467&-&-&        6541 $\pm$         126&8.9&70.9&1324&45&F-like&\\
2837475&-&-&        6710 $\pm$          61&8.5&76.0&1522&45&F-like&\\
3424541&-&-&        6460 $\pm$          55&9.7&41.3&712&37&F-like&\\
3427720&-&-&        5970 $\pm$          52&9.1&120.1&2574&30&simple&\\
3632418&94112&179070&        6235 $\pm$          70&8.2&60.7&1084&34&simple&c, d\\
3733735&94071&178971&        6720 $\pm$          56&8.4&92.4&2041&43&F-like&\\
3735871&-&-&        6220 $\pm$          61&9.7&123.1&2633&25&F-like&\\
5607242&-&-&        5680 $\pm$          51&10.7&40.6&610&39&mixed modes&$\dagger$\\
5955122&-&-&        5890 $\pm$          51&9.3&49.6&826&38&mixed modes&$\dagger$\\
6116048&-&-&        6020 $\pm$          51&8.4&100.7&2020&34&simple&c\\
6508366&-&-&        6480 $\pm$          56&9.0&51.6&959&45&F-like&\\
6603624&-&-&        5610 $\pm$          51&9.1&110.3&2339&31&simple&b, c\\
6679371&-&-&        6590 $\pm$          56&8.7&50.4&908&44&F-like&\\
6933899&-&-&        5820 $\pm$          50&9.6&72.1&1362&33&simple&c\\
7103006&-&-&        6390 $\pm$          56&8.9&59.9&1072&46&F-like&\\
7106245&-&-&        6020 $\pm$          51&10.8&111.6&2323&16&simple&\\
7206837&-&-&        6360 $\pm$          56&9.8&79.0&1556&40&simple&\\
7341231&-&-&        5091 $\pm$          91&9.9&29.0&404&44&mixed modes&e, $\dagger$\\
7747078&91918&-&        5910 $\pm$          70&9.5&53.7&977&37&mixed modes&d, $\dagger$\\
7799349&-&-&        5050 $\pm$          45&9.5&33.2&560&40&mixed modes&$\dagger$\\
7871531&-&-&        5390 $\pm$          47&9.3&151.3&3344&25&simple&\\
7976303&-&-&        6260 $\pm$          51&9.0&51.3&826&36&mixed modes&$\dagger$\\
8006161&91949&-&        5300 $\pm$          46&7.4&149.4&3444&26&simple&c\\
8026226&-&-&        6280 $\pm$          52&8.4&34.4&520&40&mixed modes&$\dagger$\\
8228742&95098&-&        6080 $\pm$          51&9.4&62.1&1153&33&simple&c\\
8379927&97321&187160&        5990 $\pm$          52&7.0&120.4&2669&37&simple&\\
8394589&-&-&        6210 $\pm$          52&9.5&109.4&2336&32&simple&\\
8524425&-&-&        5660 $\pm$          51&9.7&59.7&1078&33&mixed modes&$\dagger$\\
8694723&-&-&        6310 $\pm$          56&8.9&75.1&1384&49&simple&\\
8702606&-&-&        5500 $\pm$          51&9.3&39.7&626&40&mixed modes&$\dagger$\\
8760414&-&-&        5795 $\pm$          70&9.6&117.4&2349&31&simple&c, d\\
9025370&-&-&        5660 $\pm$          52&8.8&132.8&2864&23&simple&\\
9098294&-&-&        5960 $\pm$          51&9.8&109.1&2241&27&simple&\\
9139151&92961&-&        6090 $\pm$          52&9.2&117.0&2610&34&simple&\\
9139163&92962&-&        6370 $\pm$          56&8.3&81.4&1649&55&simple&\\
9206432&93607&-&        6470 $\pm$          56&9.1&85.1&1822&51&F-like&\\
9410862&-&-&        6180 $\pm$          51&10.7&107.5&2034&24&simple&\\
9574283&-&-&        5440 $\pm$          47&10.7&29.7&448&32&mixed modes&$\dagger$\\
9812850&-&-&        6380 $\pm$          55&9.5&65.3&1186&41&F-like&\\
9955598&-&-&        5450 $\pm$          47&9.4&153.1&3379&26&simple&\\
10018963&-&-&        6230 $\pm$          52&8.7&55.5&988&41&simple&c\\
10162436&97992&-&        6320 $\pm$          53&8.6&55.9&1004&45&simple&\\
10355856&-&-&        6540 $\pm$          56&9.2&68.3&1210&40&F-like&\\
10454113&92983&-&        6246 $\pm$          58&8.6&105.2&2313&43&simple&\\
10644253&-&-&        6020 $\pm$          51&9.2&123.0&2866&27&simple&\\
10909629&-&-&        6490 $\pm$          61&10.9&49.7&813&35&F-like&\\
10963065&-&-&        6280 $\pm$          51&8.8&102.9&2071&34&simple&c\\
11026764&-&-&        5710 $\pm$          51&9.3&50.4&885&30&mixed modes&a, b, $\dagger$\\
11081729&-&-&        6600 $\pm$          62&9.0&90.2&1820&45&F-like&\\
11193681&-&-&        5690 $\pm$          51&10.7&42.9&752&32&mixed modes&$\dagger$\\
11244118&-&-&        5620 $\pm$          51&9.7&71.4&1352&31&simple&c\\
11253226&97071&186700&        6690 $\pm$          56&8.4&76.9&1637&30&F-like&\\
11395018&-&-&        5640 $\pm$          51&10.8&47.7&840&26&mixed modes&$\dagger$\\
11414712&-&-&        5660 $\pm$          51&8.5&44.1&730&44&mixed modes&$\dagger$\\
11717120&-&-&        5220 $\pm$          48&9.3&37.6&555&42&mixed modes&$\dagger$\\
11771760&-&-&        6030 $\pm$          51&11.4&32.2&505&37&mixed modes&$\dagger$\\
11772920&-&-&        5420 $\pm$          51&9.7&157.6&3439&23&simple&\\
12009504&-&-&        6230 $\pm$          51&9.3&88.1&1768&34&simple&c\\
12258514&95568&183298&        5950 $\pm$          51&8.1&74.8&1449&34&simple&c\\
12317678&97316&-&        6540 $\pm$          55&8.7&64.1&1162&51&F-like&\\
12508433&-&-&        5280 $\pm$          47&9.5&45.0&758&36&mixed modes&$\dagger$\\
\hline
\end{tabular}
\begin{list}{}{}
\item[$^{\rm a,b,c}$] Already analysed by \citet{Metcalfe2010}, \citet{Chaplin2011}, \citet{Mathur2012}, respectively.
\item[$^{\rm d, e}$] Effective temperature from \citet{Bruntt2011} and \citet{Casa2010}, respectively
\item[$\dagger$] Star with mixed modes for which $\Delta\nu$ was derived from the $l=0$ modes only
\end{list}
\end{table*}

\begin{table*}[!]
\caption{Frequencies for KIC 9139163. The first column is the degree.  The second column is the frequency.  The third column is the 1-$\sigma$ uncertainty quoted when the mode is fitted.  The last column provides an indication of the quality of the detection: {\it OK} indicates that the mode was correctly detected and fitted; {\it Not detected} indicates that the mode was fitted but not detected by the quality assurance test and {\it Not fitted} indicates that the mode was detected with a posterior probability provided by the quality assurance test.  When an uncertainty {\it and} a posterior probability are quoted, it means that the mode is fitted but detected using the quality assurance test with a probability lower than 90\%.}
\centering
\begin{tabular}{c c c c} 
\hline
\hline
Degree&Frequency ($\mu$Hz)&1-$\sigma$ error ($\mu$Hz)&Comment\\
\hline
\hline
       0&986.105&1.130&Not detected\\
       0&1064.982&0.690&0.703\\
       0&1142.941&0.230&OK\\
       0&1221.476&0.544&OK\\
       0&1301.395&0.332&OK\\
       0&1383.093&0.366&OK\\
       0&1464.189&0.381&OK\\
       0&1544.456&0.317&OK\\
       0&1623.952&0.380&OK\\
       0&1703.1000&0.340&OK\\
       0&1785.675&0.330&OK\\
       0&1866.729&0.420&OK\\
       0&1949.424&0.391&OK\\
       0&2031.407&0.706&OK\\
       0&2114.451&0.607&OK\\
       0&2195.335&1.219&OK\\
       0&2276.836&0.928&OK\\
       0&2359.243&1.229&OK\\
       0&2444.022&1.734&OK\\
       0&2689.590&Not fitted&0.873\\
\hline
       1&1023.888&0.576&0.705\\
       1&1102.258&0.427&OK\\
       1&1179.797&0.181&OK\\
       1&1258.873&0.451&OK\\
       1&1340.250&0.280&OK\\
       1&1421.435&0.305&OK\\
       1&1502.036&0.292&OK\\
       1&1581.801&0.270&OK\\
       1&1661.689&0.269&OK\\
       1&1742.582&0.261&OK\\
       1&1824.248&0.271&OK\\
       1&1905.932&0.376&OK\\
       1&1989.005&0.401&OK\\
       1&2071.453&0.469&OK\\
       1&2153.260&0.526&OK\\
       1&2235.598&0.876&OK\\
       1&2319.330&0.806&OK\\
       1&2399.494&0.844&OK\\
       1&2485.579&1.291&OK\\
       1&2553.800&Not fitted&0.866\\
\hline
       2&982.173&1.756&Not detected\\
       2&1057.448&0.955&Not detected\\
       2&1135.556&3.432&OK\\
       2&1216.915&1.141&OK\\
       2&1294.411&0.609&OK\\
       2&1377.256&0.924&OK\\
       2&1458.740&0.863&OK\\
       2&1538.360&0.805&OK\\
       2&1620.012&1.032&OK\\
       2&1696.864&0.684&OK\\
       2&1779.203&0.601&OK\\
       2&1860.596&1.060&OK\\
       2&1942.470&0.960&OK\\
       2&2025.300&1.424&OK\\
       2&2105.809&1.195&OK\\
       2&2189.691&2.425&OK\\
       2&2265.891&1.291&OK\\
       2&2352.782&3.939&OK\\
       2&2437.430&2.378&OK\\
\hline
\hline
\end{tabular}
\label{9139163}
\end{table*}	
\begin{table*}[!]
\caption{Frequencies for KIC 7106245. The first column is the degree.  The second column is the frequency.  The third column is the 1-$\sigma$ uncertainty quoted when the mode is fitted.  The last column provides an indication of the quality of the detection: {\it OK} indicates that the mode was correctly detected and fitted; {\it Not detected} indicates that the mode was fitted but not detected by the quality assurance test and {\it Not fitted} indicates that the mode was detected with a posterior probability provided by the quality assurance test.  When an uncertainty {\it and} a posterior probability are quoted, it means that the mode is fitted but detected using the quality assurance test with a probability lower than 90\%.}
\centering
\begin{tabular}{c c c c} 
\hline
\hline
Degree&Frequency ($\mu$Hz)&1-$\sigma$ error ($\mu$Hz)&Comment\\
\hline
\hline
       0&1718.954&0.529&Not detected\\
       0&1939.538&0.042&Not detected\\
       0&2049.668&0.383&Not detected\\
       0&2159.780&0.251&OK\\
       0&2271.402&0.207&OK\\
       0&2382.167&0.235&OK\\
       0&2494.757&0.336&OK\\
       0&2605.011&0.688&OK\\
\hline
       1&1770.100&0.248&Not detected\\
       1&1989.784&0.038&0.703\\
       1&2100.640&0.426&OK\\
       1&2211.749&0.270&OK\\
       1&2323.982&0.281&OK\\
       1&2434.871&0.192&OK\\
       1&2546.874&0.278&OK\\
       1&2659.004&0.864&OK\\
\hline
       2&1707.675&0.600&Not detected\\
       2&1931.915&0.041&Not detected\\
       2&2043.194&0.678&Not detected\\
       2&2152.546&0.441&OK\\
       2&2264.302&0.380&OK\\
       2&2375.609&0.398&OK\\
       2&2487.192&1.003&OK\\
       2&2598.758&1.726&OK\\
\hline
\hline
\end{tabular}
\label{7106245}
\end{table*}
\begin{table*}[!]
\caption{Frequencies for KIC 7976303. The first column is the degree.  The second column is the frequency.  The third column is the 1-$\sigma$ uncertainty quoted when the mode is fitted.  The last column provides an indication of the quality of the detection: {\it OK} indicates that the mode was correctly detected and fitted; {\it Not detected} indicates that the mode was fitted but not detected by the quality assurance test and {\it Not fitted} indicates that the mode was detected with a posterior probability provided by the quality assurance test.  When an uncertainty {\it and} a posterior probability are quoted, it means that the mode is fitted but detected using the quality assurance test with a probability lower than 90\%.}
\centering
\begin{tabular}{c c c c} 
\hline
\hline
Degree&Frequency ($\mu$Hz)&1-$\sigma$ error ($\mu$Hz)&Comment\\
\hline
\hline
       0&571.580&Not fitted&0.700\\
       0&628.566&0.283&OK\\
       0&679.655&0.119&OK\\
       0&729.203&0.173&OK\\
       0&778.696&0.173&OK\\
       0&830.027&0.142&OK\\
       0&881.747&0.136&OK\\
       0&933.089&0.153&OK\\
       0&985.019&0.242&OK\\
       0&1036.294&0.263&OK\\
       0&1087.945&0.403&OK\\
       0&1140.957&1.088&OK\\
       0&1193.380&Not fitted&0.628\\
       0&1240.840&Not fitted&0.869\\
       0&1296.350&Not fitted&0.864\\
\hline
       1&544.749&Not fitted&0.697\\
       1&571.580&Not fitted&0.700\\
       1&585.461&0.130&OK\\
       1&612.258&0.192&OK\\
       1&649.628&0.210&OK\\
       1&692.135&0.145&OK\\
       1&724.920&0.314&OK\\
       1&759.883&0.098&OK\\
       1&804.981&0.109&OK\\
       1&853.894&0.105&OK\\
       1&903.255&0.127&OK\\
       1&950.341&0.140&OK\\
       1&980.670&0.255&OK\\
       1&1013.362&0.163&OK\\
       1&1061.529&0.178&OK\\
       1&1112.412&0.323&OK\\
       1&1165.116&0.506&OK\\
       1&1213.050&Not fitted&0.748\\
       1&1219.290&Not fitted&0.862\\
       1&1320.830&Not fitted&0.869\\
\hline
       2&571.580&Not fitted&0.700\\
       2&622.955&0.694&OK\\
       2&725.384&0.659&OK\\
       2&773.602&0.193&OK\\
       2&825.606&0.143&OK\\
       2&877.049&0.165&OK\\
       2&928.254&0.217&OK\\
       2&980.674&0.625&OK\\
       2&1031.753&0.385&OK\\
       2&1084.067&0.719&OK\\
       2&1137.327&0.634&OK\\
       2&1240.840&Not fitted&0.869\\
       2&1388.240&Not fitted&0.786\\
\hline
\hline
\end{tabular}
\label{7976303}
\end{table*}	
\begin{table*}[!]
\caption{Frequencies for KIC 9139151. The first column is the degree.  The second column is the frequency.  The third column is the 1-$\sigma$ uncertainty quoted when the mode is fitted.  The last column provides an indication of the quality of the detection: {\it OK} indicates that the mode was correctly detected and fitted; {\it Not detected} indicates that the mode was fitted but not detected by the quality assurance test and {\it Not fitted} indicates that the mode was detected with a posterior probability provided by the quality assurance test.  When an uncertainty {\it and} a posterior probability are quoted, it means that the mode is fitted but detected using the quality assurance test with a probability lower than 90\%.}
\centering
\begin{tabular}{c c c c} 
\hline
\hline
Degree&Frequency ($\mu$Hz)&1-$\sigma$ error ($\mu$Hz)&Comment\\
\hline
\hline
       0&2038.621&0.910&Not detected\\
       0&2154.355&0.498&Not detected\\
       0&2269.980&0.382&OK\\
       0&2385.860&0.317&OK\\
       0&2502.911&0.272&OK\\
       0&2620.348&0.219&OK\\
       0&2737.332&0.295&OK\\
       0&2855.191&0.380&OK\\
       0&2972.734&0.355&OK\\
       0&3090.440&0.947&OK\\
       0&3205.579&1.475&OK\\
\hline
       1&1976.567&0.588&0.679\\
       1&2091.662&0.916&0.686\\
       1&2208.146&0.456&OK\\
       1&2324.004&0.363&OK\\
       1&2440.178&0.285&OK\\
       1&2557.906&0.254&OK\\
       1&2675.122&0.306&OK\\
       1&2793.160&0.247&OK\\
       1&2909.913&0.353&OK\\
       1&3028.277&0.428&OK\\
       1&3146.702&0.893&OK\\
       1&3266.749&1.555&OK\\
\hline
       2&2027.493&1.783&Not detected\\
       2&2142.205&0.686&0.698\\
       2&2257.832&1.365&OK\\
       2&2374.529&1.419&OK\\
       2&2492.946&0.519&OK\\
       2&2609.584&0.425&OK\\
       2&2728.805&1.021&OK\\
       2&2845.599&0.513&OK\\
       2&2961.734&0.592&OK\\
       2&3081.247&2.382&OK\\
       2&3200.368&2.253&OK\\
\hline
\hline
\end{tabular}
\label{9139151}
\end{table*}
\begin{table*}[!]
\caption{Frequencies for KIC 10162436. The first column is the degree.  The second column is the frequency.  The third column is the 1-$\sigma$ uncertainty quoted when the mode is fitted.  The last column provides an indication of the quality of the detection: {\it OK} indicates that the mode was correctly detected and fitted; {\it Not detected} indicates that the mode was fitted but not detected by the quality assurance test and {\it Not fitted} indicates that the mode was detected with a posterior probability provided by the quality assurance test.  When an uncertainty {\it and} a posterior probability are quoted, it means that the mode is fitted but detected using the quality assurance test with a probability lower than 90\%.}
\centering
\begin{tabular}{c c c c} 
\hline
\hline
Degree&Frequency ($\mu$Hz)&1-$\sigma$ error ($\mu$Hz)&Comment\\
\hline
\hline
       0&505.001&Not fitted&0.553\\
       0&623.951&0.439&Not detected\\
       0&678.865&0.224&OK\\
       0&733.125&0.323&OK\\
       0&788.276&0.619&OK\\
       0&844.092&0.284&OK\\
       0&898.933&0.380&OK\\
       0&953.132&0.240&OK\\
       0&1008.093&0.276&OK\\
       0&1064.704&0.260&OK\\
       0&1121.035&0.287&OK\\
       0&1176.898&0.478&OK\\
       0&1233.595&0.372&OK\\
       0&1290.285&0.526&OK\\
       0&1344.962&1.219&OK\\
       0&1401.148&3.386&OK\\
       0&1458.427&1.058&OK\\
       0&1513.630&Not fitted&0.832\\
\hline
       1&596.328&0.558&Not detected\\
       1&649.673&0.445&0.658\\
       1&702.876&0.251&OK\\
       1&756.638&0.266&OK\\
       1&812.507&0.254&OK\\
       1&867.878&0.166&OK\\
       1&922.993&0.219&OK\\
       1&976.540&0.171&OK\\
       1&1032.384&0.187&OK\\
       1&1088.880&0.207&OK\\
       1&1145.493&0.232&OK\\
       1&1201.864&0.345&OK\\
       1&1258.261&0.304&OK\\
       1&1314.613&0.448&OK\\
       1&1370.593&0.655&OK\\
       1&1427.560&1.164&OK\\
       1&1485.568&1.553&OK\\
\hline
       2&505.001&Not fitted&0.553\\
       2&565.319&1.059&Not detected\\
       2&618.665&1.674&Not detected\\
       2&675.087&0.366&OK\\
       2&728.565&0.514&OK\\
       2&783.952&0.434&OK\\
       2&839.653&0.385&OK\\
       2&895.445&0.344&OK\\
       2&947.503&0.449&OK\\
       2&1004.016&0.336&OK\\
       2&1061.103&0.492&OK\\
       2&1116.703&0.401&OK\\
       2&1174.210&0.674&OK\\
       2&1229.375&0.505&OK\\
       2&1286.456&0.770&OK\\
       2&1345.021&2.270&OK\\
       2&1399.581&7.090&OK\\
       2&1453.386&2.094&OK\\
       2&1513.630&Not fitted&0.832\\
\hline
\hline
\end{tabular}
\label{10162436}
\end{table*}

\appendix

\section{Expressions for $\lambda$ and $\nu$}
The expressions that we use to derive $\lambda$ and $\nu$ are related to the power spectrum, which is binned over $n$ bins.  The approximation of the probability density function of the mode profile in a binned power spectrum is therefore related to the mean power in a mode and its rms deviation.  This approximation is given in more detail in \citet{Appourchaux2004}.  The mean power in a mode of the binned power spectrum ${\cal S}$ is given by
\begin{equation}
	E[{\cal S}]=\sum_{i=1}^{i=n} f(\nu_i)
\end{equation}
\begin{equation}
	\sigma=\sqrt{\sum_{i=1}^{i=n} f(\nu_i)^2}
\end{equation}
where $f$ is the mode profile given at frequency $\nu$ by
\begin{equation}
	f(\nu)=\frac{h}{1+\frac{4\nu^2}{W^2}}+1
\end{equation}
and $\nu$ is the frequency measured with respect to the central mode frequency, which is omitted, $h$ is the mode height in units of the background noise (hence the additional 1), and $W$ is the mode linewidth.  The summation is done over frequency $\nu_i$ given by
\begin{equation}
	\nu_i=\left(\frac{i-1}{n-1}-\frac{1}{2}\right) n \Delta \nu,
\end{equation}
where $\Delta\nu$ is the frequency resolution of the original power spectrum and $\nu_i$ varies between $-n\Delta\nu/2$ and $+n\Delta\nu/2$, spanning $n \Delta \nu$, which is the resolution of the binned power spectrum.  Finally the expressions for $\lambda$ and $\nu$ are given by
\begin{equation}
	\lambda=\frac{E[{\cal S}]}{\sigma^2},
\end{equation}

\begin{equation}
	\nu=\frac{E[{\cal S}]^2}{\sigma^2},
\end{equation}
which are both implicitly functions of $h$ and $W$.

\Online

\begin{figure*}[!]
\centering
\hbox{
\includegraphics[angle=90,width=9.cm]{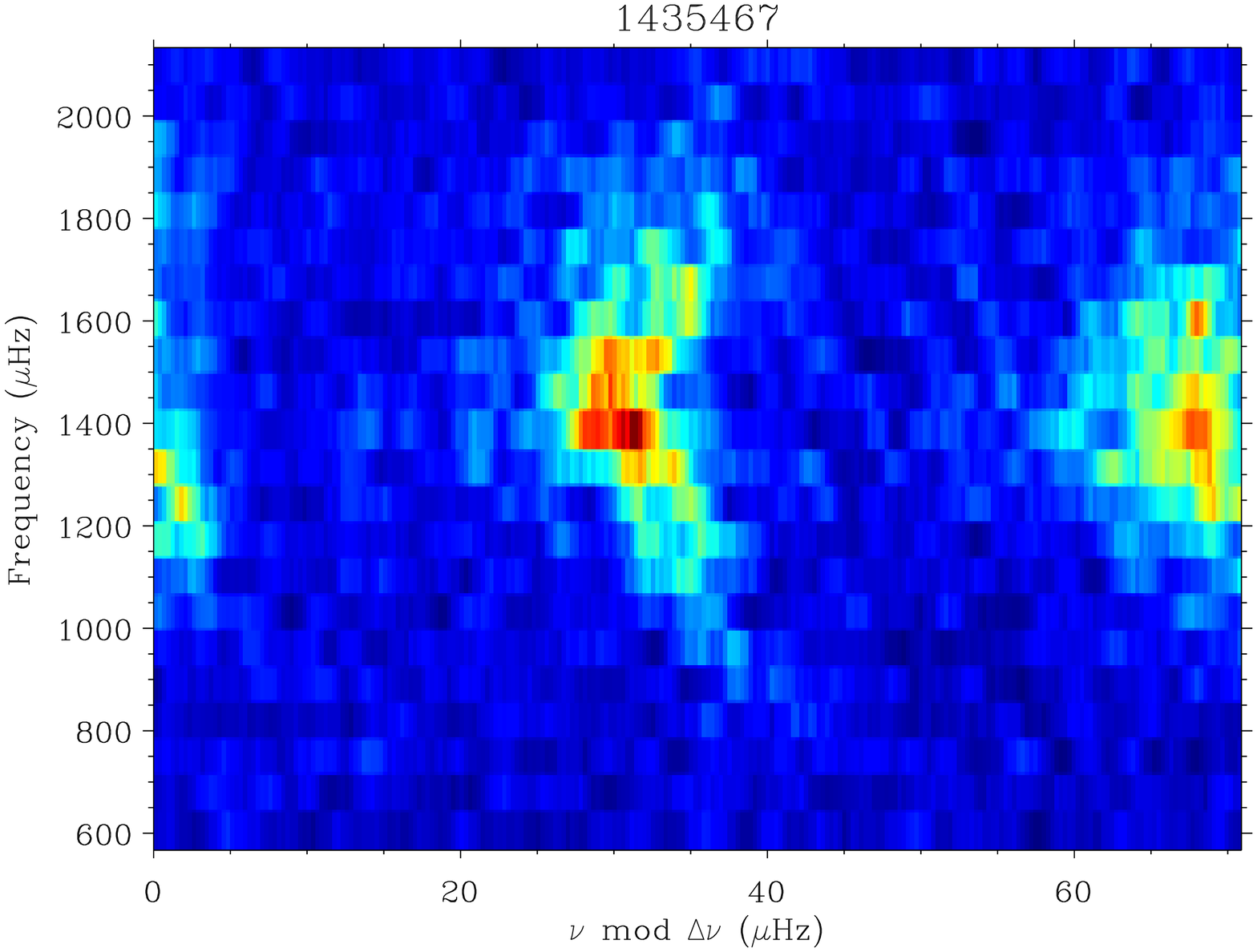}
\includegraphics[angle=90,width=9.cm]{echelle_diagram_2837475.ps}
}
\hbox{
\includegraphics[angle=90,width=9.cm]{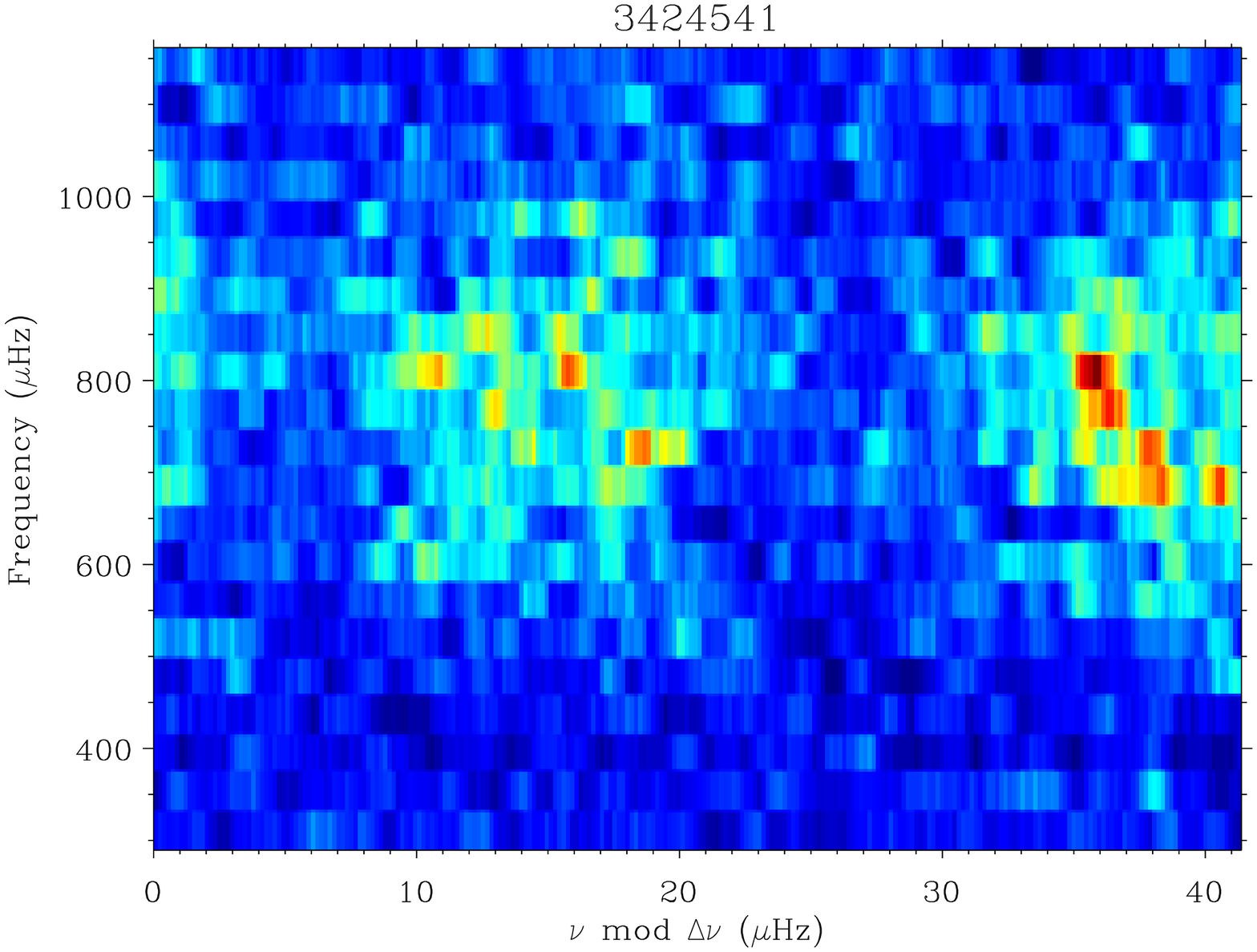}
\includegraphics[angle=90,width=9.cm]{echelle_diagram_3427720.ps}
}
\hbox{
\includegraphics[angle=90,width=9.cm]{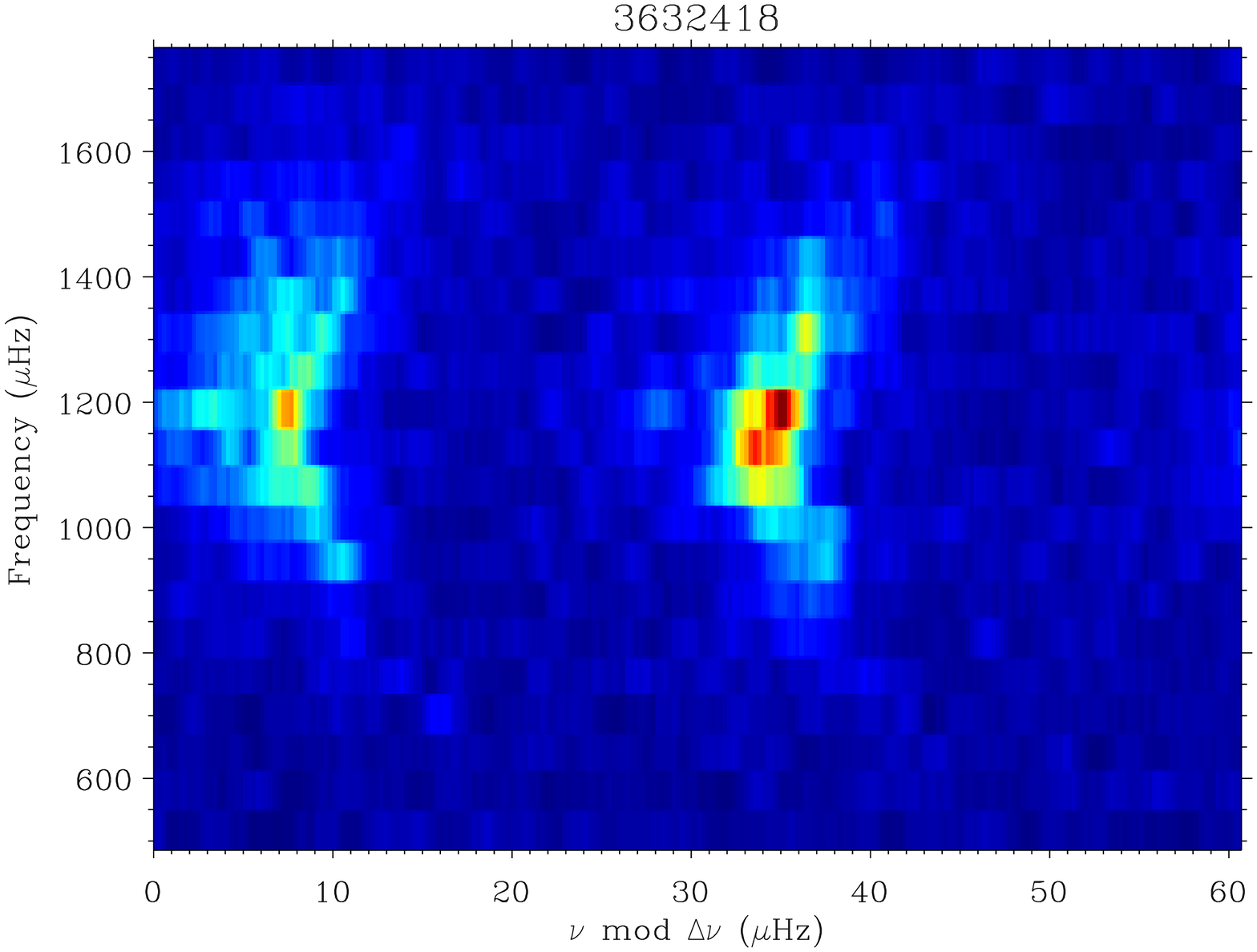}
\includegraphics[angle=90,width=9.cm]{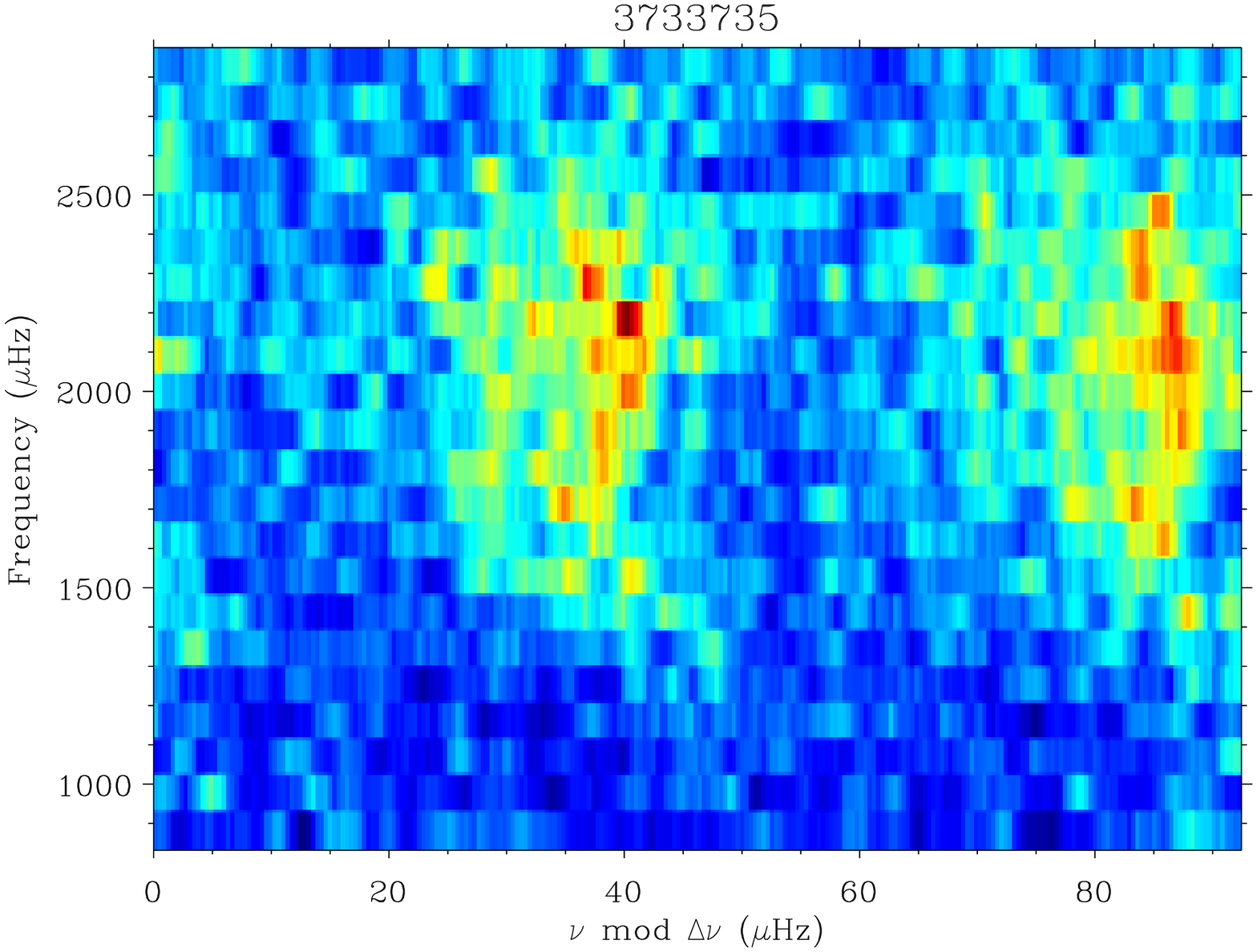}
}
\caption{Echelle diagrams of the power spectra of KIC 1435467, KIC 2837475, KIC 3424541, KIC 3427720, KIC 3632418 and KIC 3733735.  The power spectra are normalised by the background and then smoothed over 3 $\mu$Hz.}
\end{figure*}
\begin{figure*}[!]
\centering
\hbox{
\includegraphics[angle=90,width=9.cm]{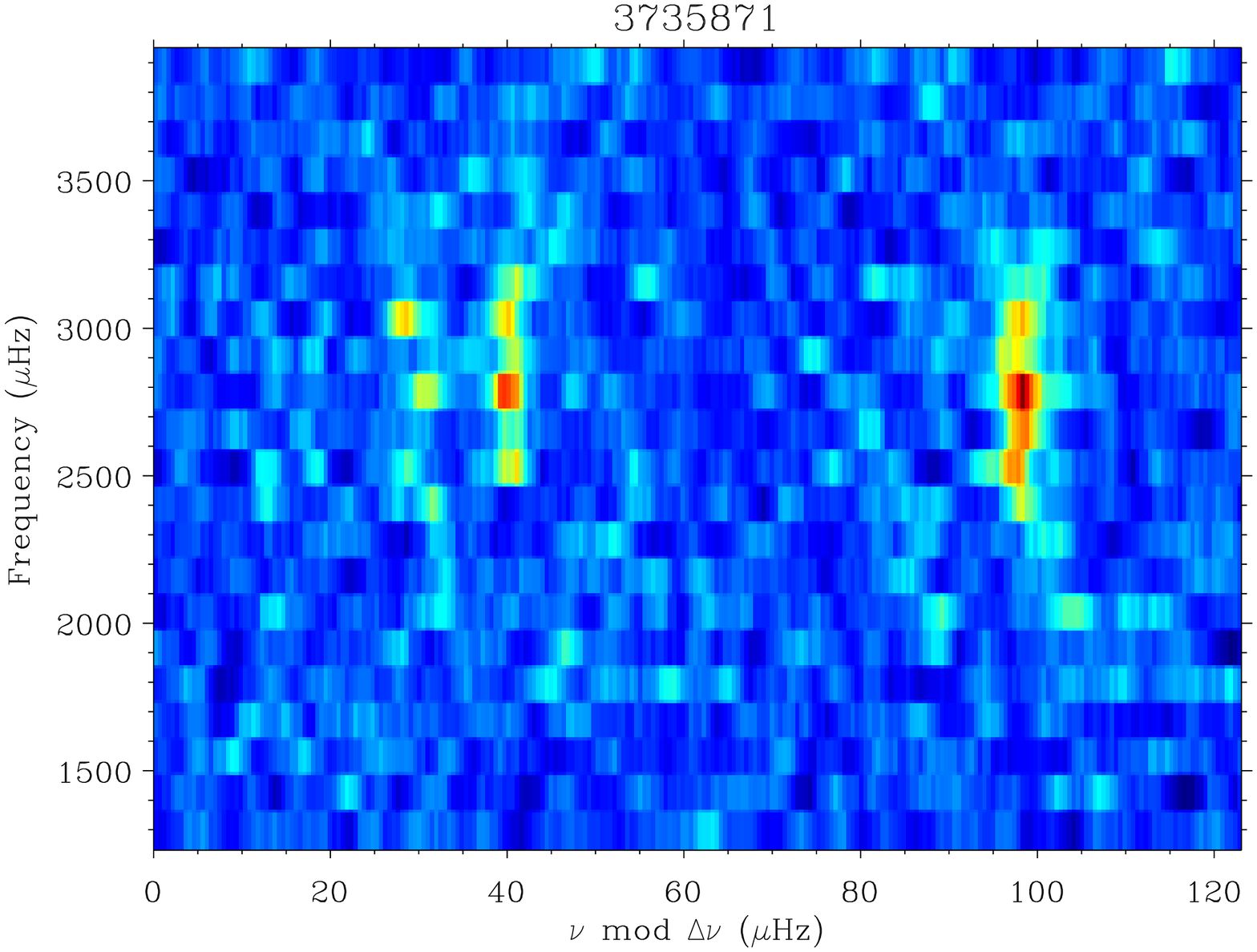}
\includegraphics[angle=90,width=9.cm]{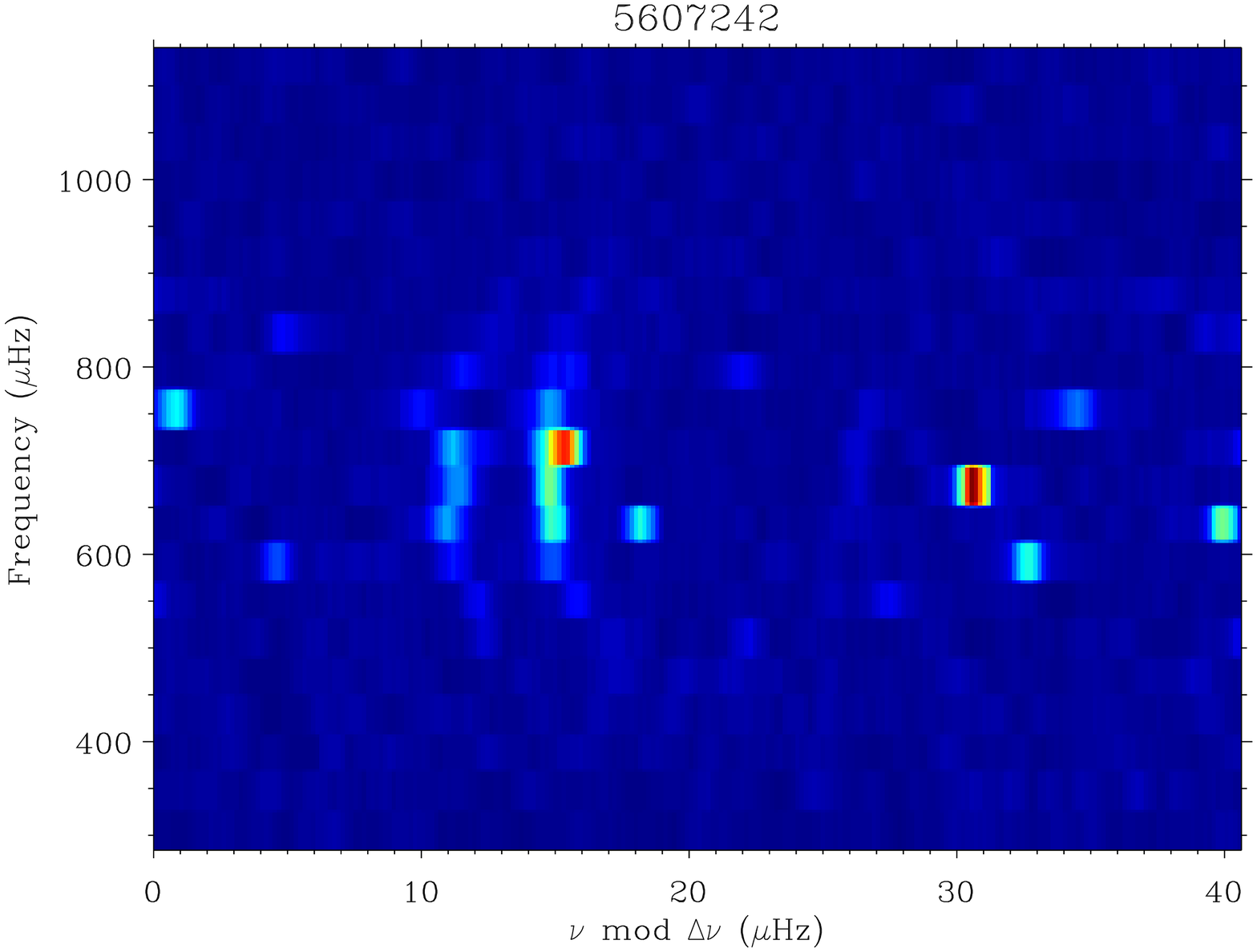}
}
\hbox{
\includegraphics[angle=90,width=9.cm]{echelle_diagram_5955122.ps}
\includegraphics[angle=90,width=9.cm]{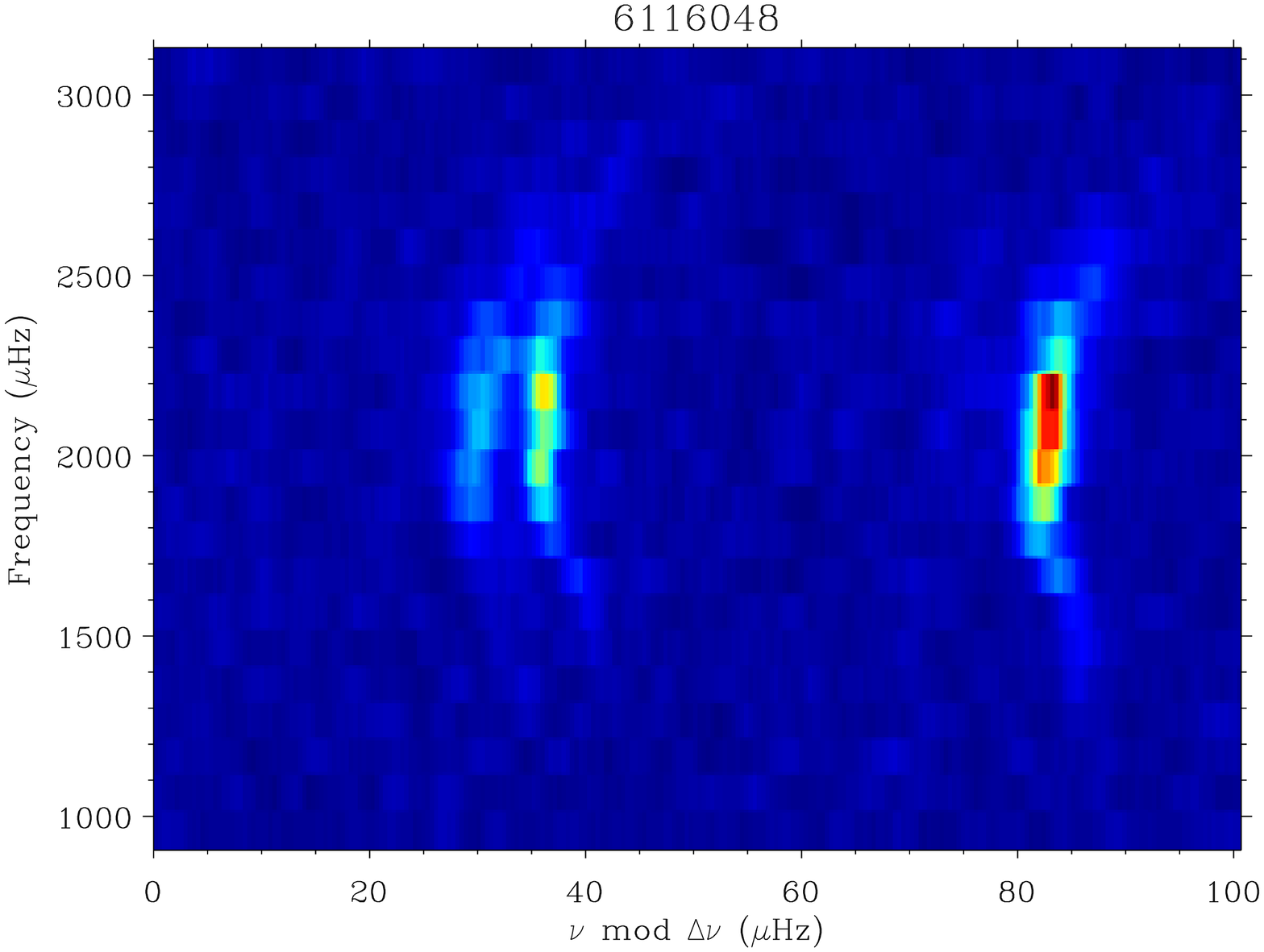}
}
\hbox{
\includegraphics[angle=90,width=9.cm]{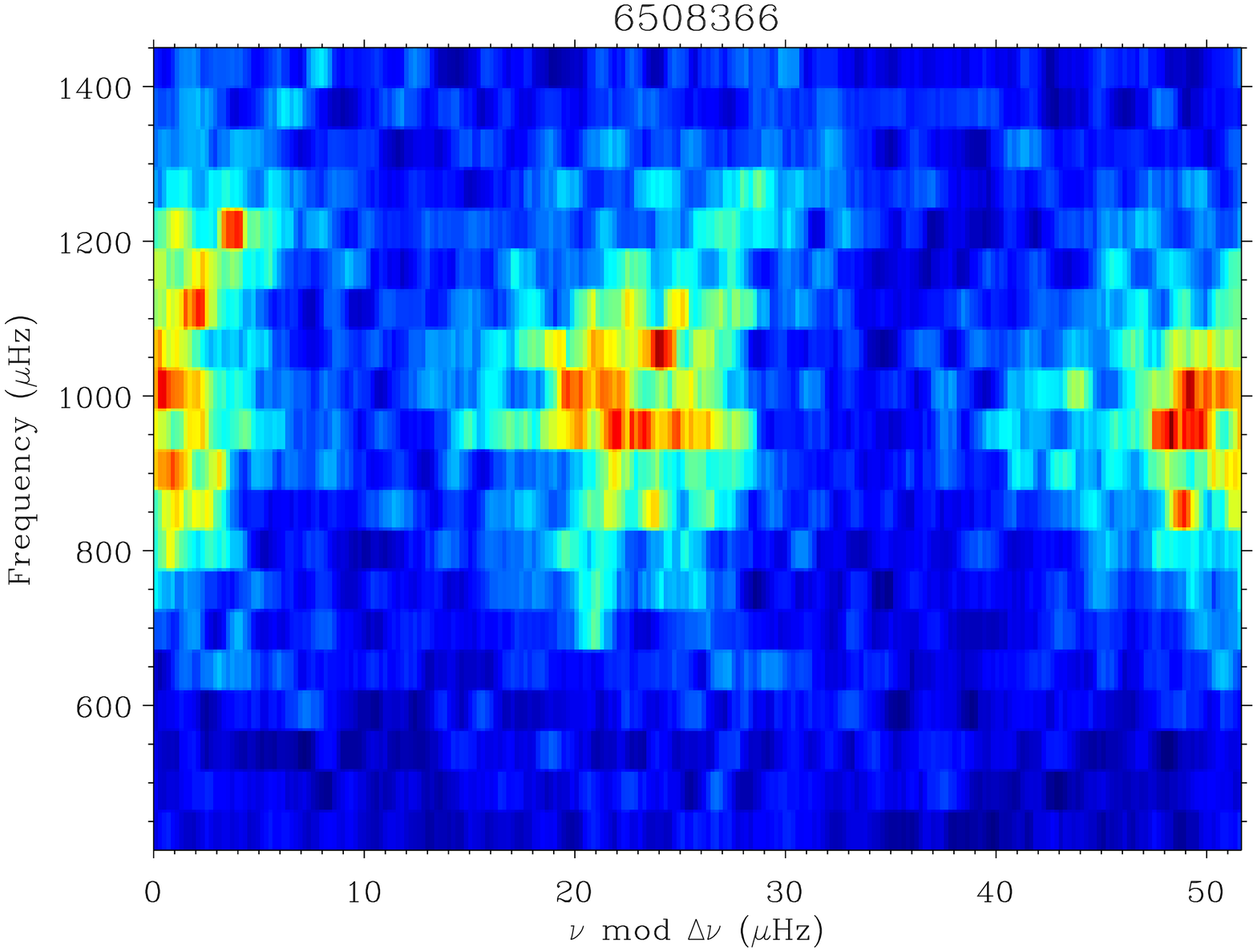}
\includegraphics[angle=90,width=9.cm]{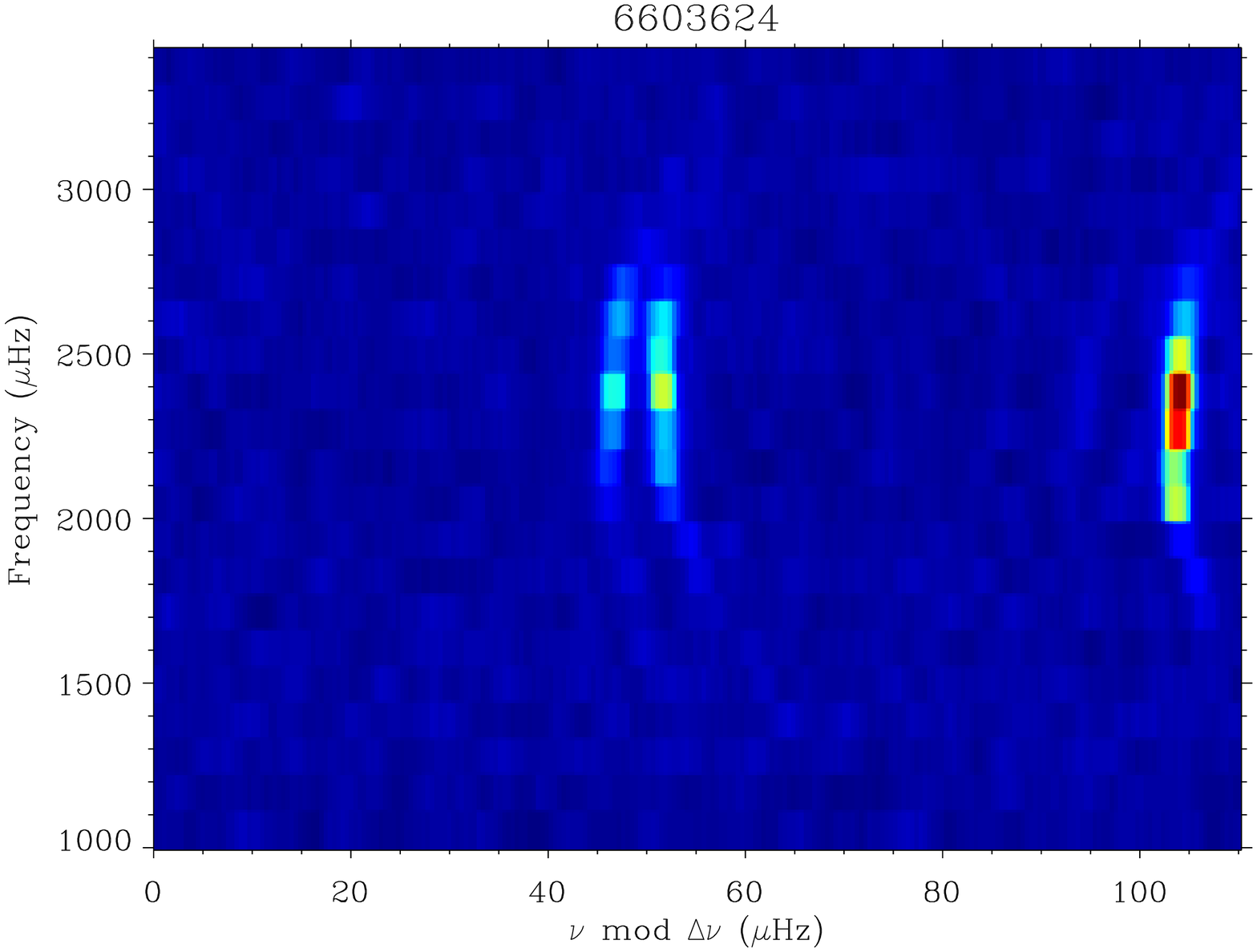}
}
\caption{Echelle diagrams of the power spectra of KIC 3735871, KIC 5607242, KIC 5955122, KIC 6116048, KIC 6508366 and KIC 6603624.  The power spectra are normalised by the background and then smoothed over 3 $\mu$Hz.}
\end{figure*}
\begin{figure*}[!]
\centering
\hbox{
\includegraphics[angle=90,width=9.cm]{echelle_diagram_6679371.ps}
\includegraphics[angle=90,width=9.cm]{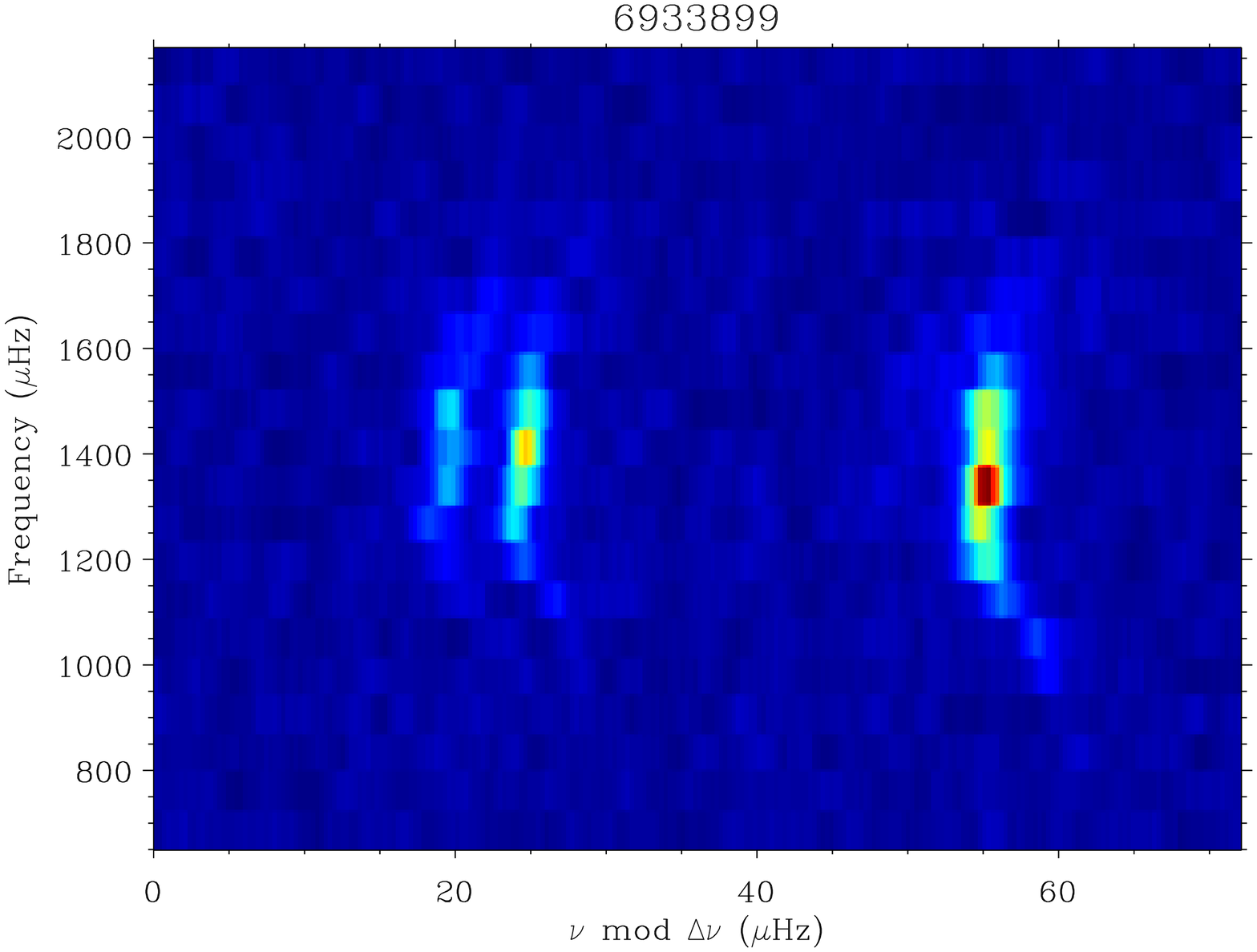}
}
\hbox{
\includegraphics[angle=90,width=9.cm]{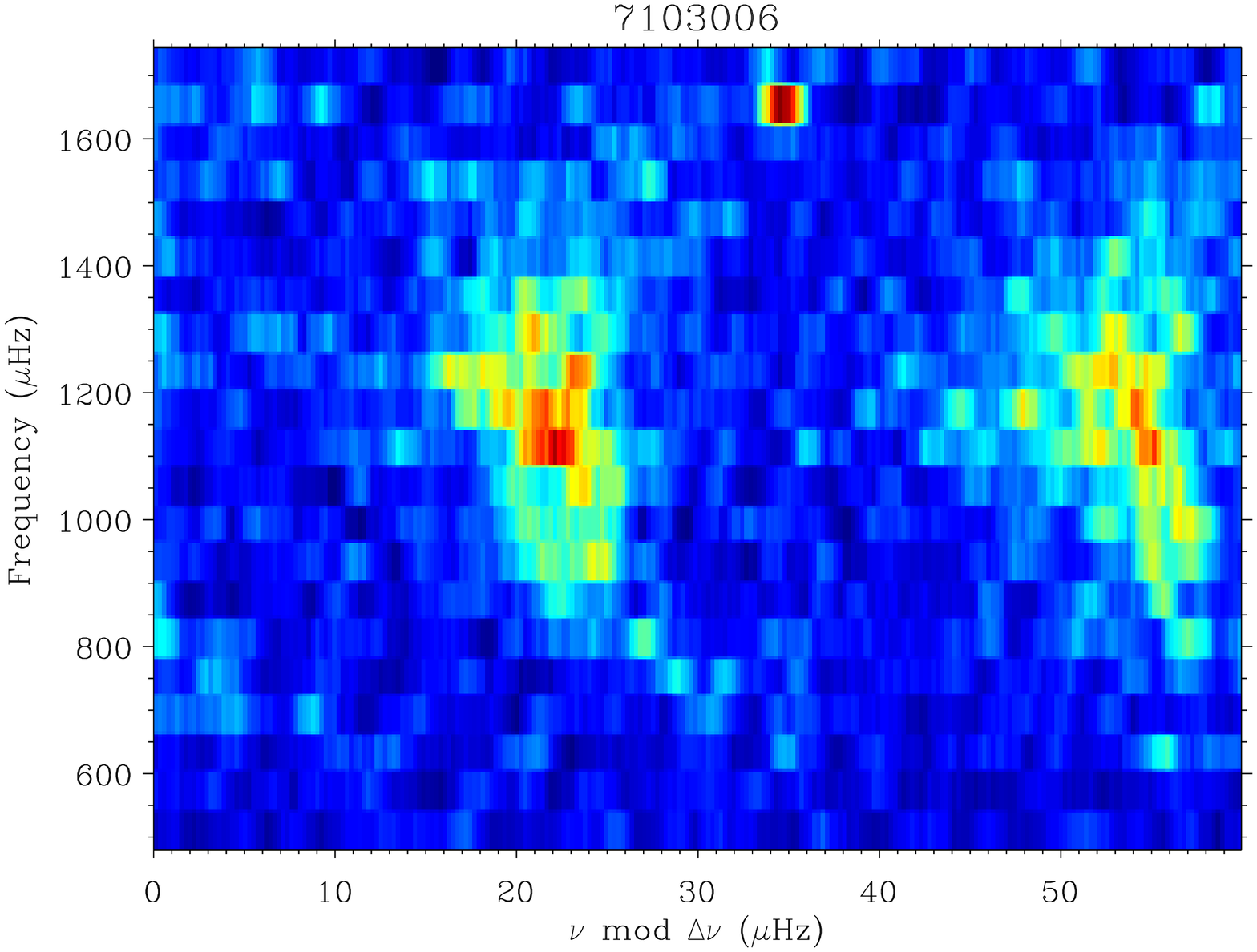}
\includegraphics[angle=90,width=9.cm]{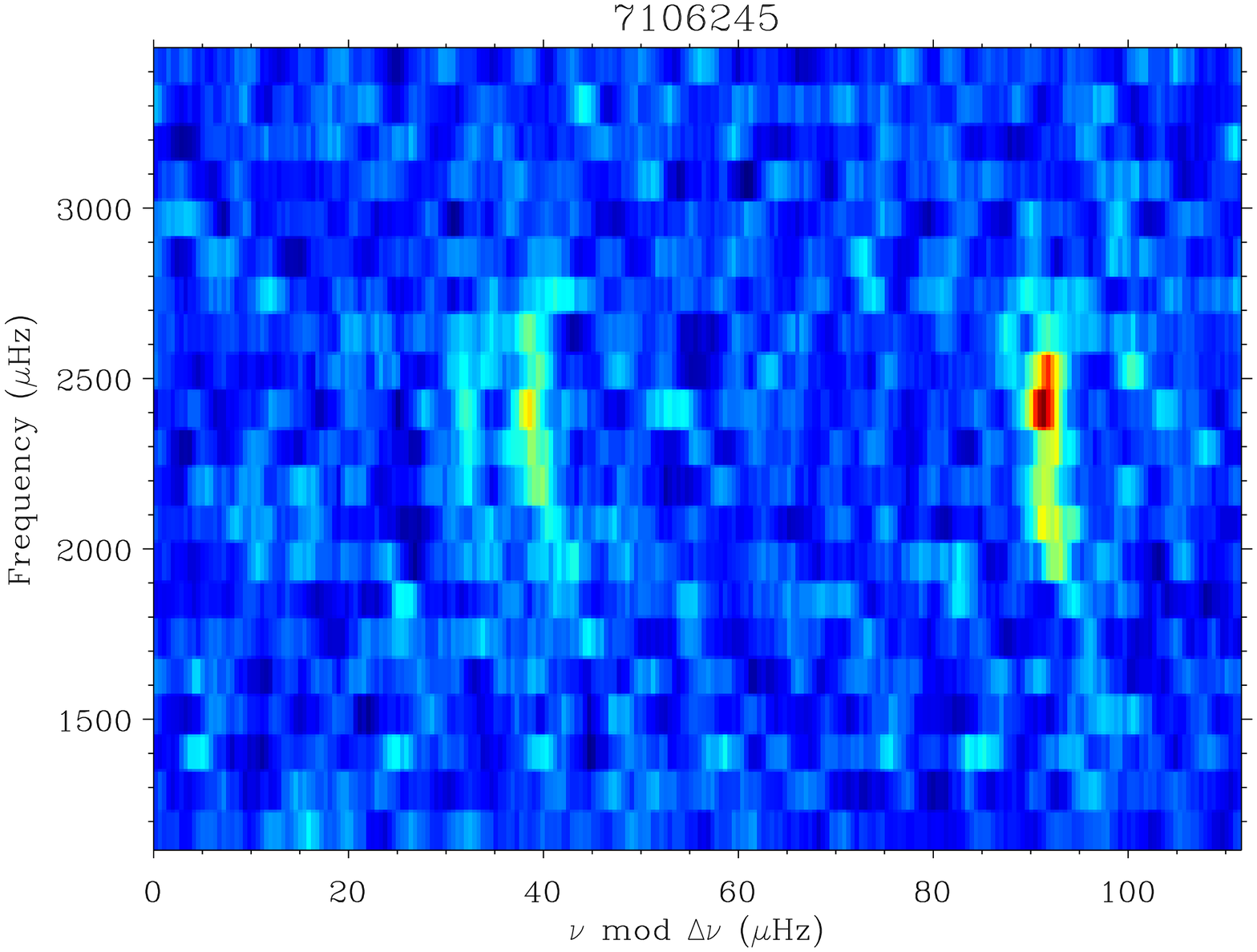}
}
\hbox{
\includegraphics[angle=90,width=9.cm]{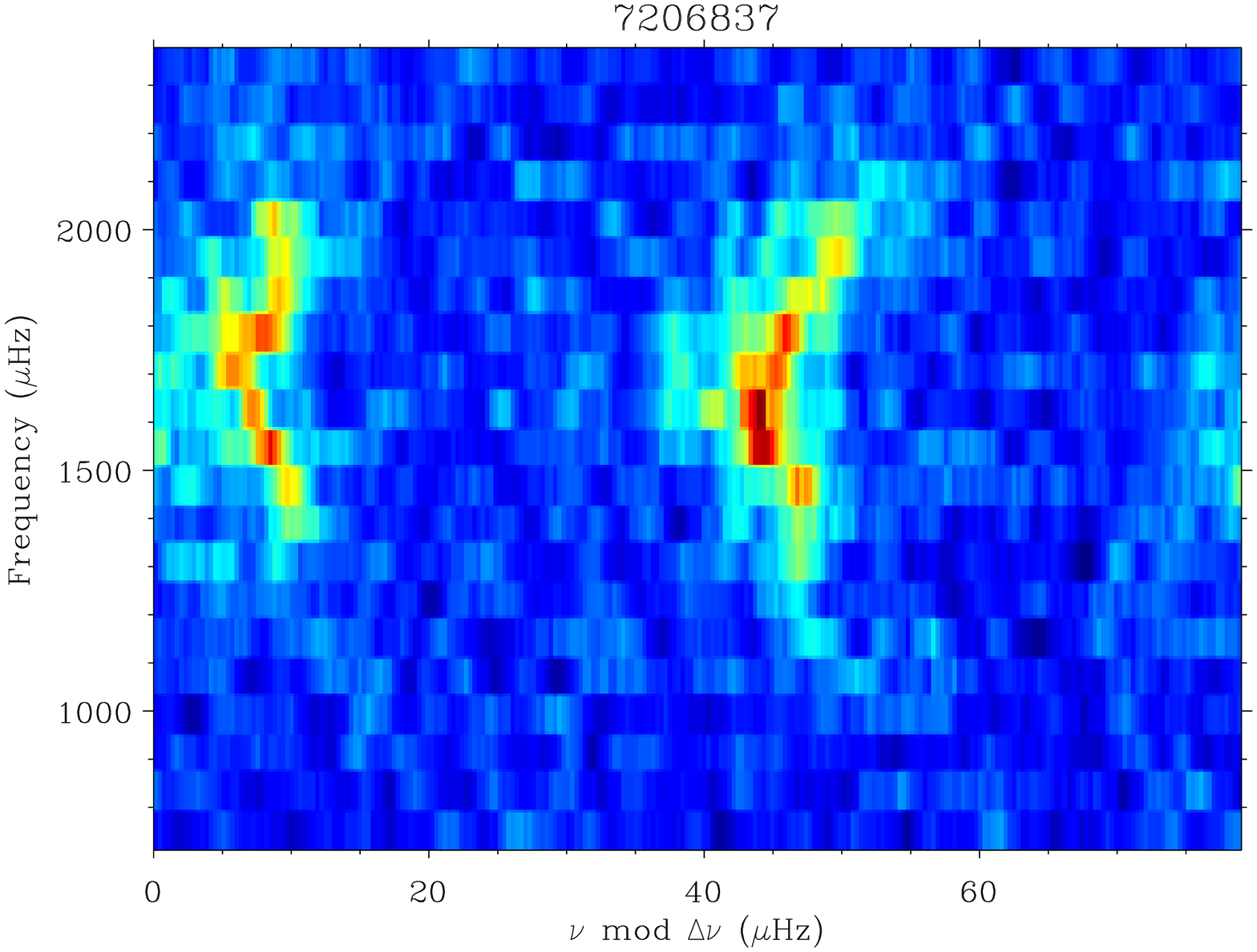}
\includegraphics[angle=90,width=9.cm]{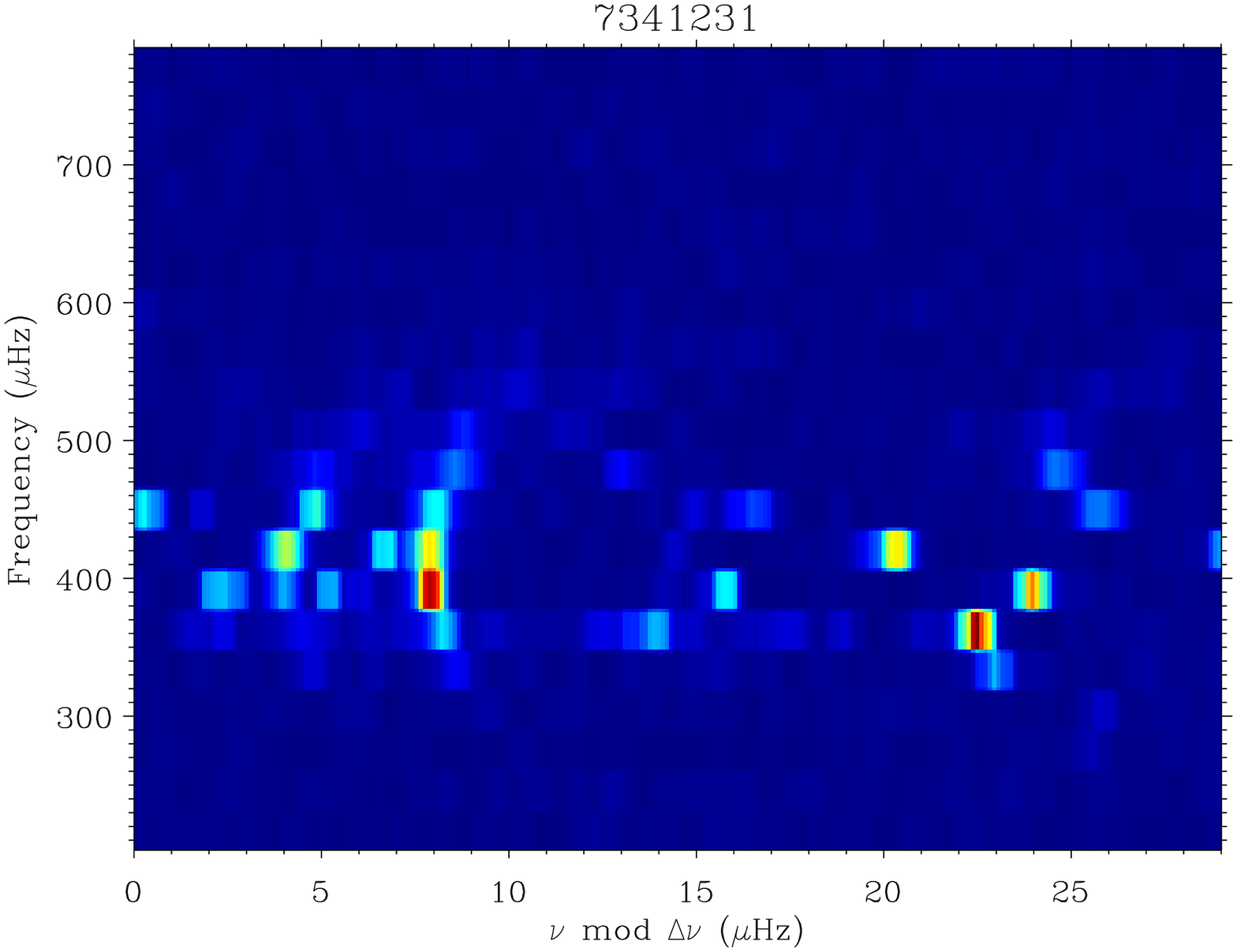}
}
\caption{Echelle diagrams of the power spectra of KIC 6679371, KIC 6933899, KIC 7103006, KIC 7106245, KIC 7206837 and KIC 7341231.  The power spectra are normalised by the background and then smoothed over 3 $\mu$Hz.}
\end{figure*}
\begin{figure*}[!]
\centering
\hbox{
\includegraphics[angle=90,width=9.cm]{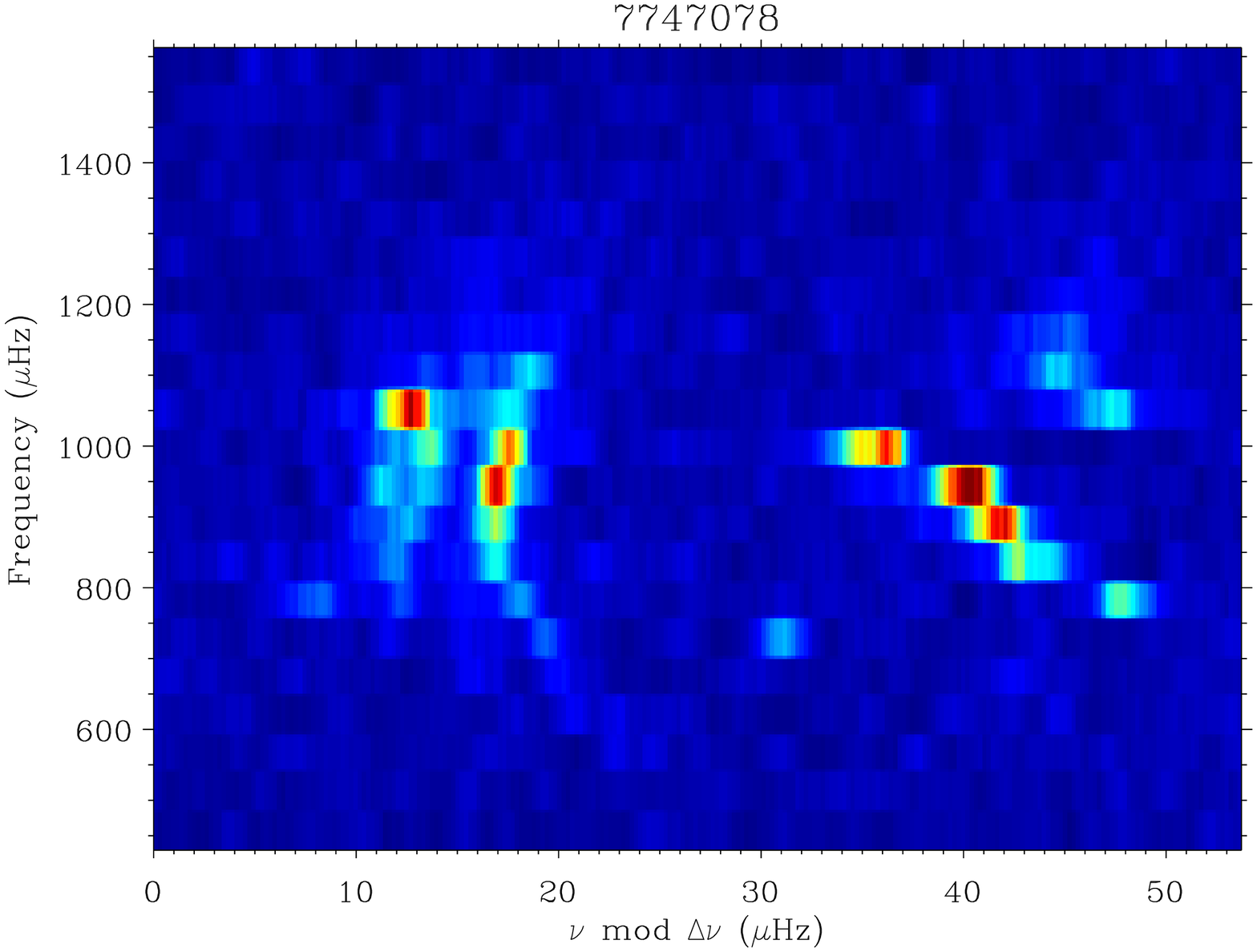}
\includegraphics[angle=90,width=9.cm]{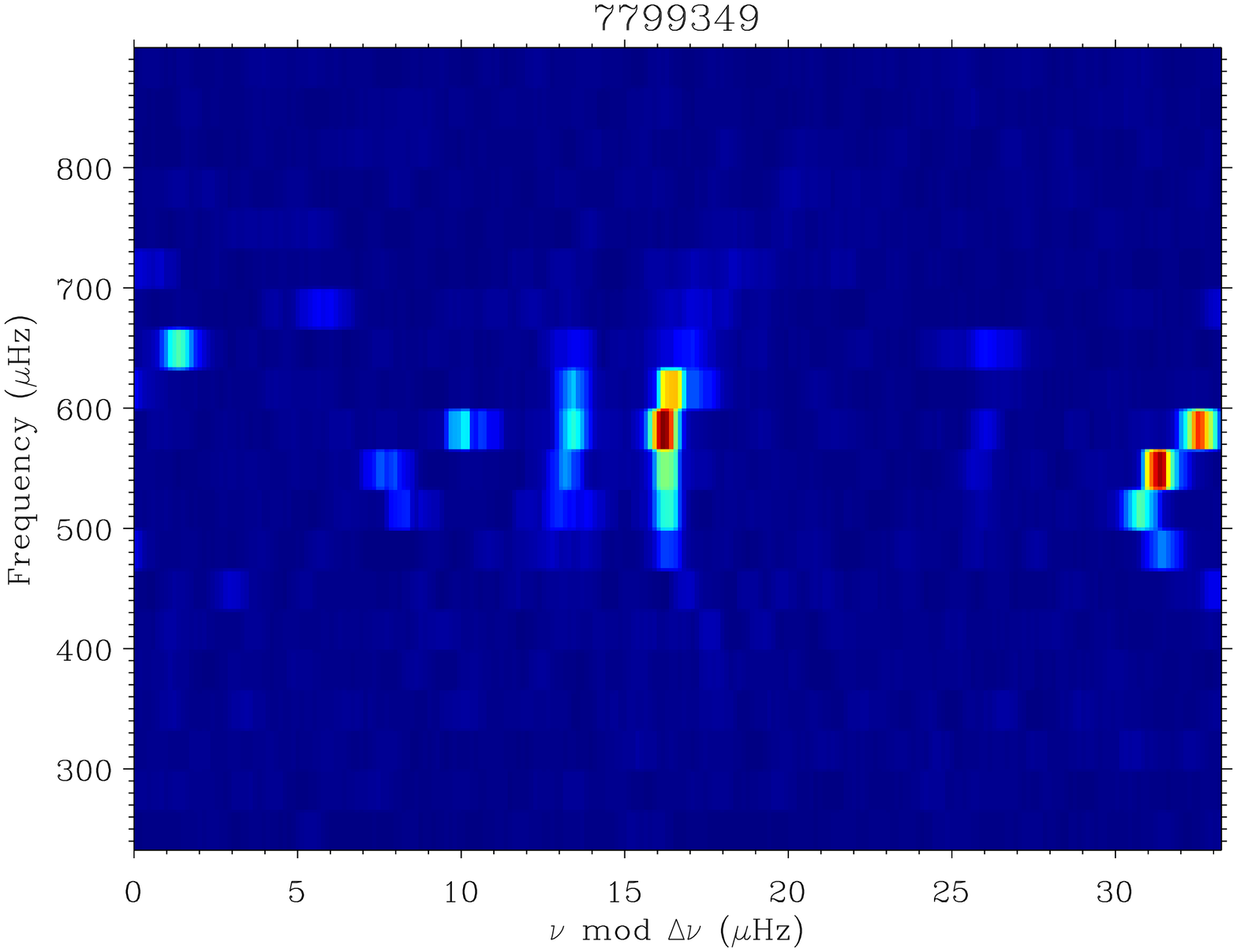}
}
\hbox{
\includegraphics[angle=90,width=9.cm]{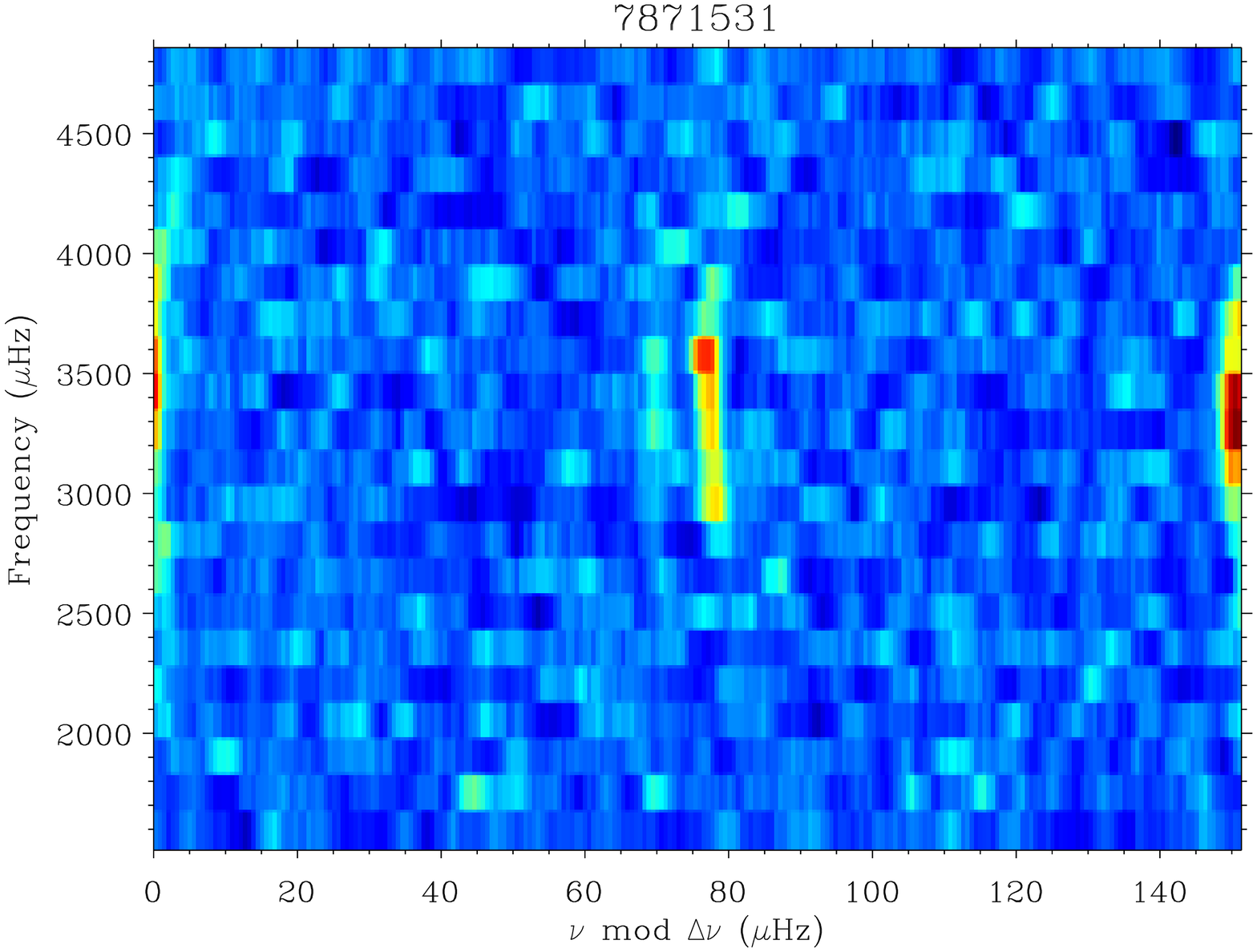}
\includegraphics[angle=90,width=9.cm]{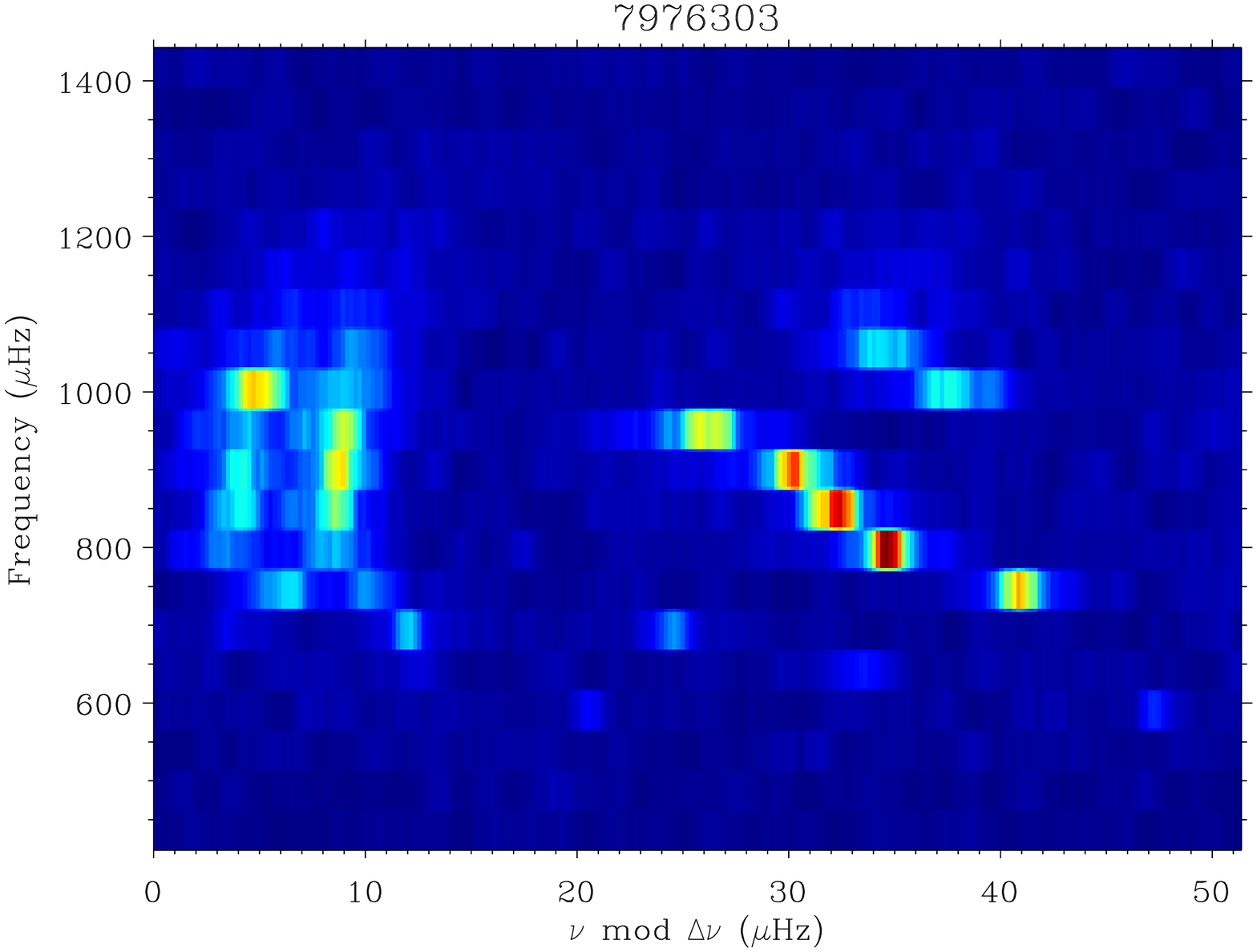}
}
\hbox{
\includegraphics[angle=90,width=9.cm]{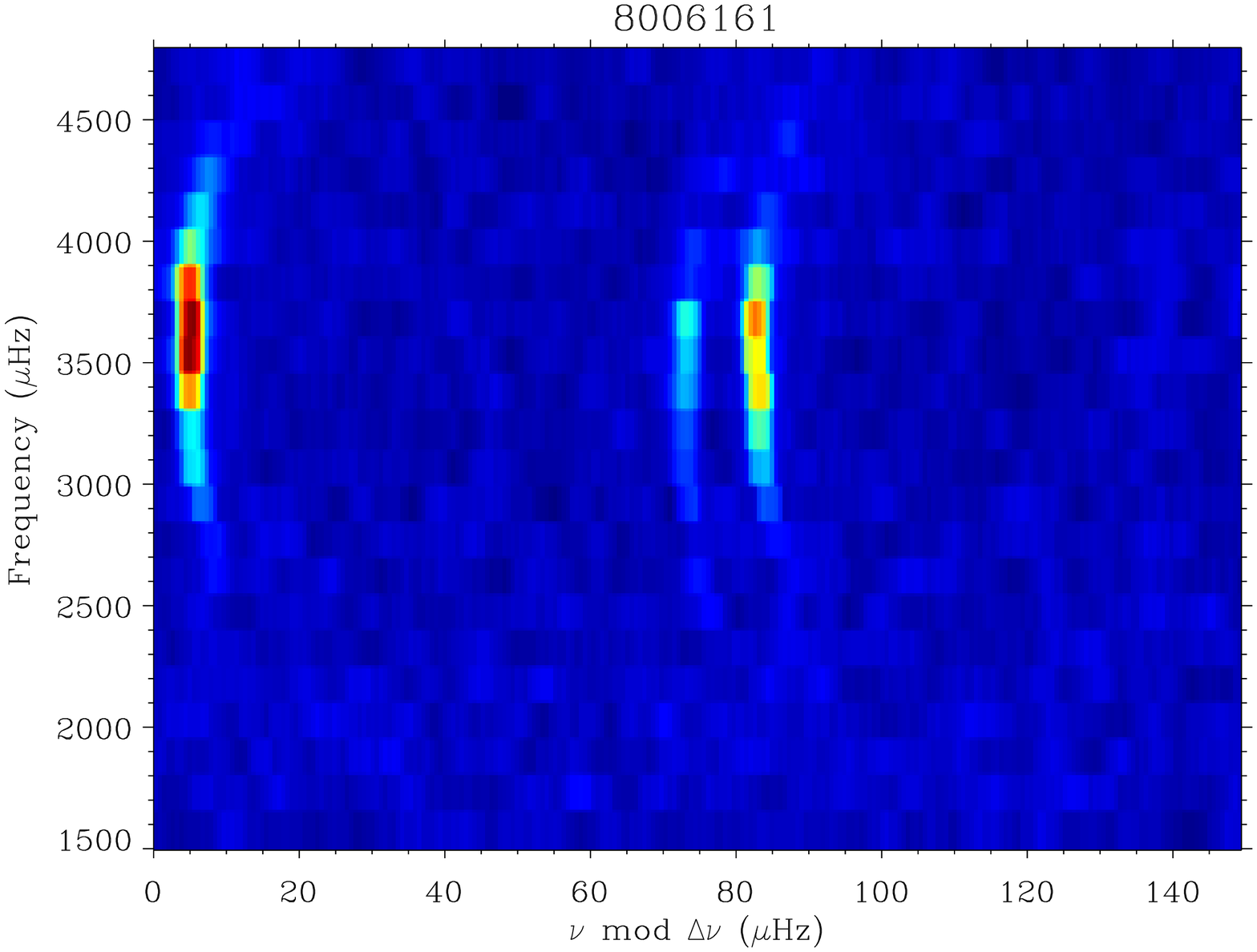}
\includegraphics[angle=90,width=9.cm]{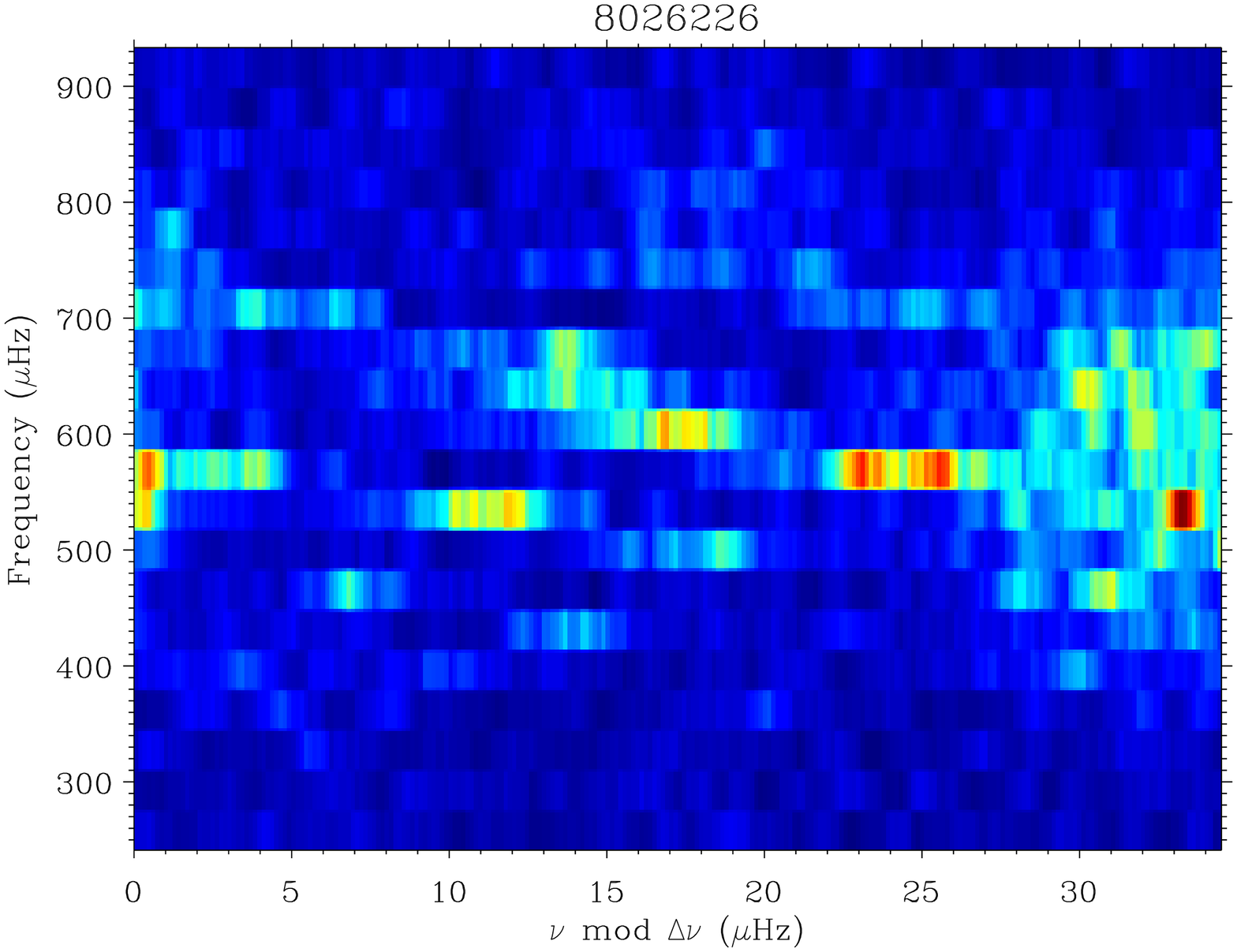}
}
\caption{Echelle diagrams of the power spectra of KIC 7747078, KIC 7799349, KIC 7871531, KIC 7976303, KIC 8006161 and KIC 8026226.  The power spectra are normalised by the background and then smoothed over 3 $\mu$Hz.}
\end{figure*}
\begin{figure*}[!]
\centering
\hbox{
\includegraphics[angle=90,width=9.cm]{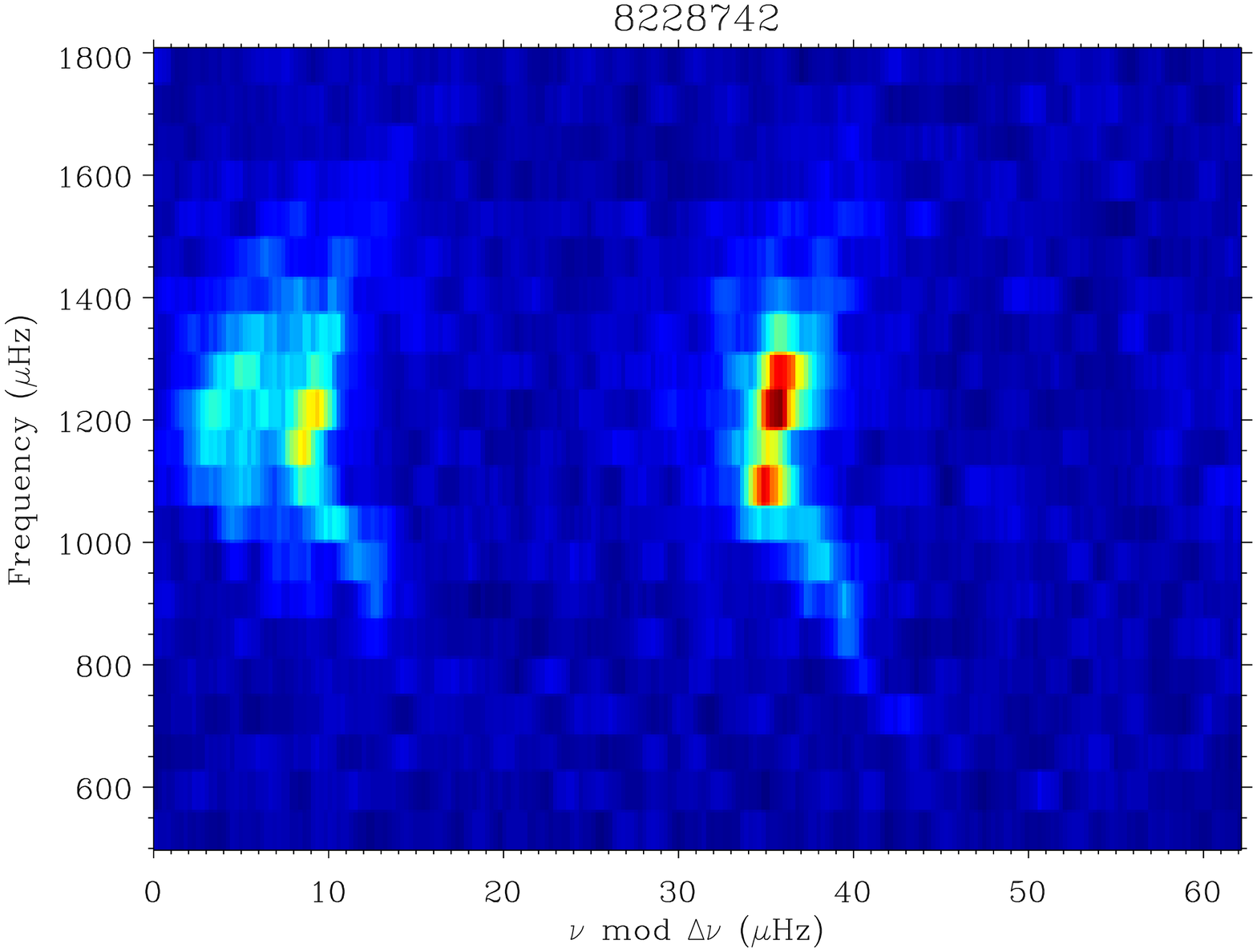}
\includegraphics[angle=90,width=9.cm]{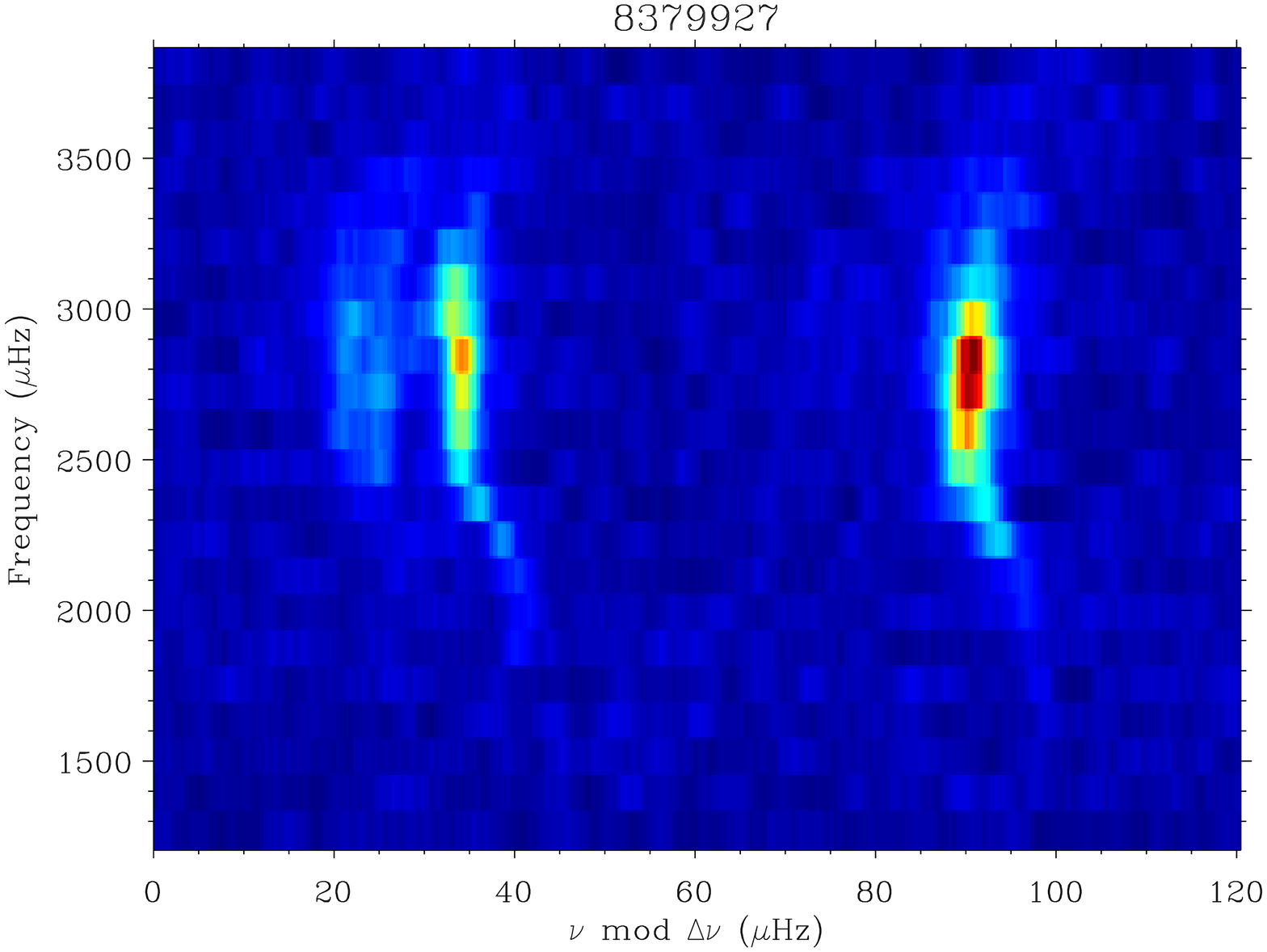}
}
\hbox{
\includegraphics[angle=90,width=9.cm]{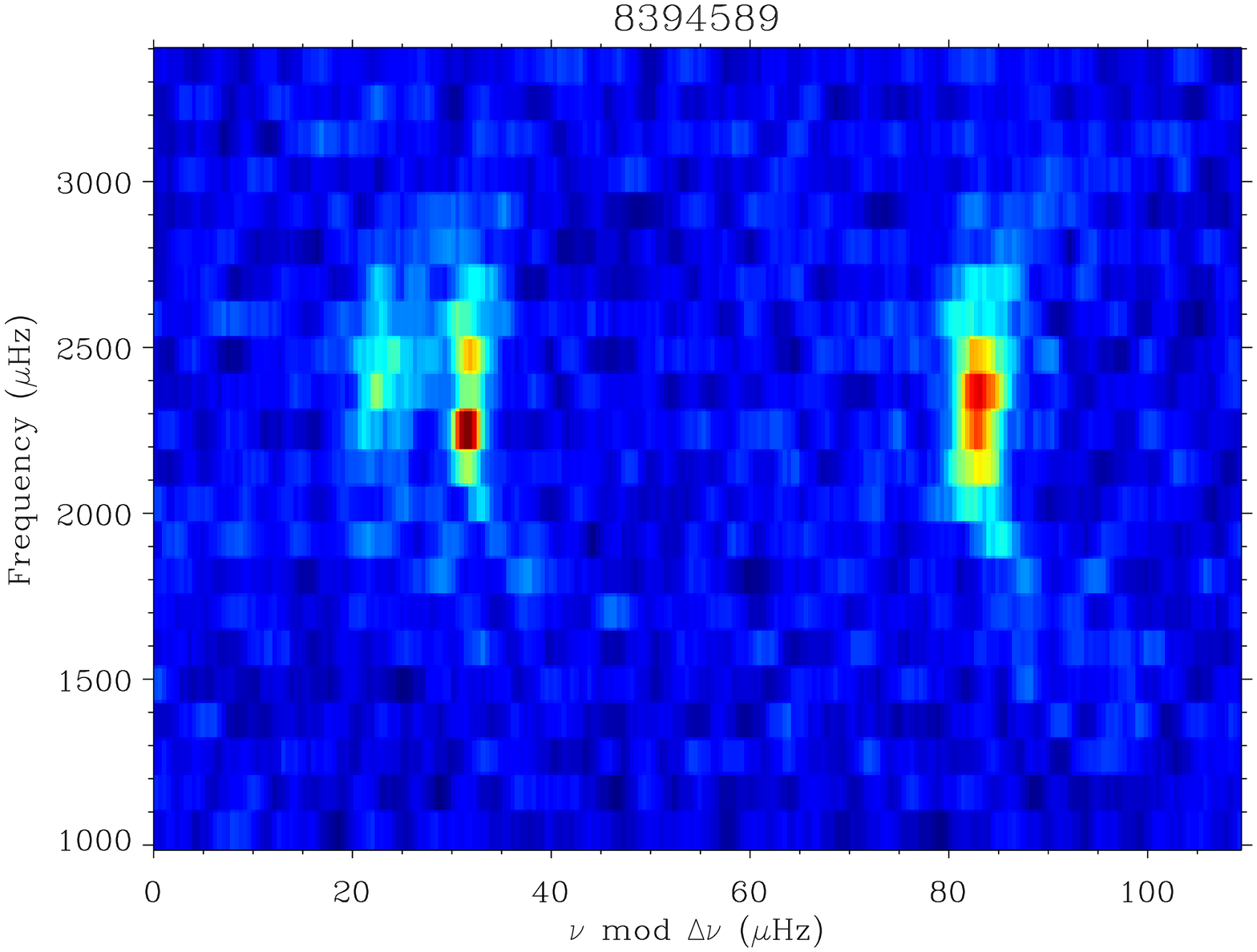}
\includegraphics[angle=90,width=9.cm]{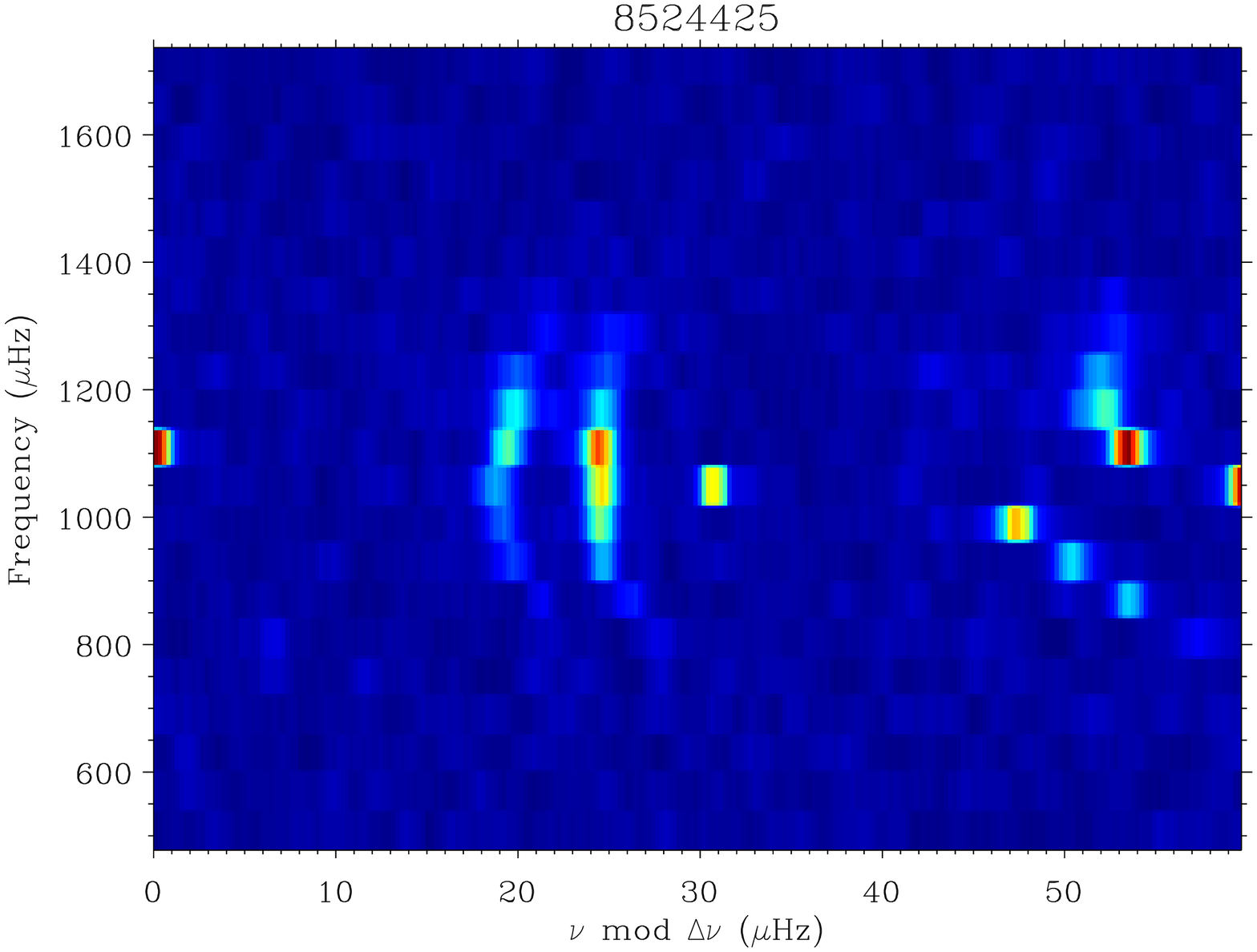}
}
\hbox{
\includegraphics[angle=90,width=9.cm]{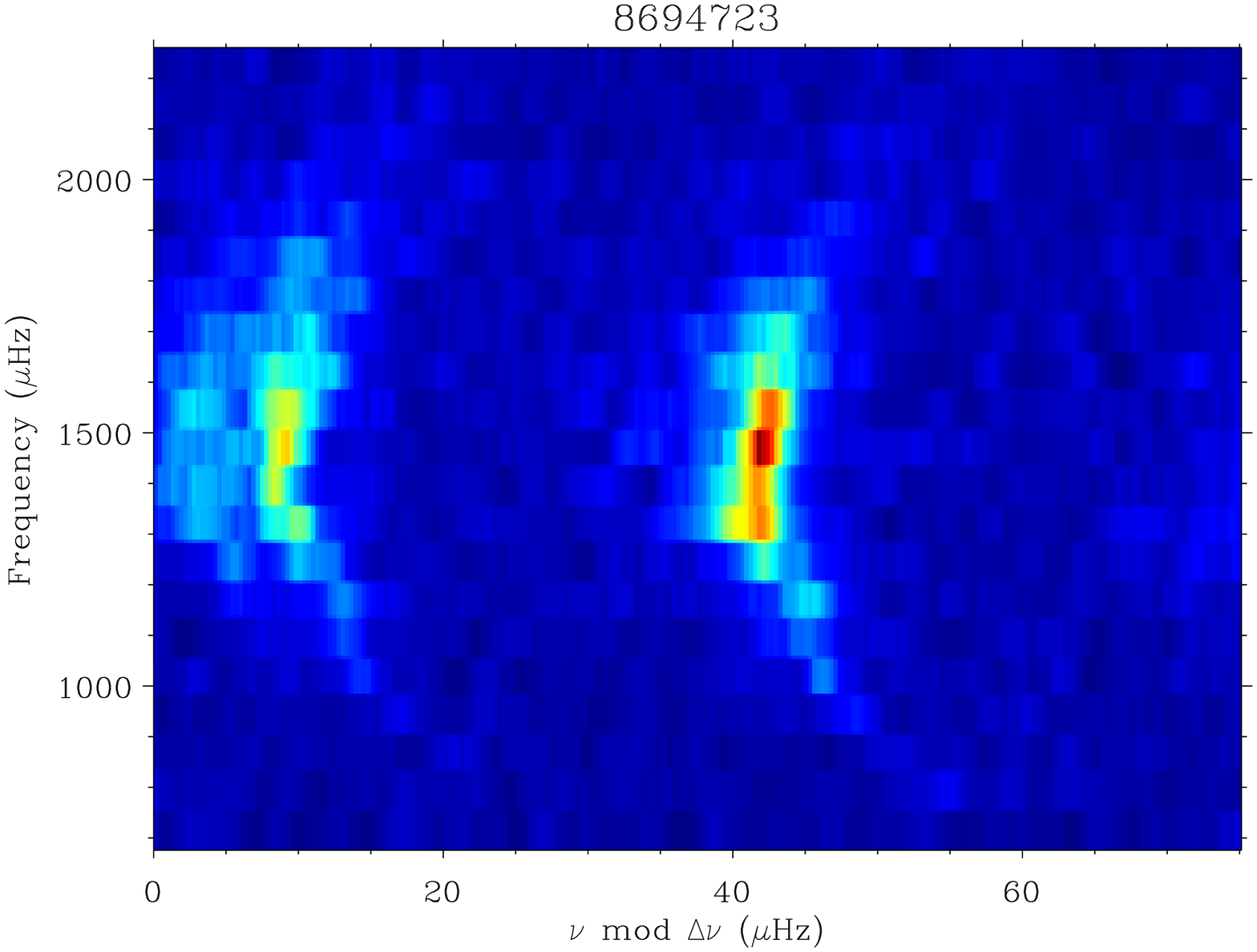}
\includegraphics[angle=90,width=9.cm]{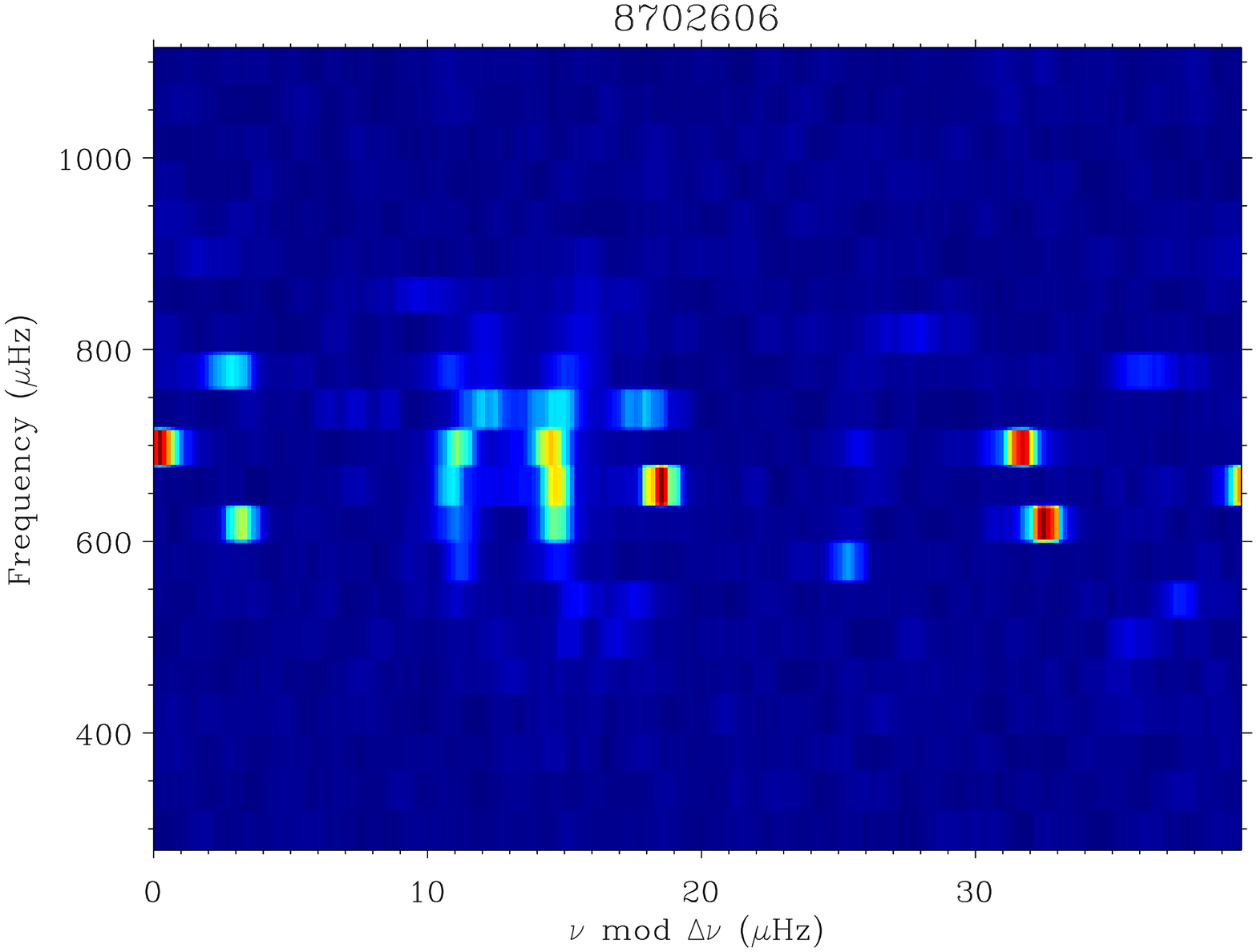}
}
\caption{Echelle diagrams of the power spectra of KIC 8228742, KIC 8379927, KIC 8394589, KIC 8524425, KIC 8694723 and KIC 8702606.  The power spectra are normalised by the background and then smoothed over 3 $\mu$Hz.}
\end{figure*}
\begin{figure*}[!]
\centering
\hbox{
\includegraphics[angle=90,width=9.cm]{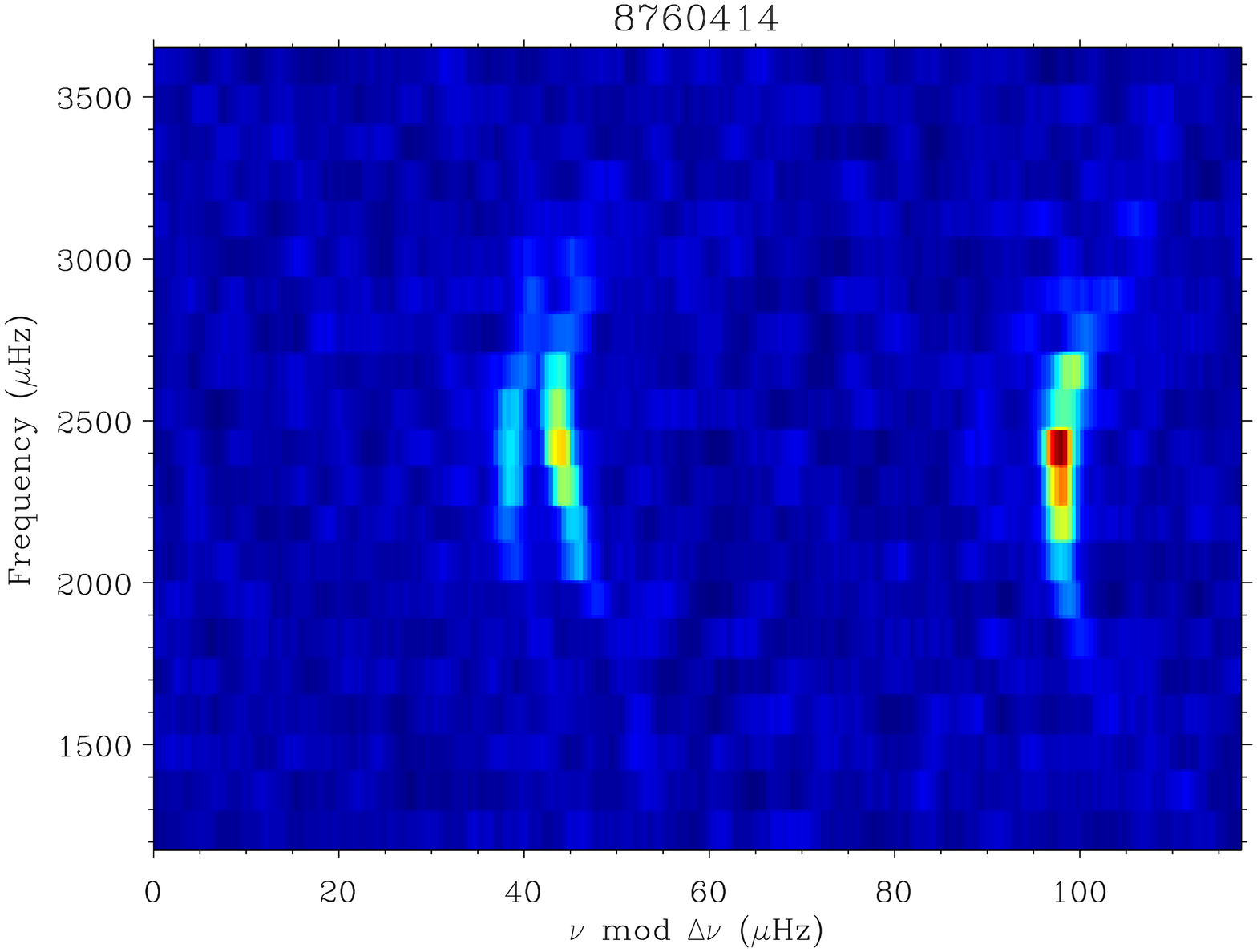}
\includegraphics[angle=90,width=9.cm]{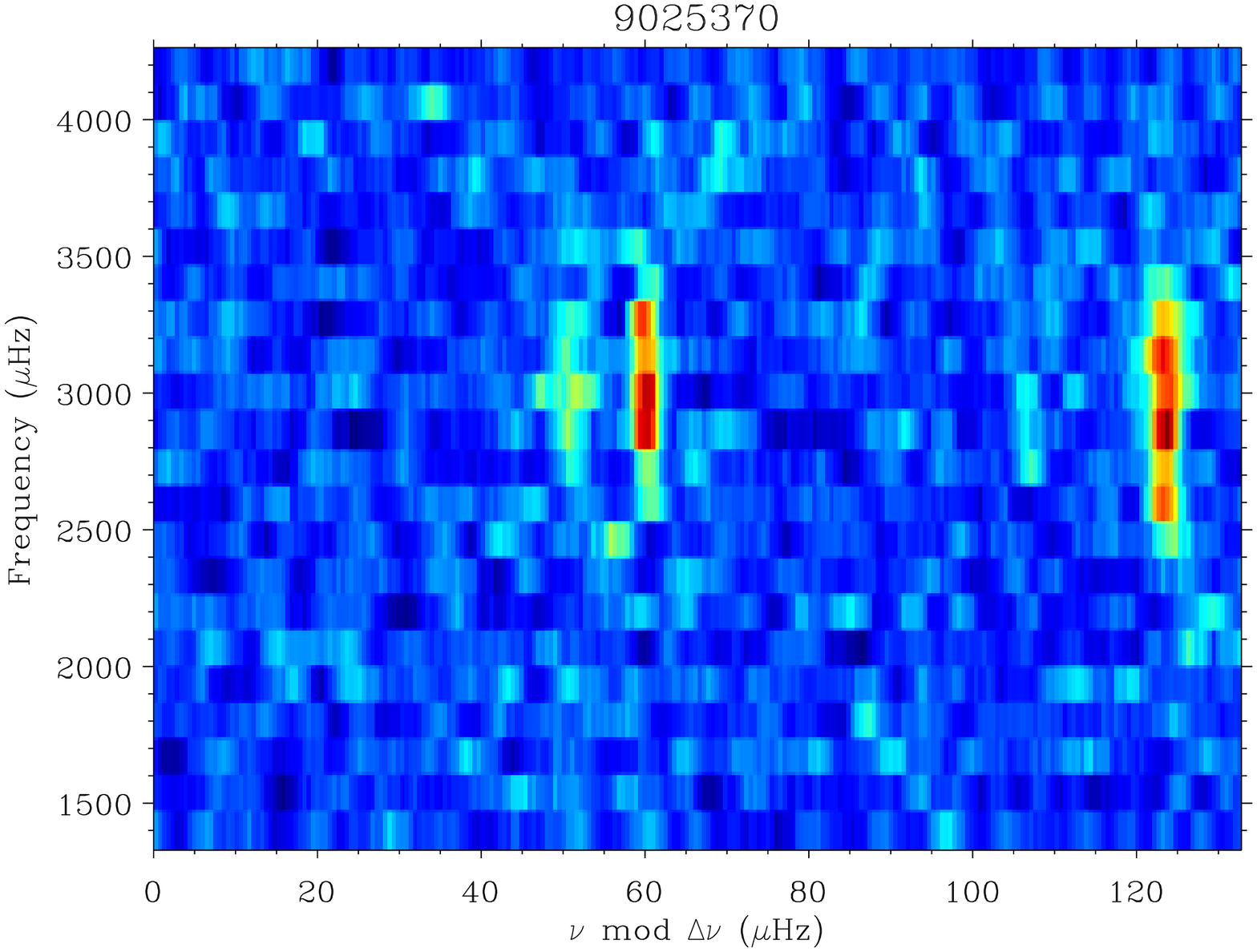}
}
\hbox{
\includegraphics[angle=90,width=9.cm]{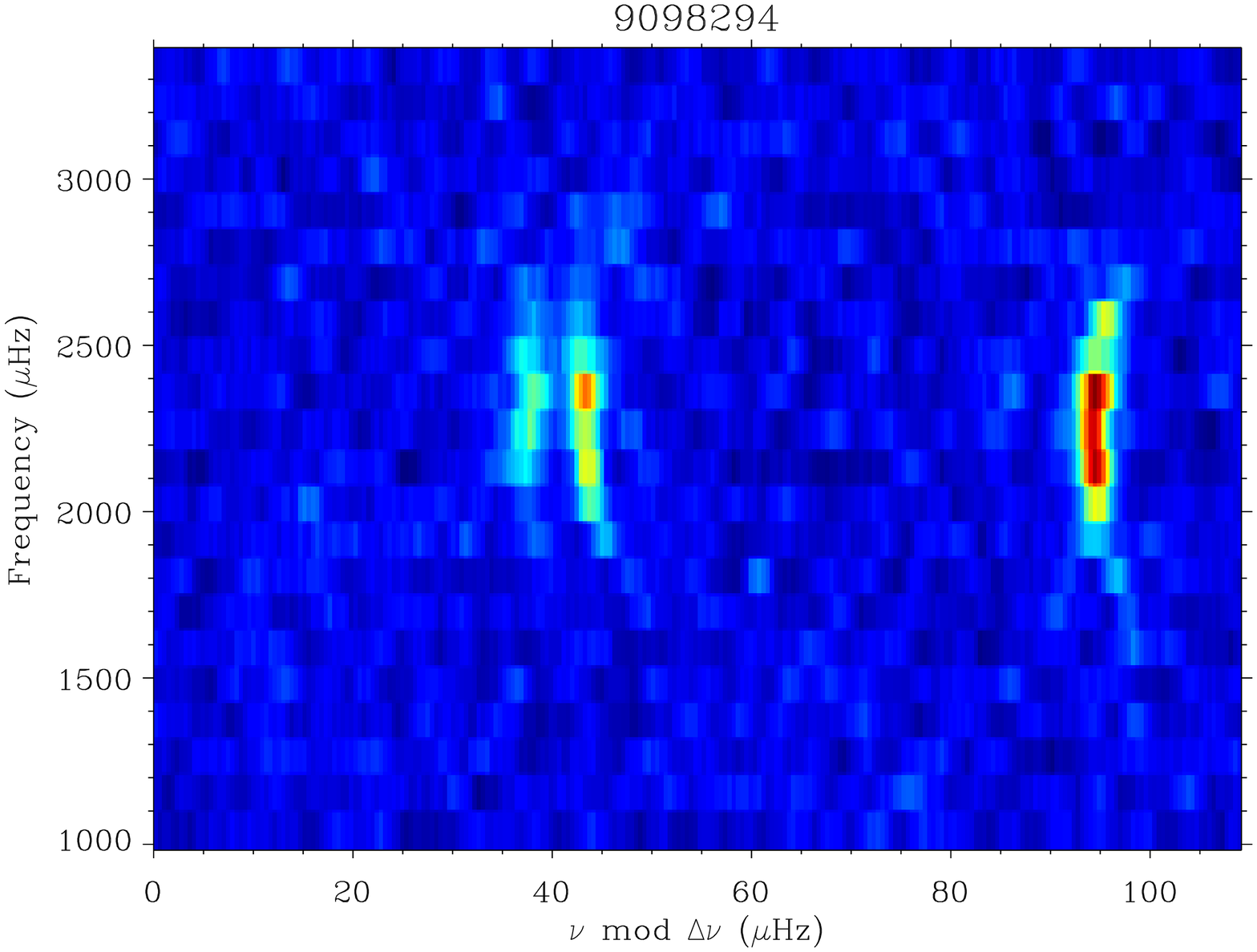}
\includegraphics[angle=90,width=9.cm]{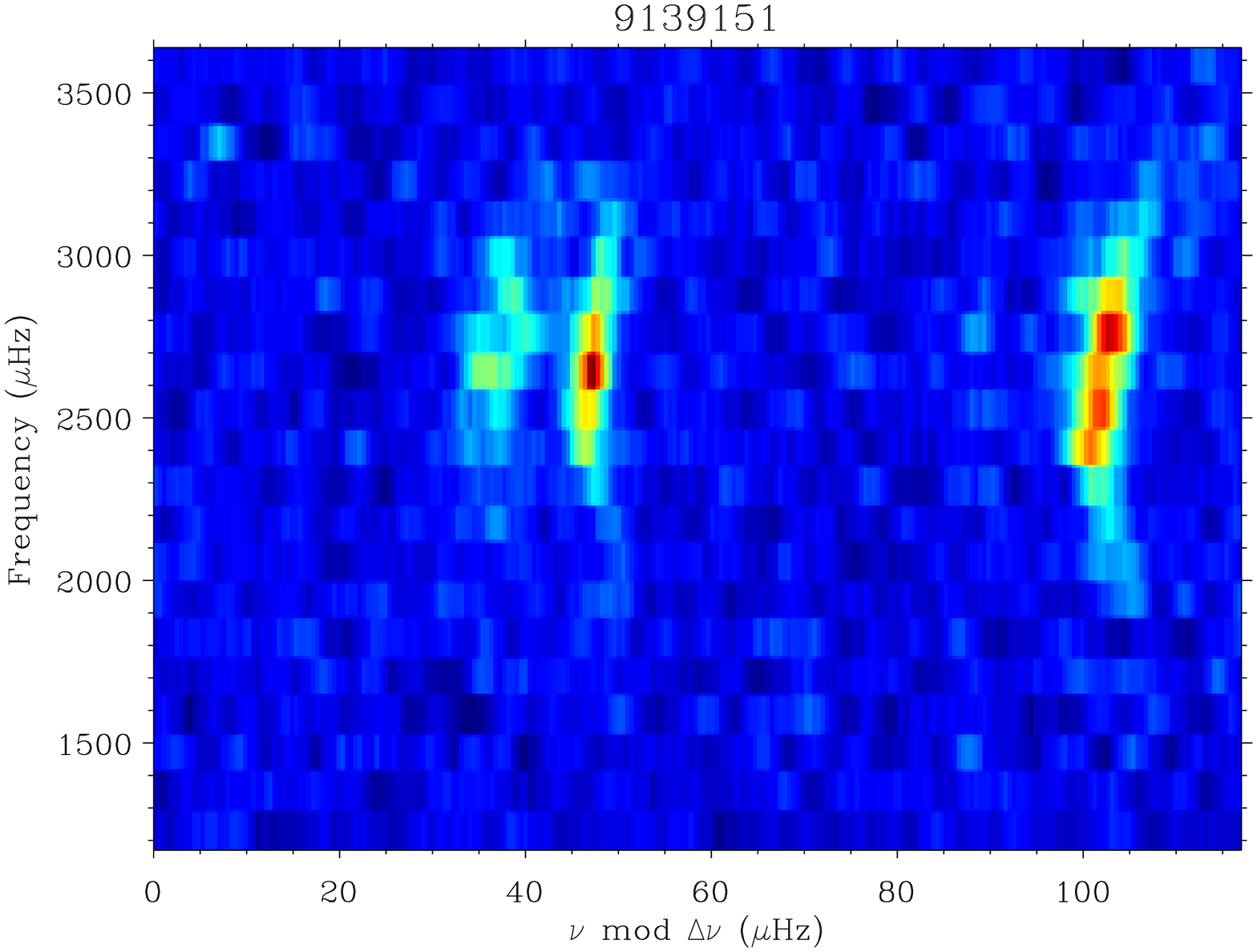}
}
\hbox{
\includegraphics[angle=90,width=9.cm]{echelle_diagram_9139163.ps}
\includegraphics[angle=90,width=9.cm]{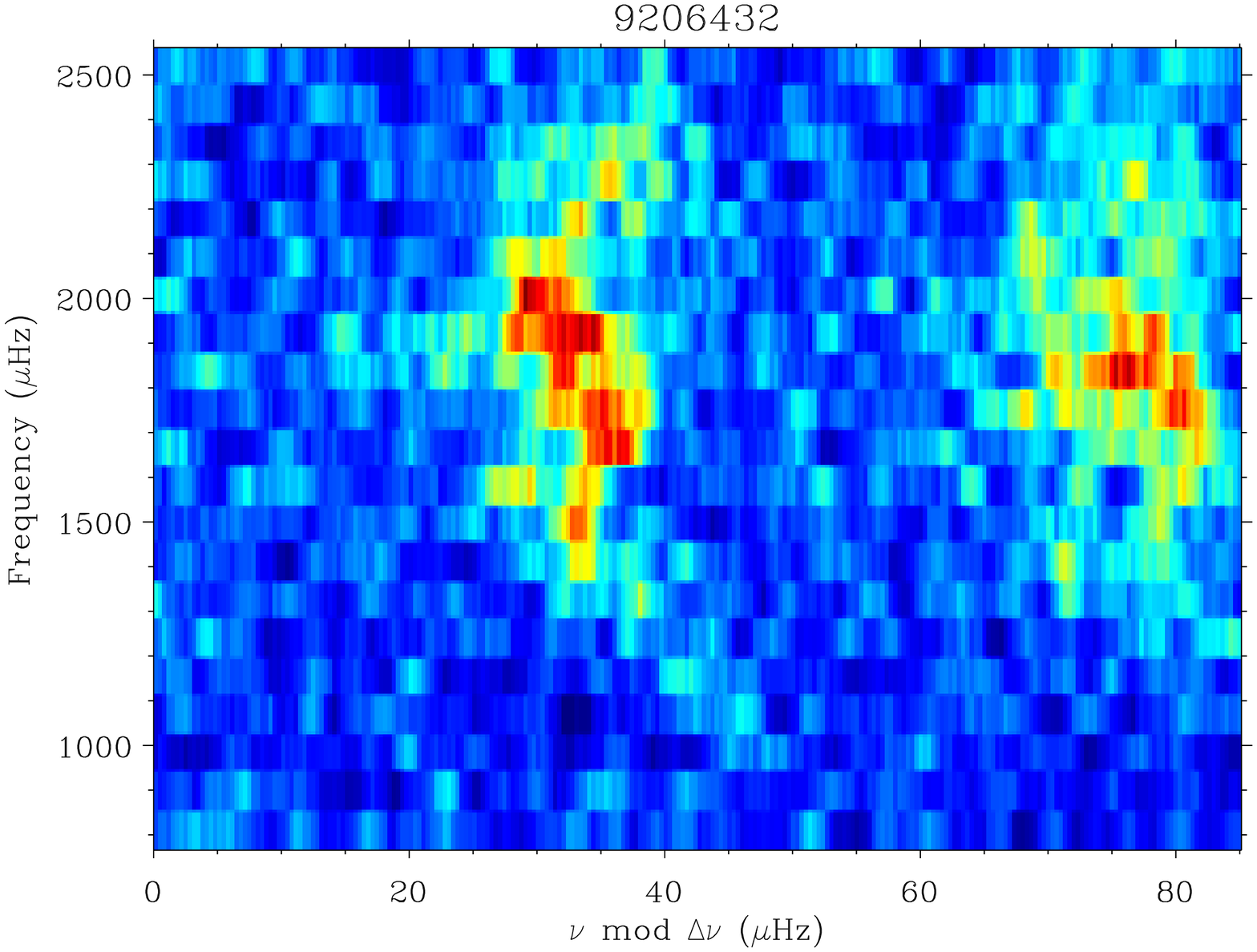}
}
\caption{Echelle diagrams of the power spectra of KIC 8760414, KIC 9025370, KIC 9098294, KIC 9139151, KIC 9139163 and KIC 9206432.  The power spectra are normalised by the background and then smoothed over 3 $\mu$Hz.}
\end{figure*}
\begin{figure*}[!]
\centering
\hbox{
\includegraphics[angle=90,width=9.cm]{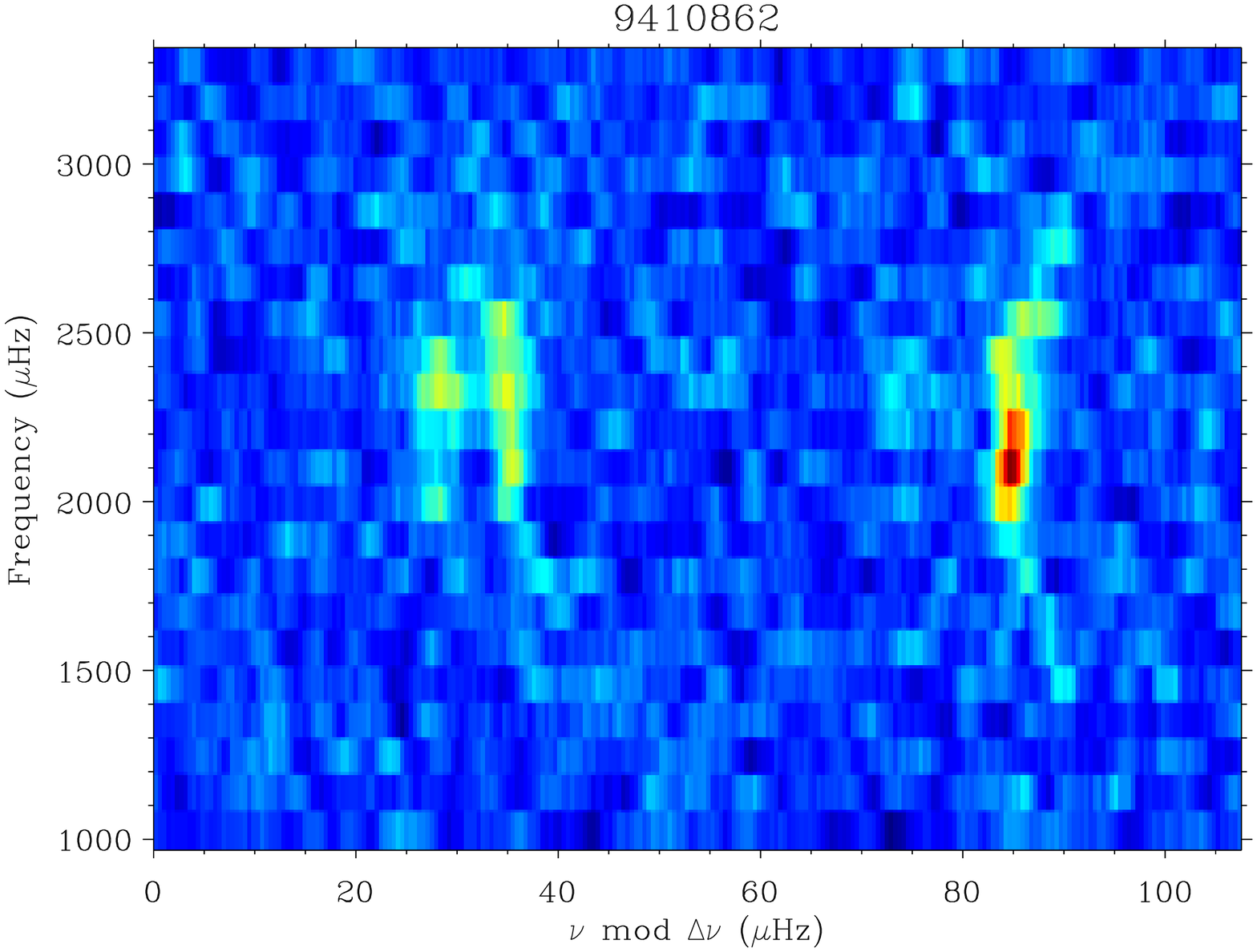}
\includegraphics[angle=90,width=9.cm]{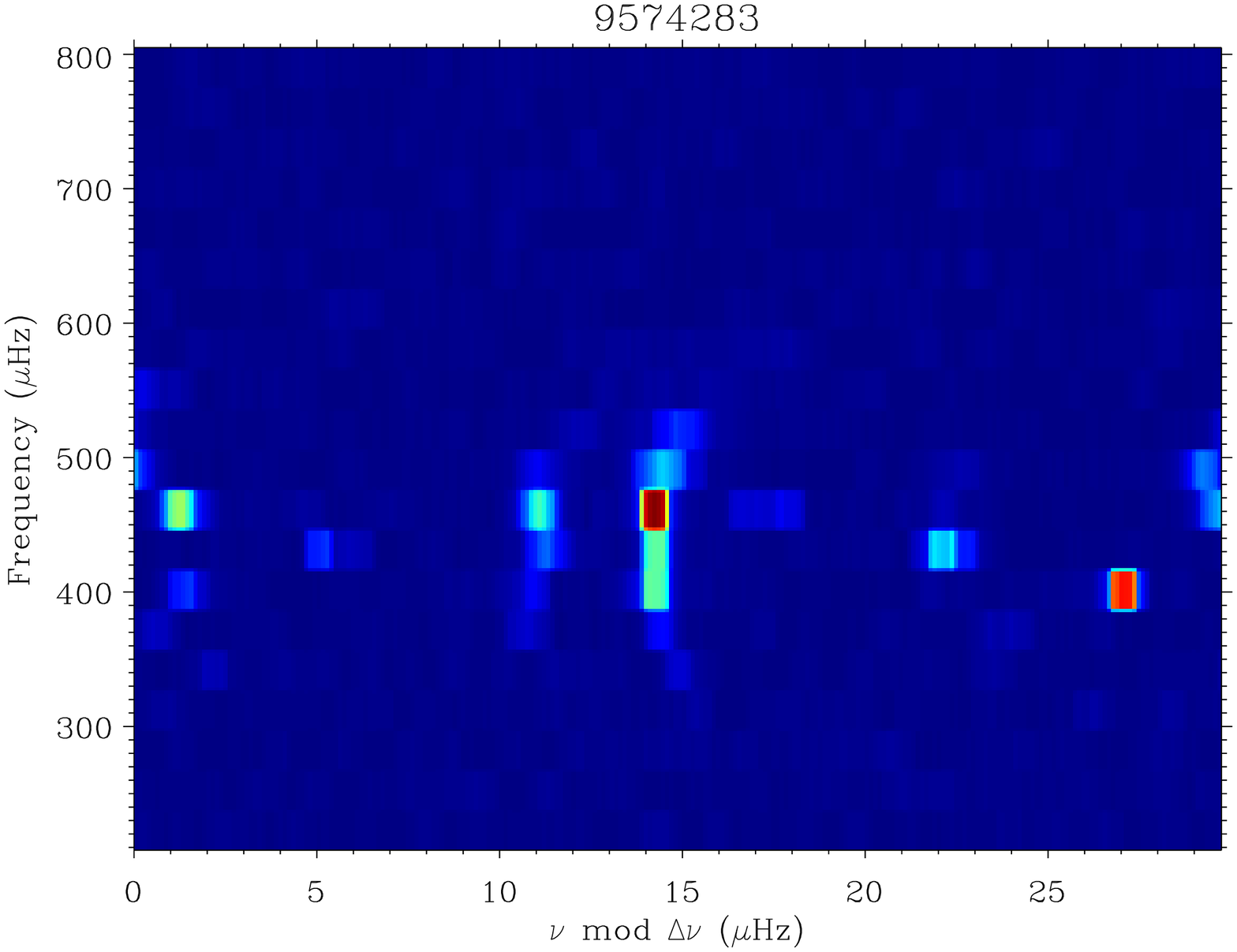}
}
\hbox{
\includegraphics[angle=90,width=9.cm]{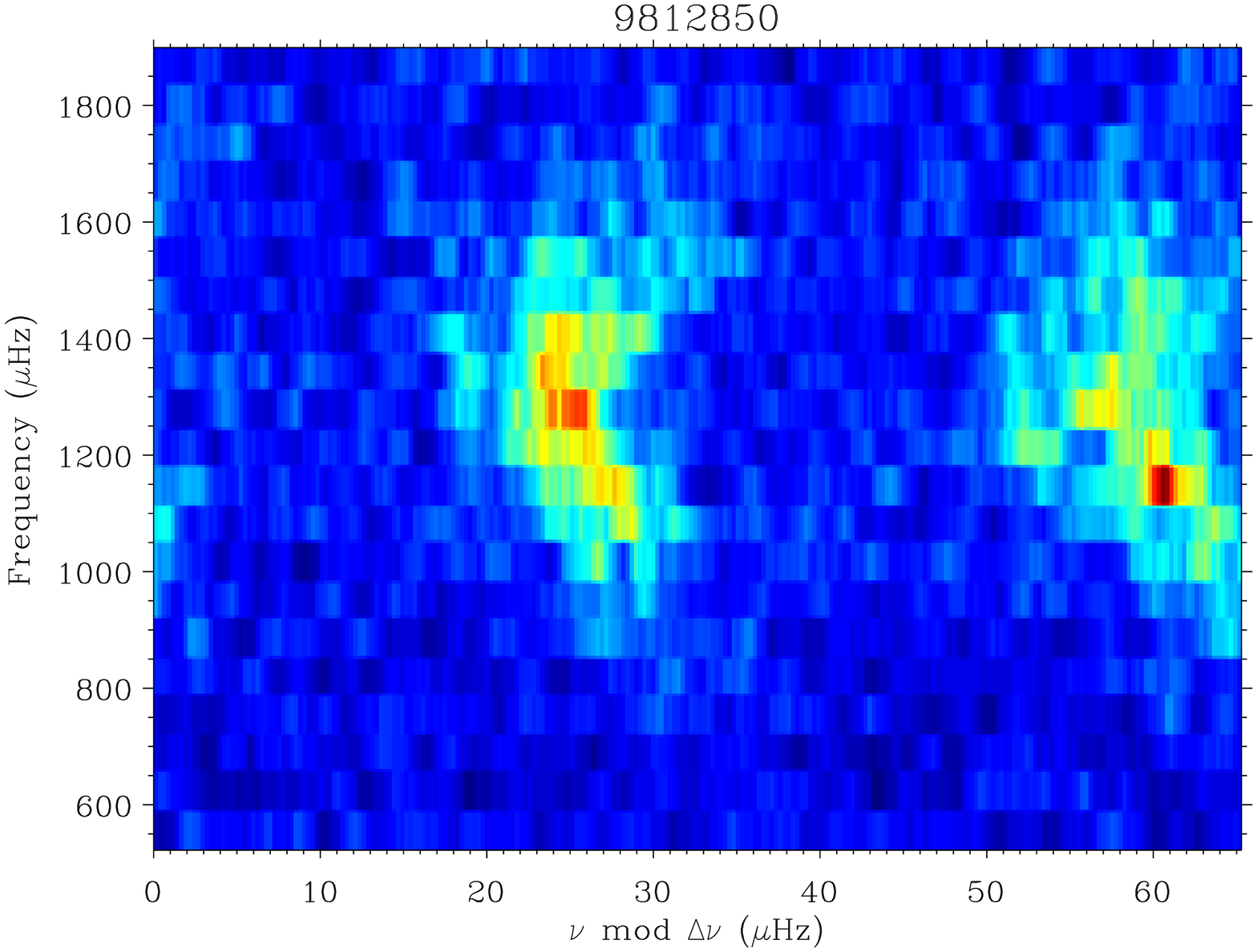}
\includegraphics[angle=90,width=9.cm]{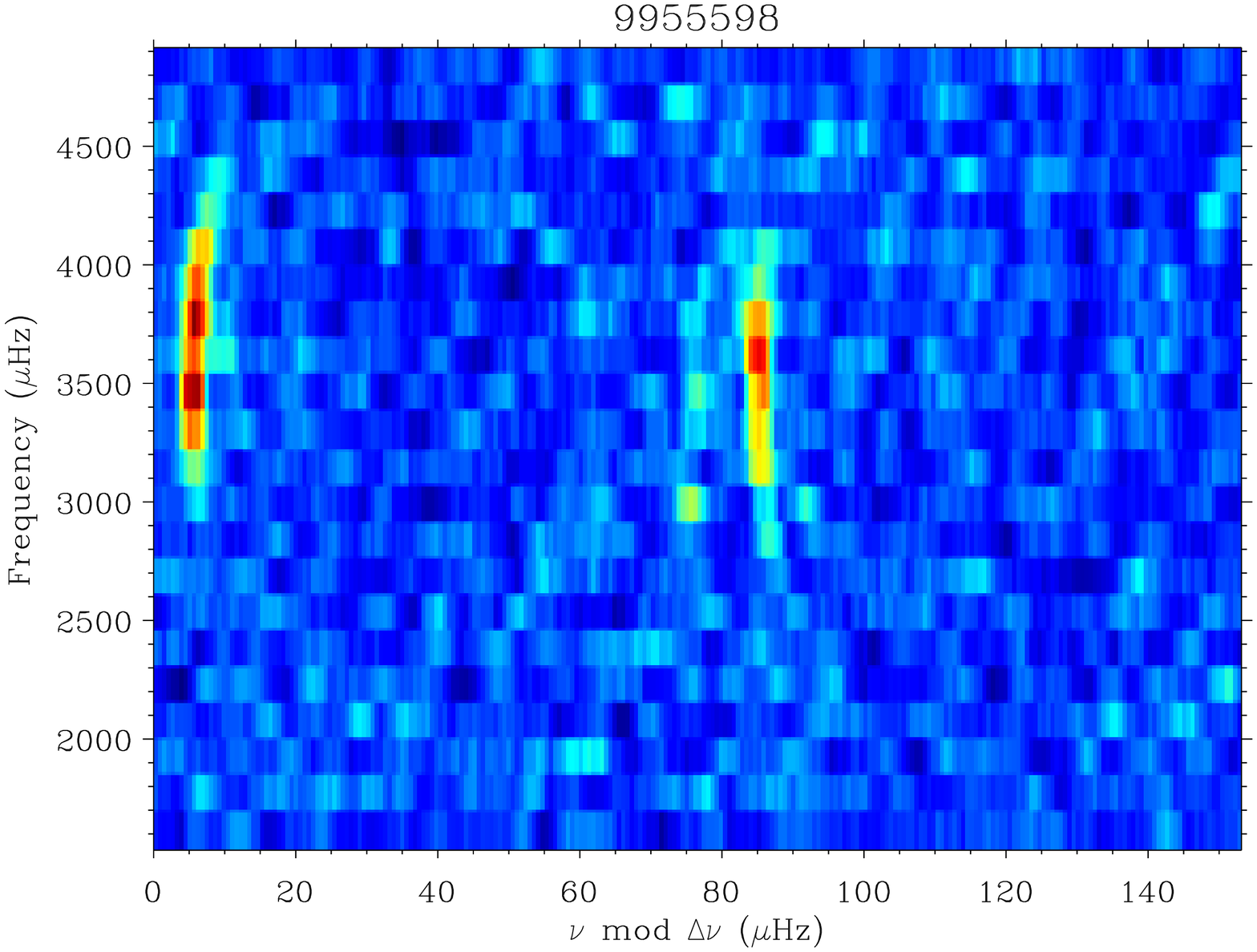}
}
\hbox{
\includegraphics[angle=90,width=9.cm]{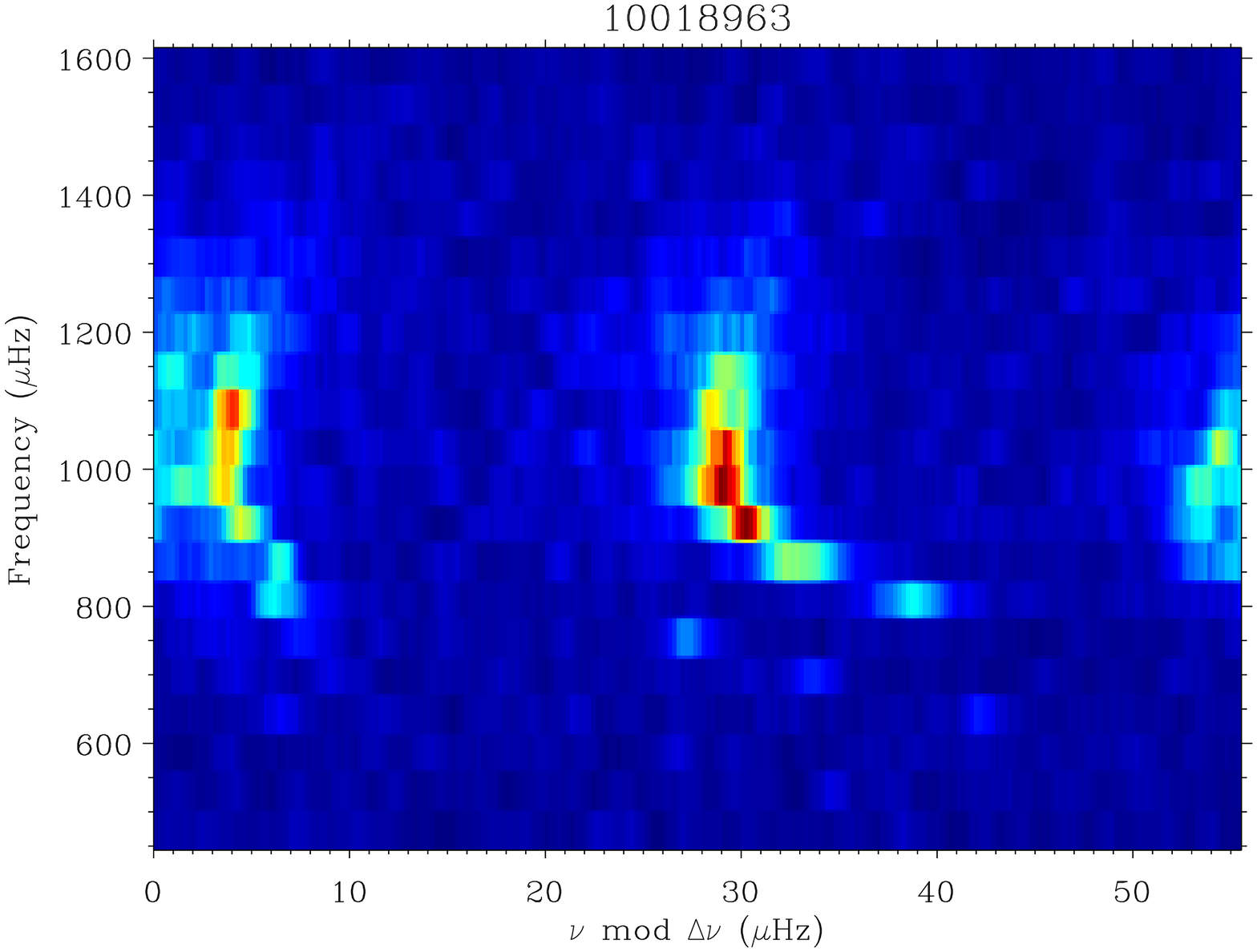}
\includegraphics[angle=90,width=9.cm]{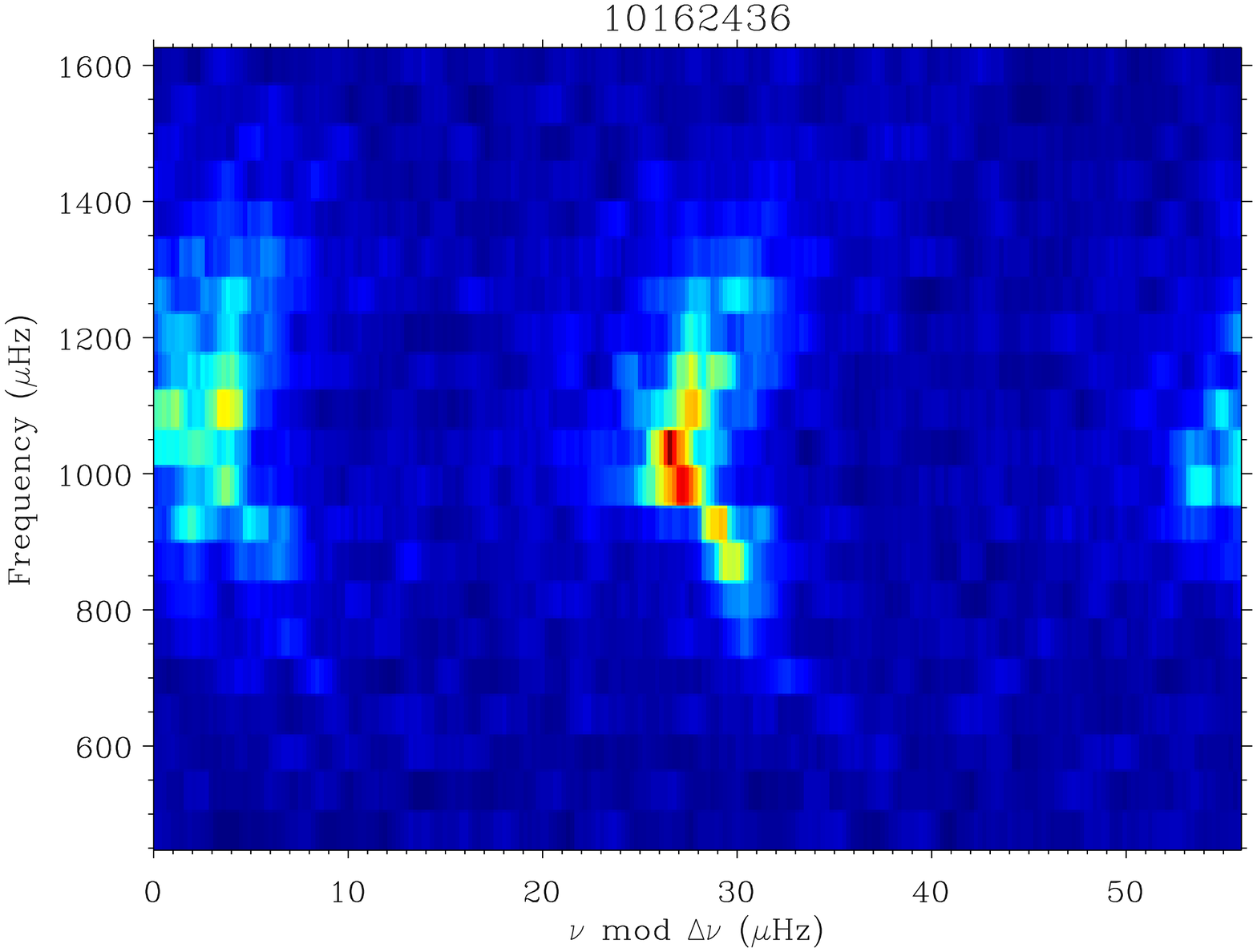}
}
\caption{Echelle diagrams of the power spectra of KIC 9410862, KIC 9574283, KIC 9812850, KIC 9955598, KIC 10018963 and KIC 10162436.  The power spectra are normalised by the background and then smoothed over 3 $\mu$Hz.}
\end{figure*}
\begin{figure*}[!]
\centering
\hbox{
\includegraphics[angle=90,width=9.cm]{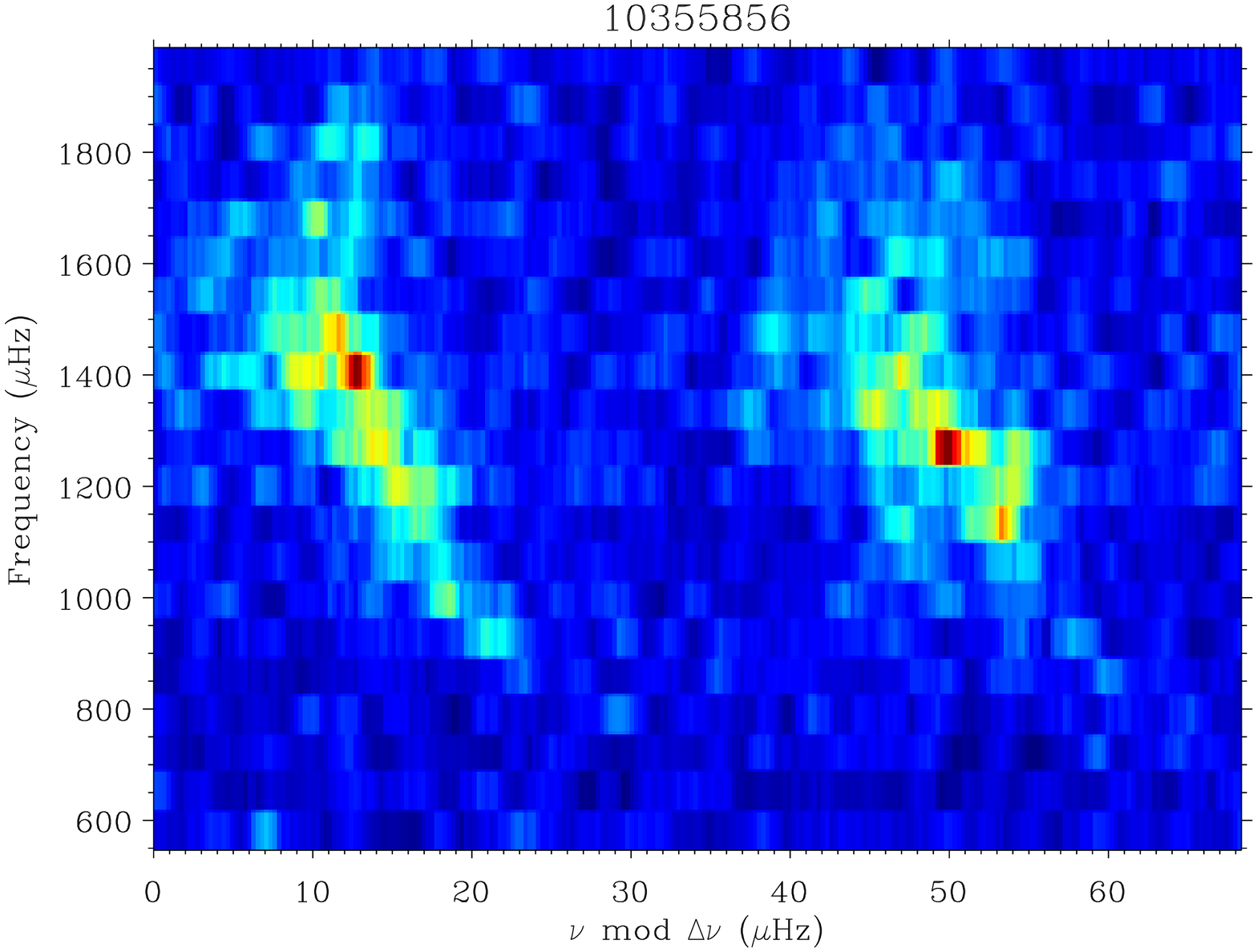}
\includegraphics[angle=90,width=9.cm]{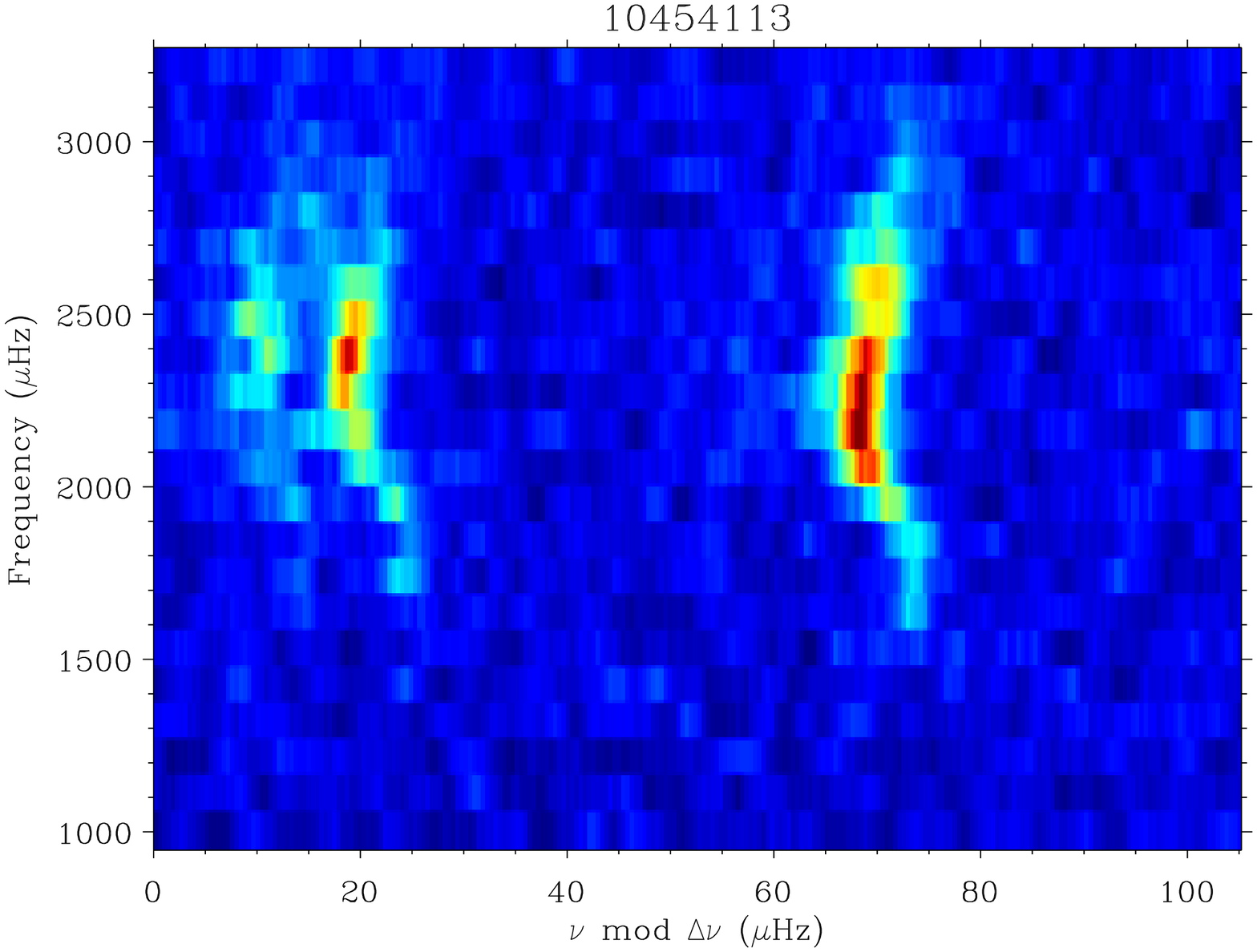}
}
\hbox{
\includegraphics[angle=90,width=9.cm]{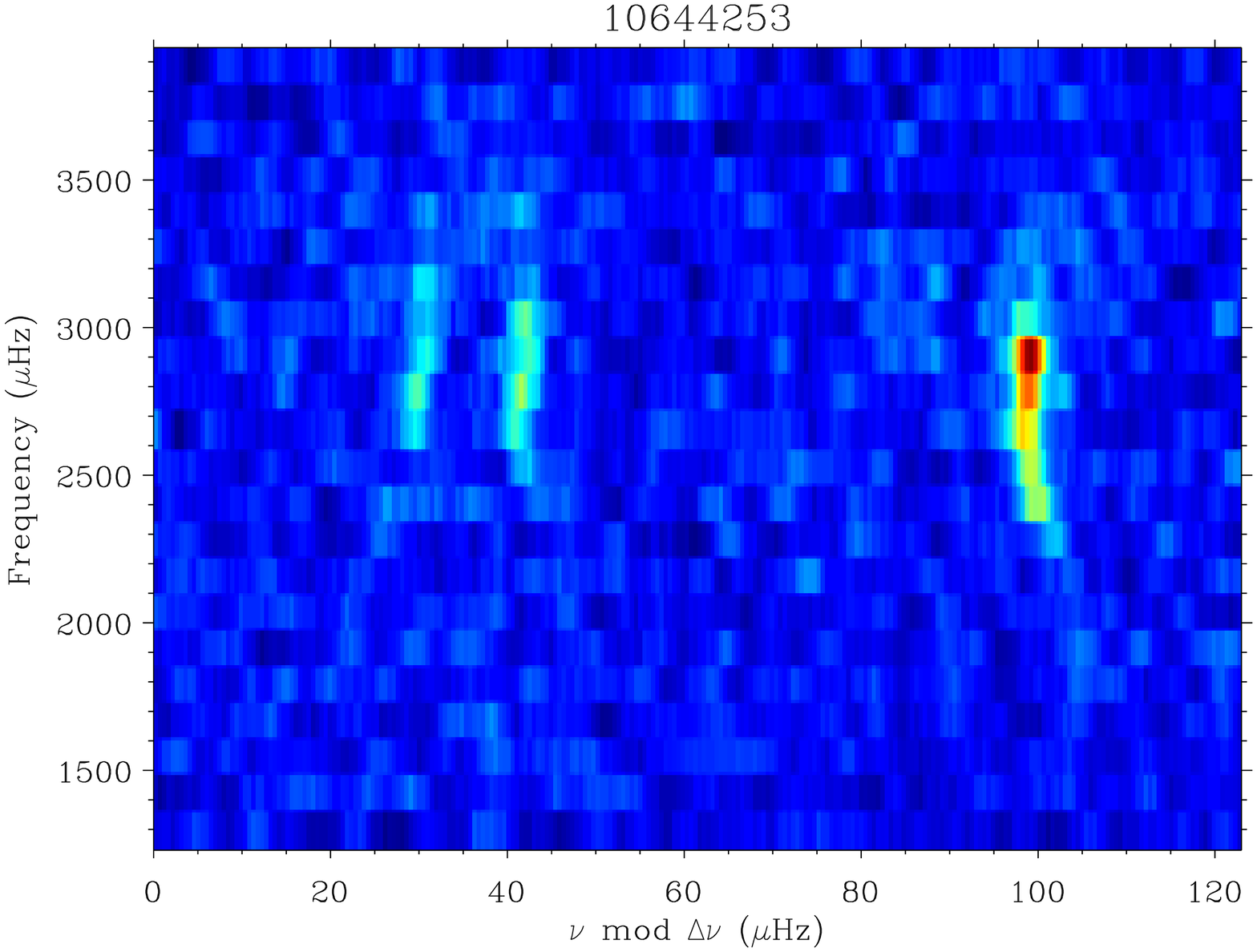}
\includegraphics[angle=90,width=9.cm]{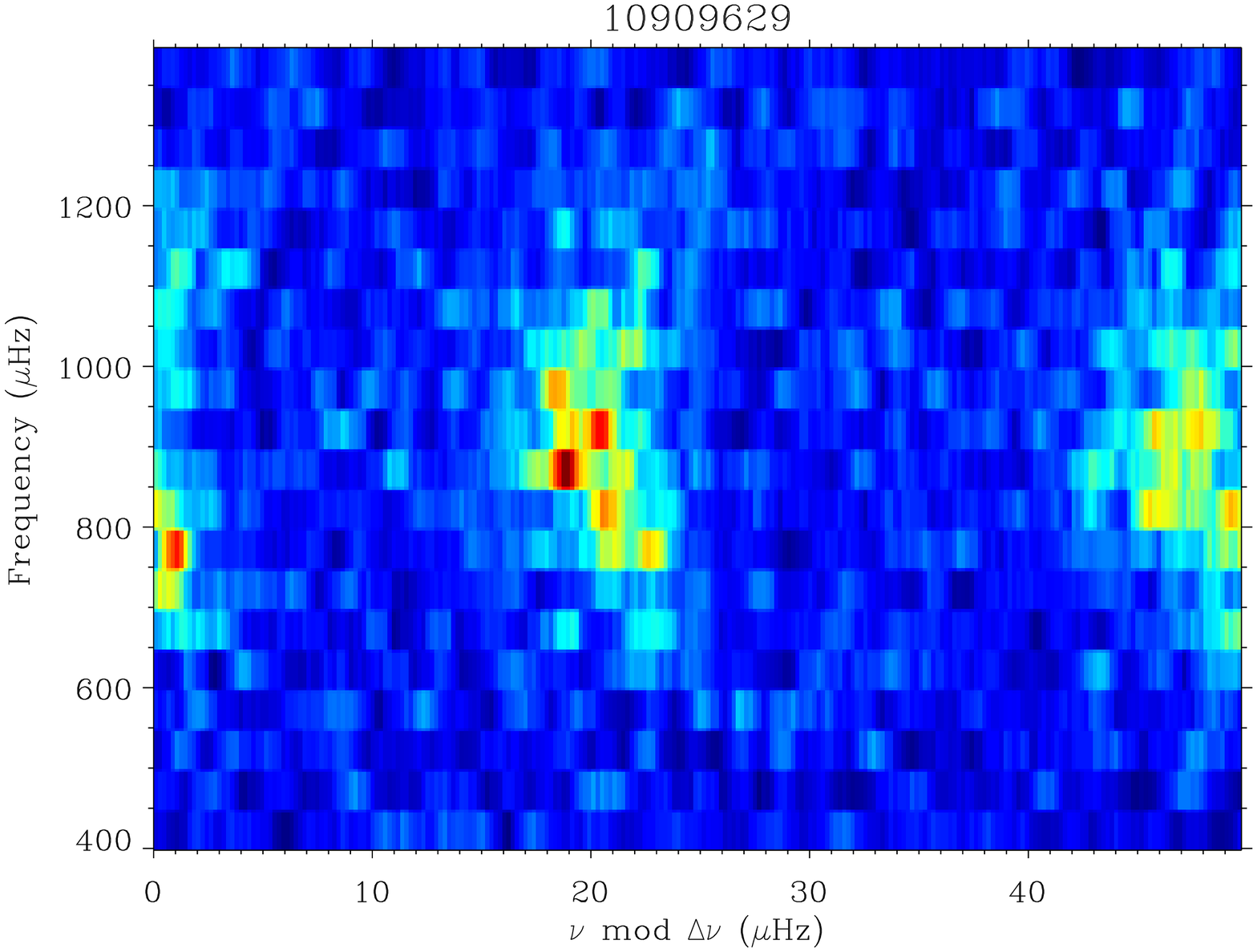}
}
\hbox{
\includegraphics[angle=90,width=9.cm]{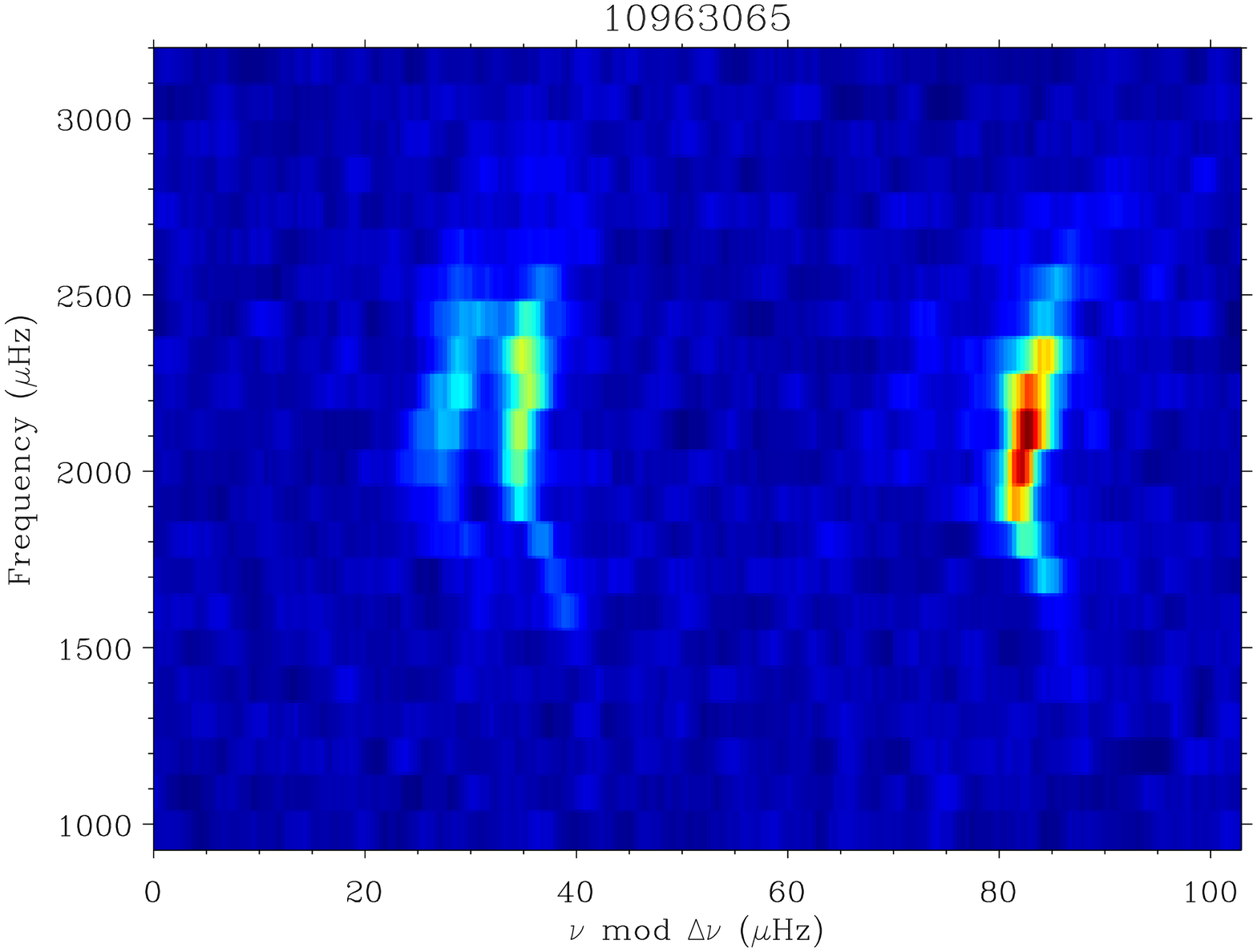}
\includegraphics[angle=90,width=9.cm]{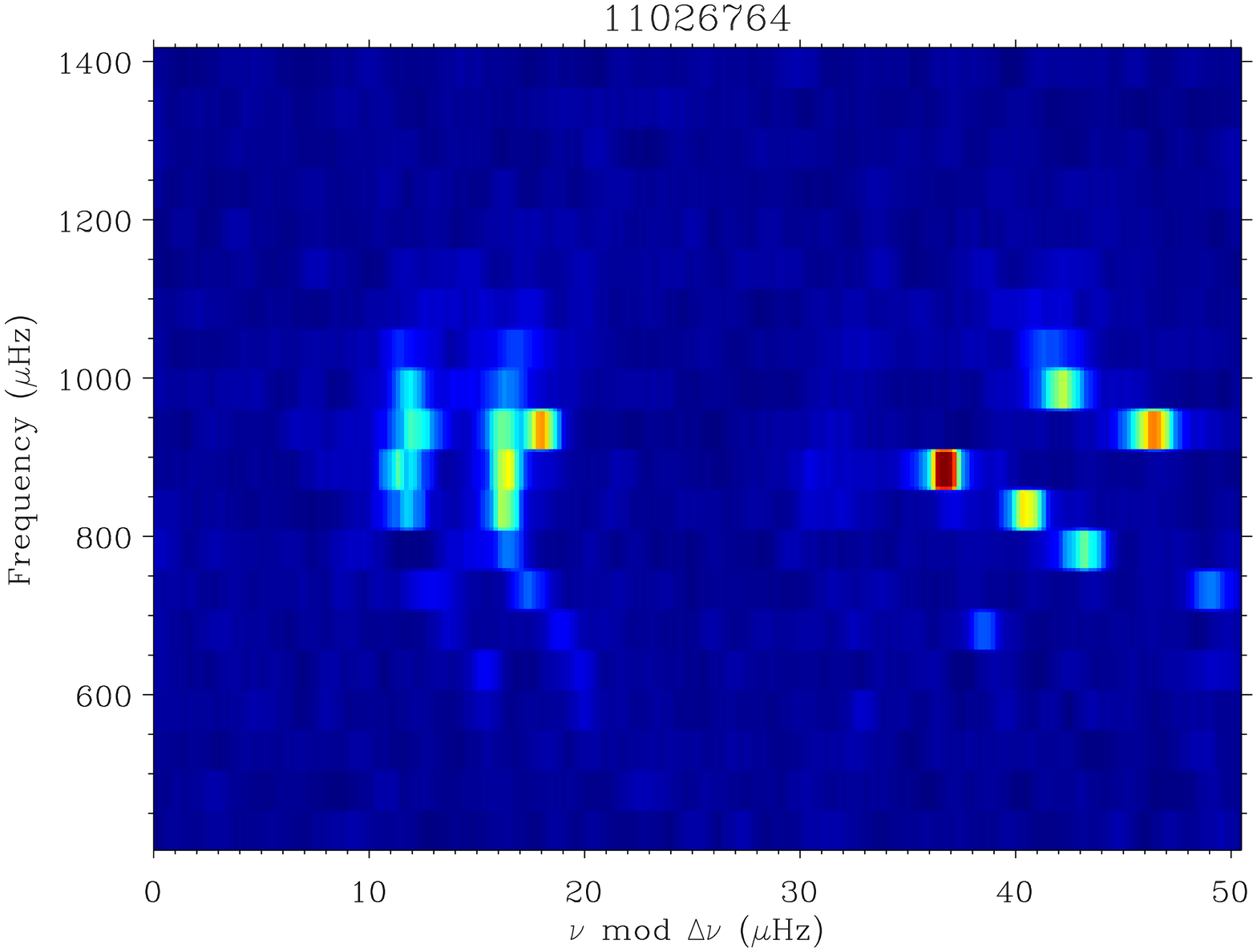}
}
\caption{Echelle diagrams of the power spectra of KIC 10355856, KIC 10454113, KIC 10644253, KIC 10909629, KIC 10963065 and KIC 11026764.  The power spectra are normalised by the background and then smoothed over 3 $\mu$Hz.}
\end{figure*}
\begin{figure*}[!]
\centering
\hbox{
\includegraphics[angle=90,width=9.cm]{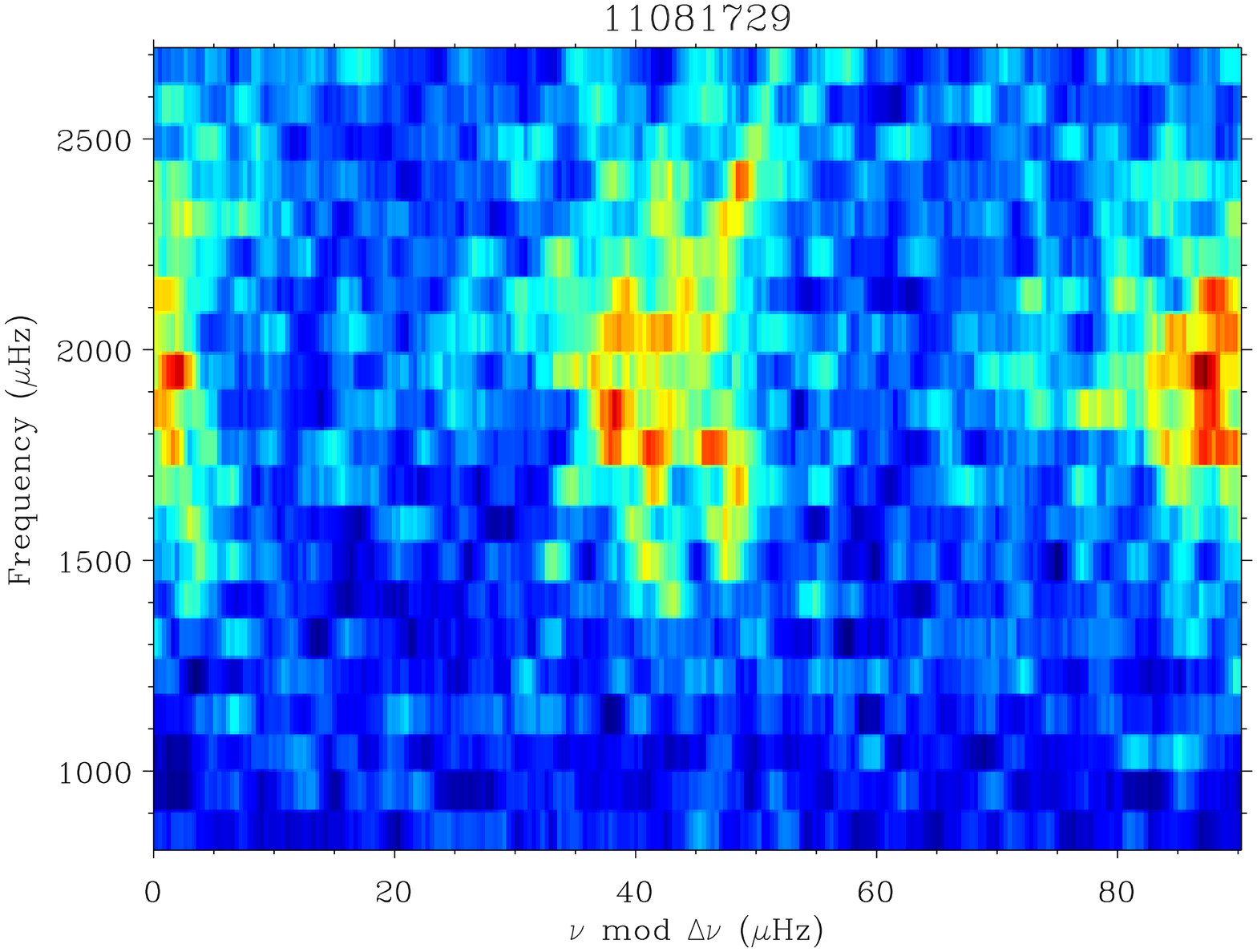}
\includegraphics[angle=90,width=9.cm]{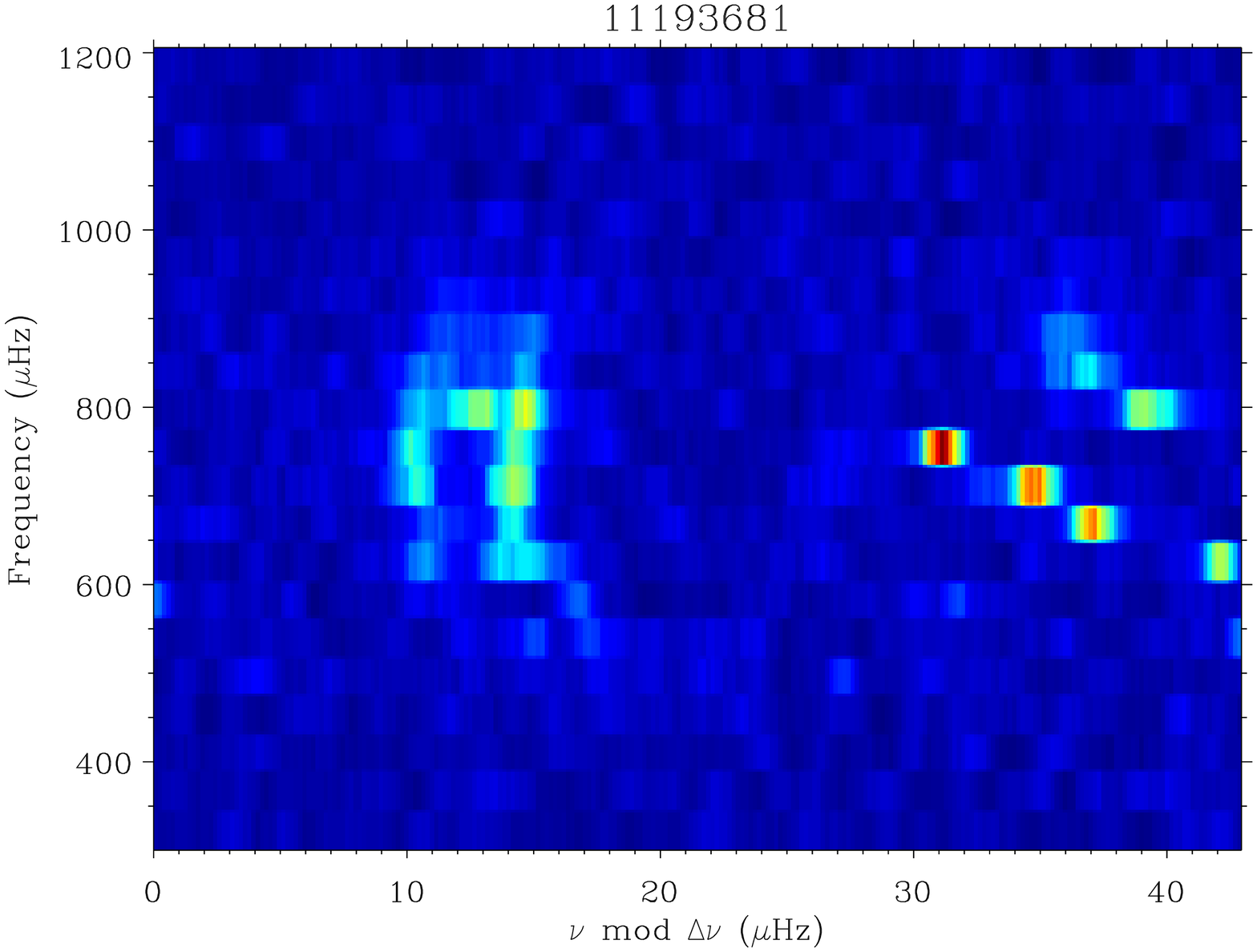}
}
\hbox{
\includegraphics[angle=90,width=9.cm]{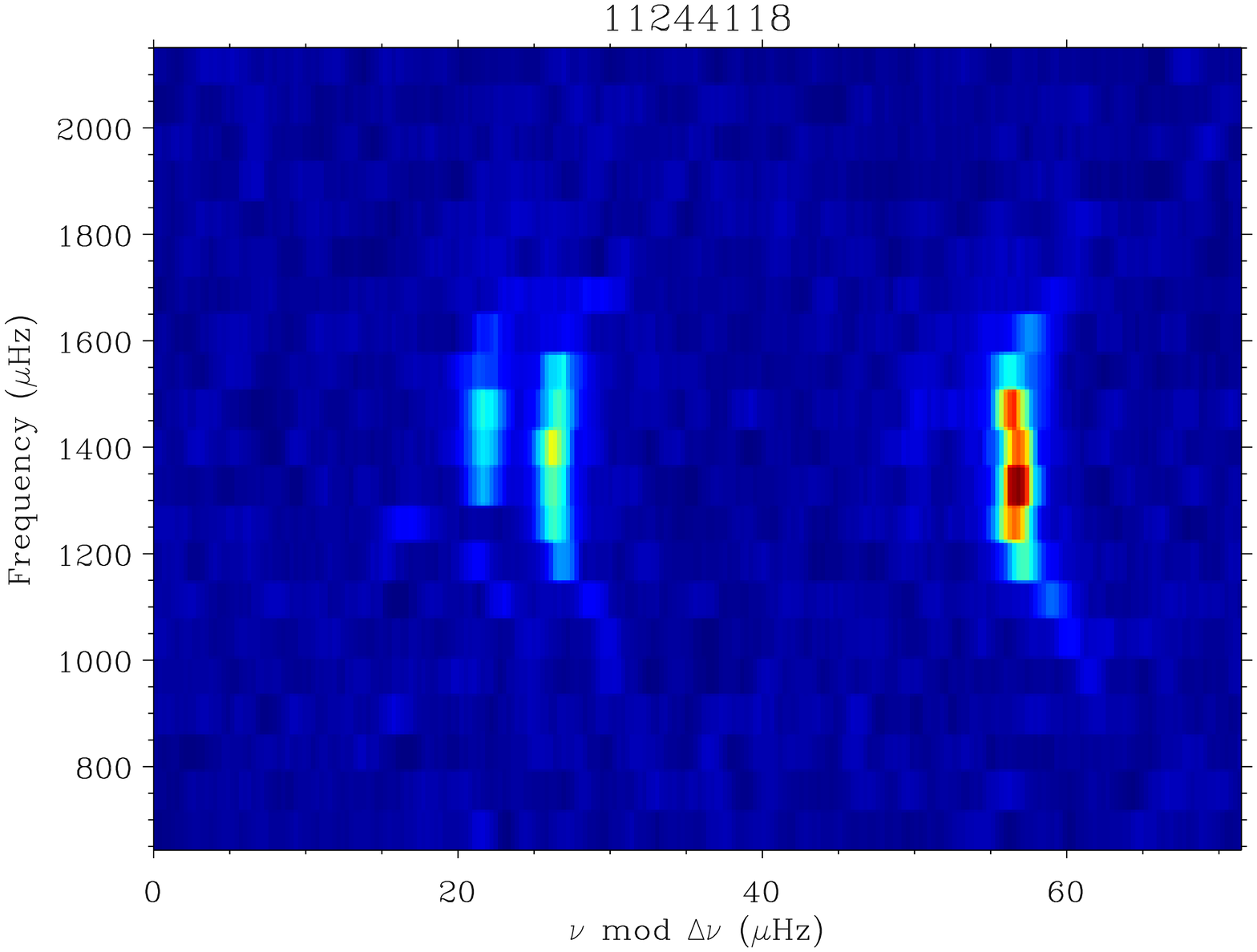}
\includegraphics[angle=90,width=9.cm]{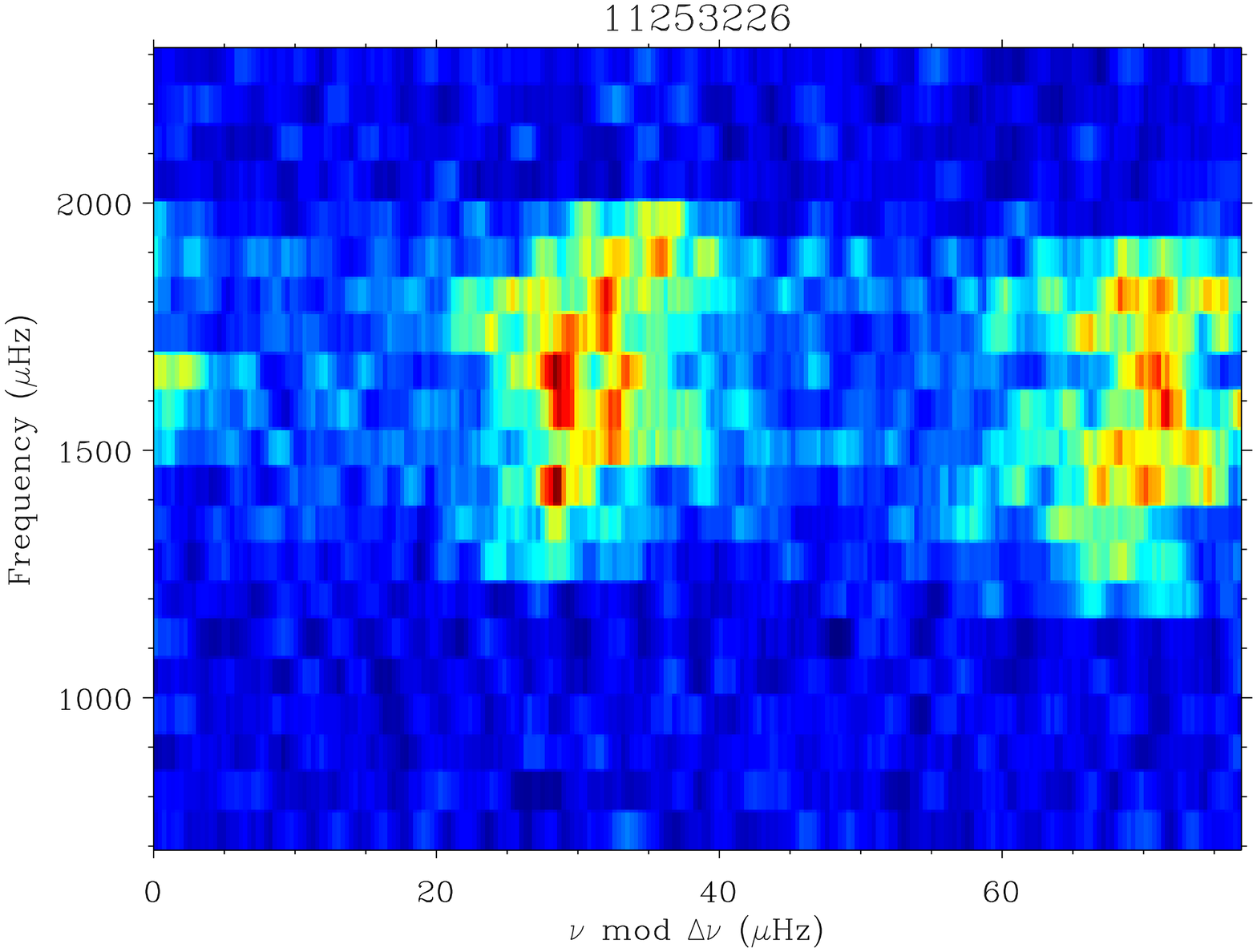}
}
\hbox{
\includegraphics[angle=90,width=9.cm]{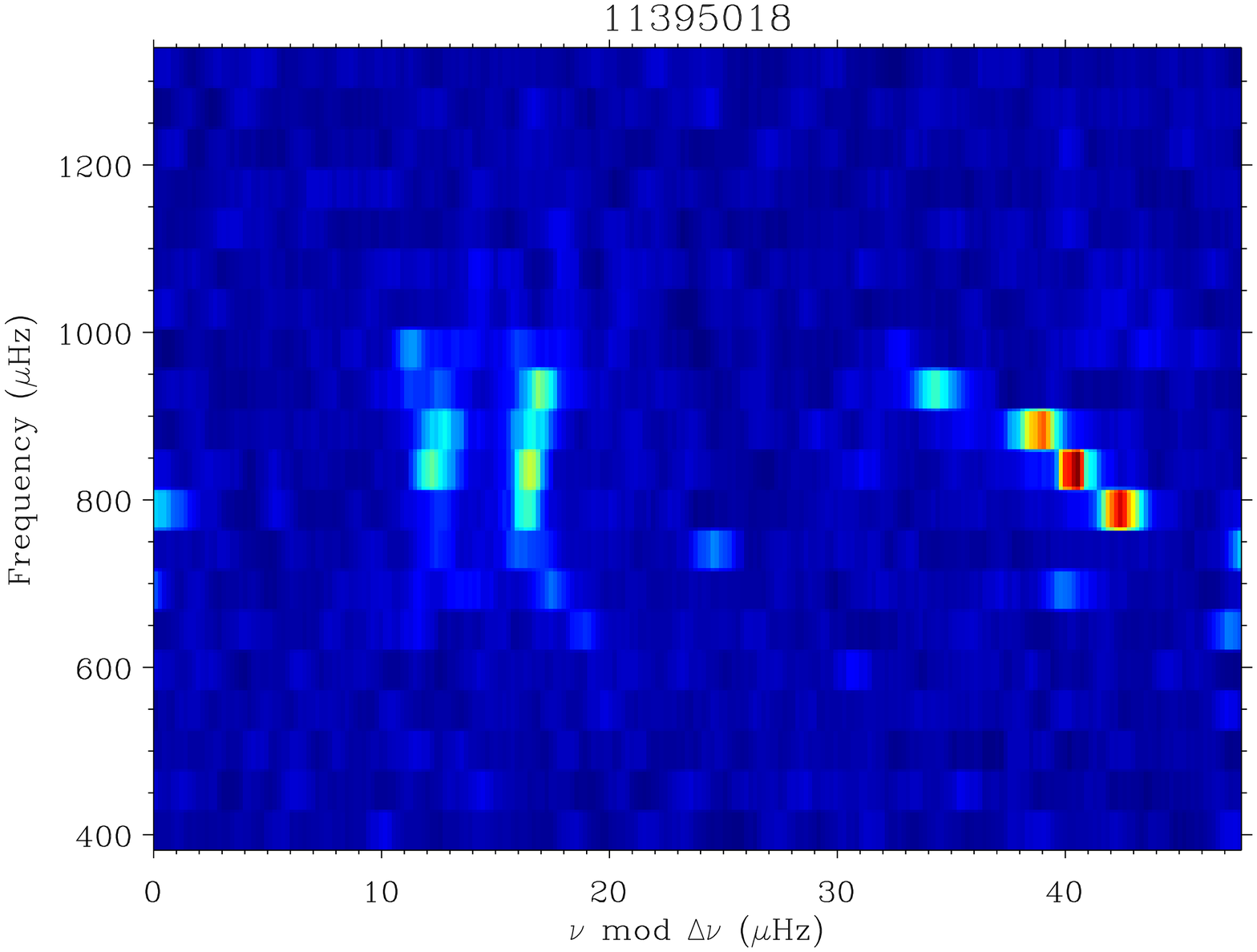}
\includegraphics[angle=90,width=9.cm]{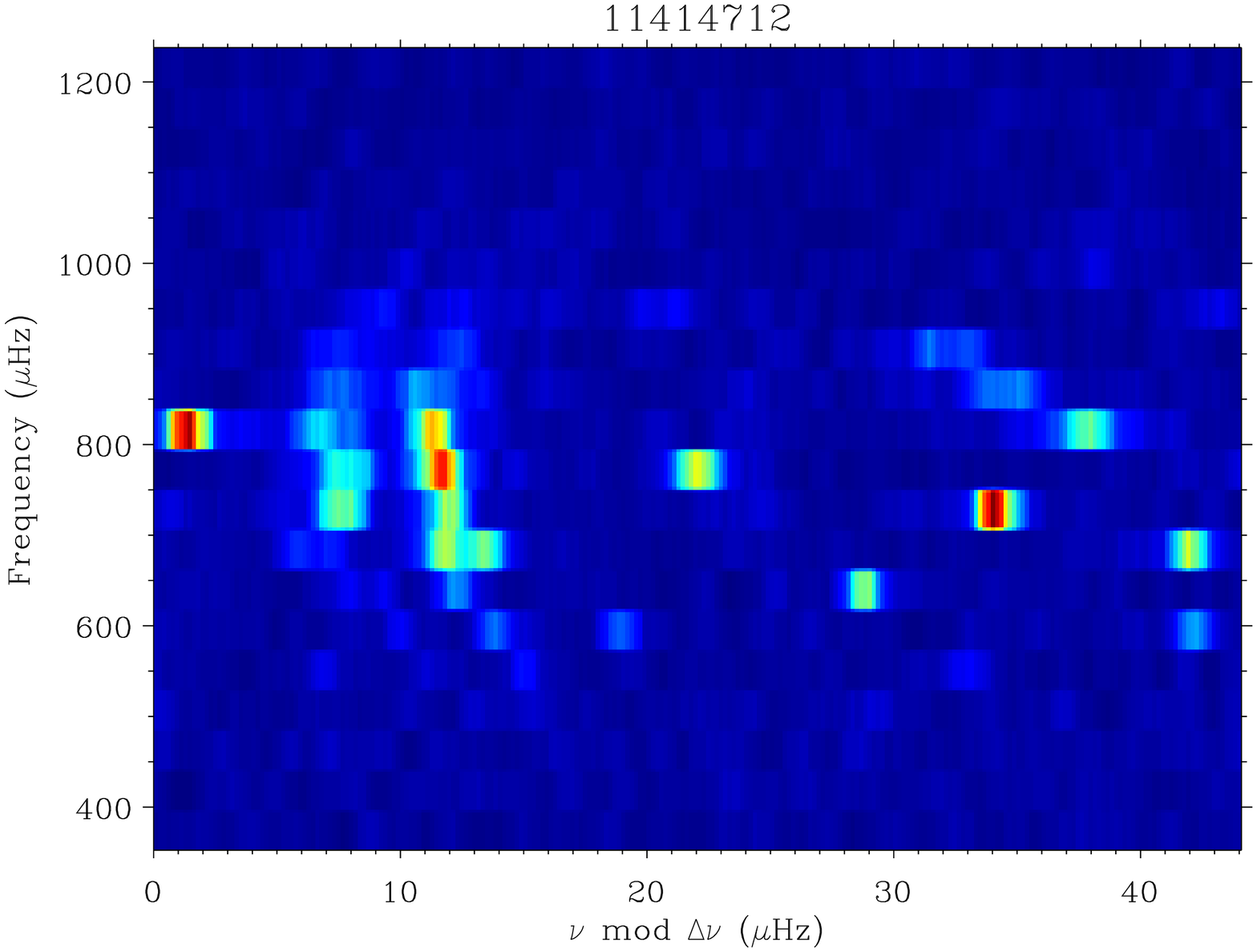}
}
\caption{Echelle diagrams of the power spectra of KIC 11081729, KIC 11193681, KIC 11244118, KIC 11253226, KIC 11395018 and KIC 11414712.  The power spectra are normalised by the background and then smoothed over 3 $\mu$Hz.}
\end{figure*}
\begin{figure*}[!]
\centering
\hbox{
\includegraphics[angle=90,width=9.cm]{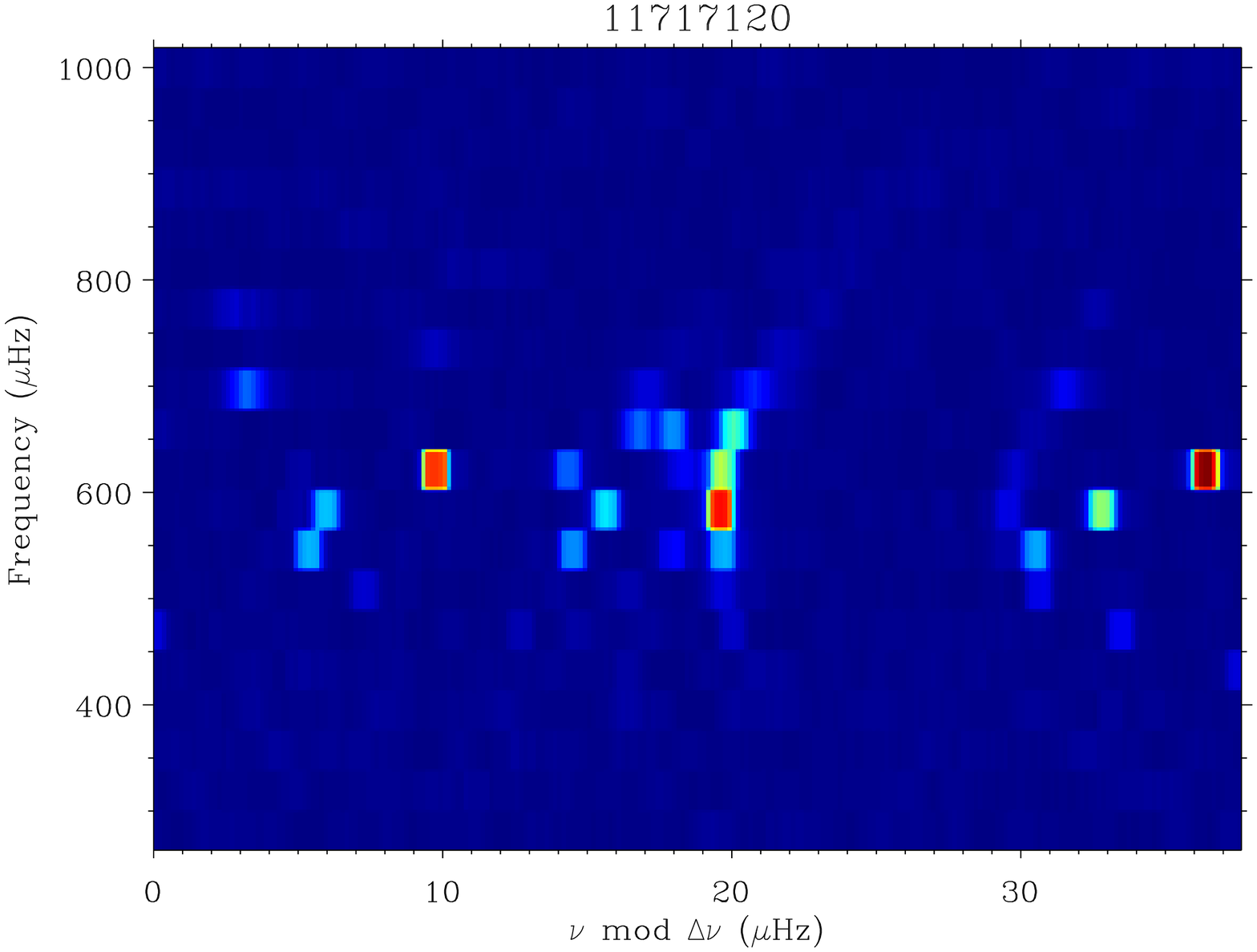}
\includegraphics[angle=90,width=9.cm]{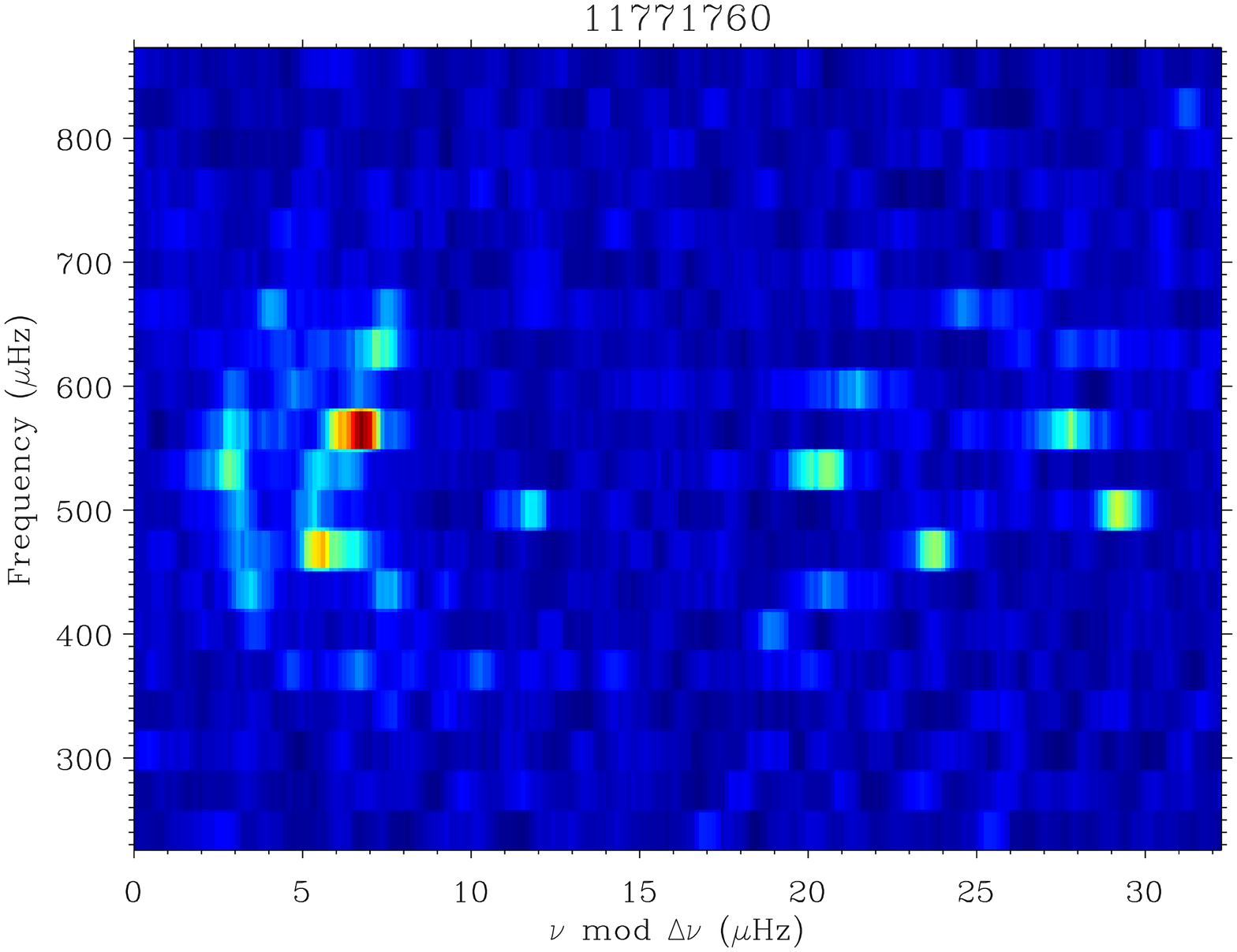}
}
\hbox{
\includegraphics[angle=90,width=9.cm]{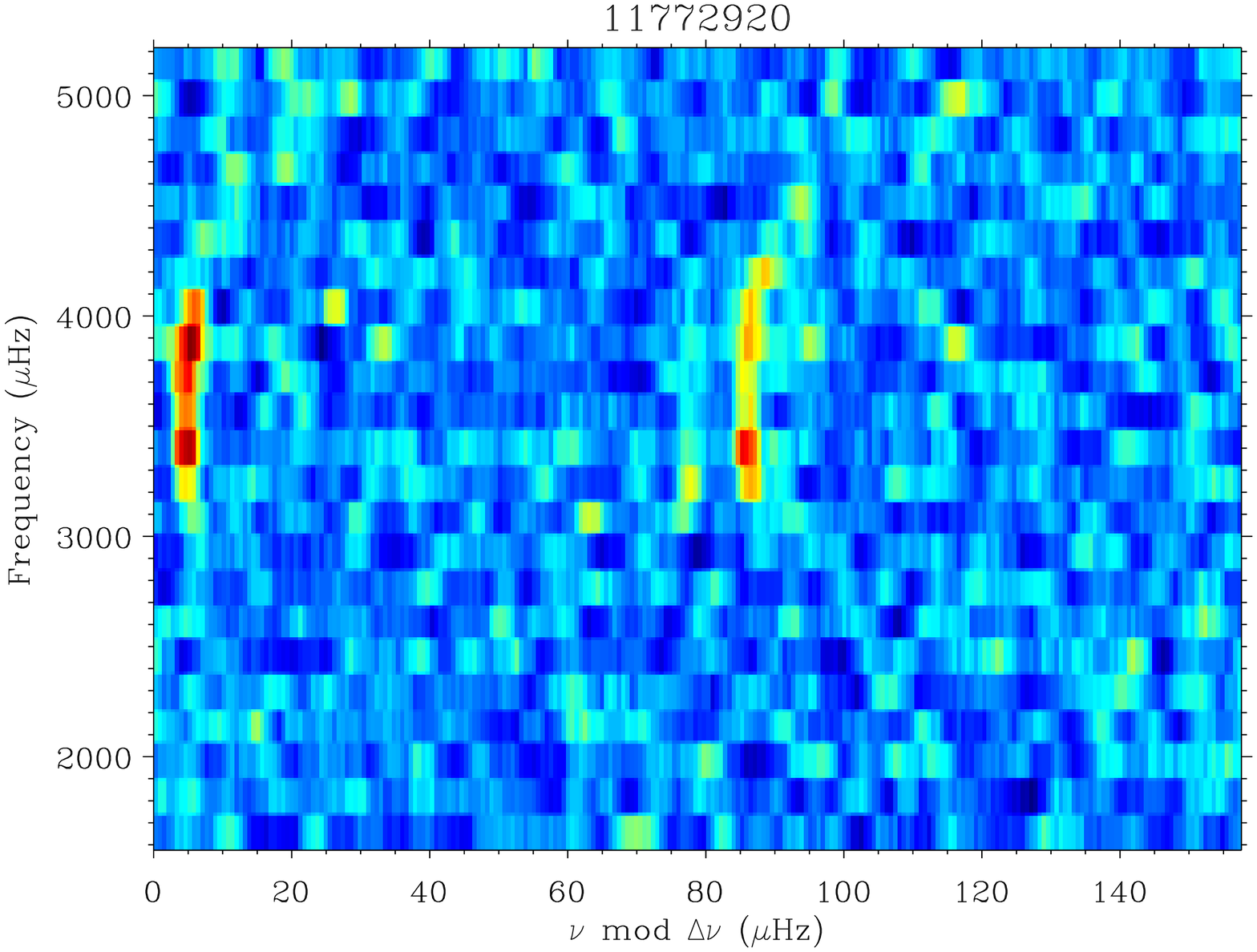}
\includegraphics[angle=90,width=9.cm]{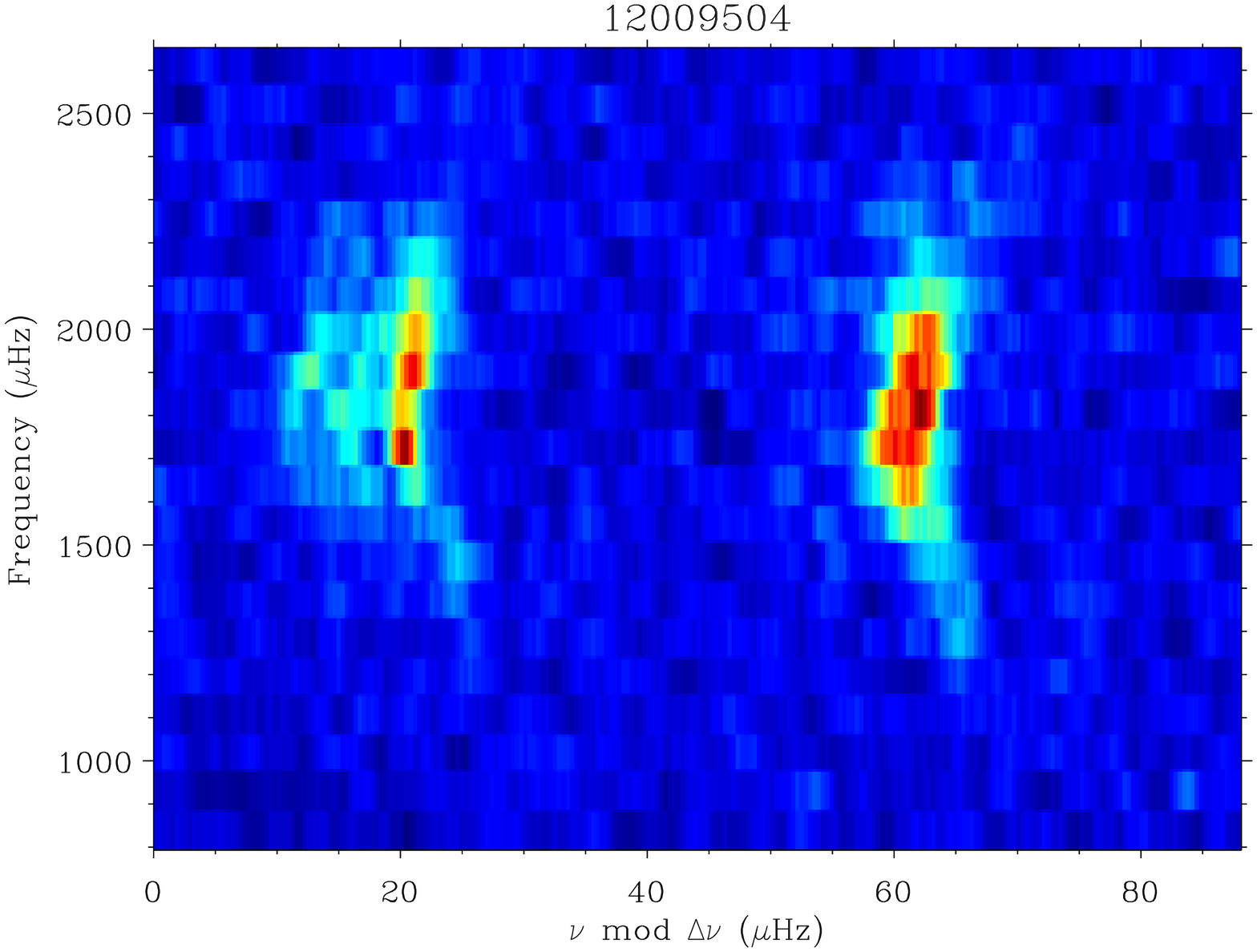}
}
\hbox{
\includegraphics[angle=90,width=9.cm]{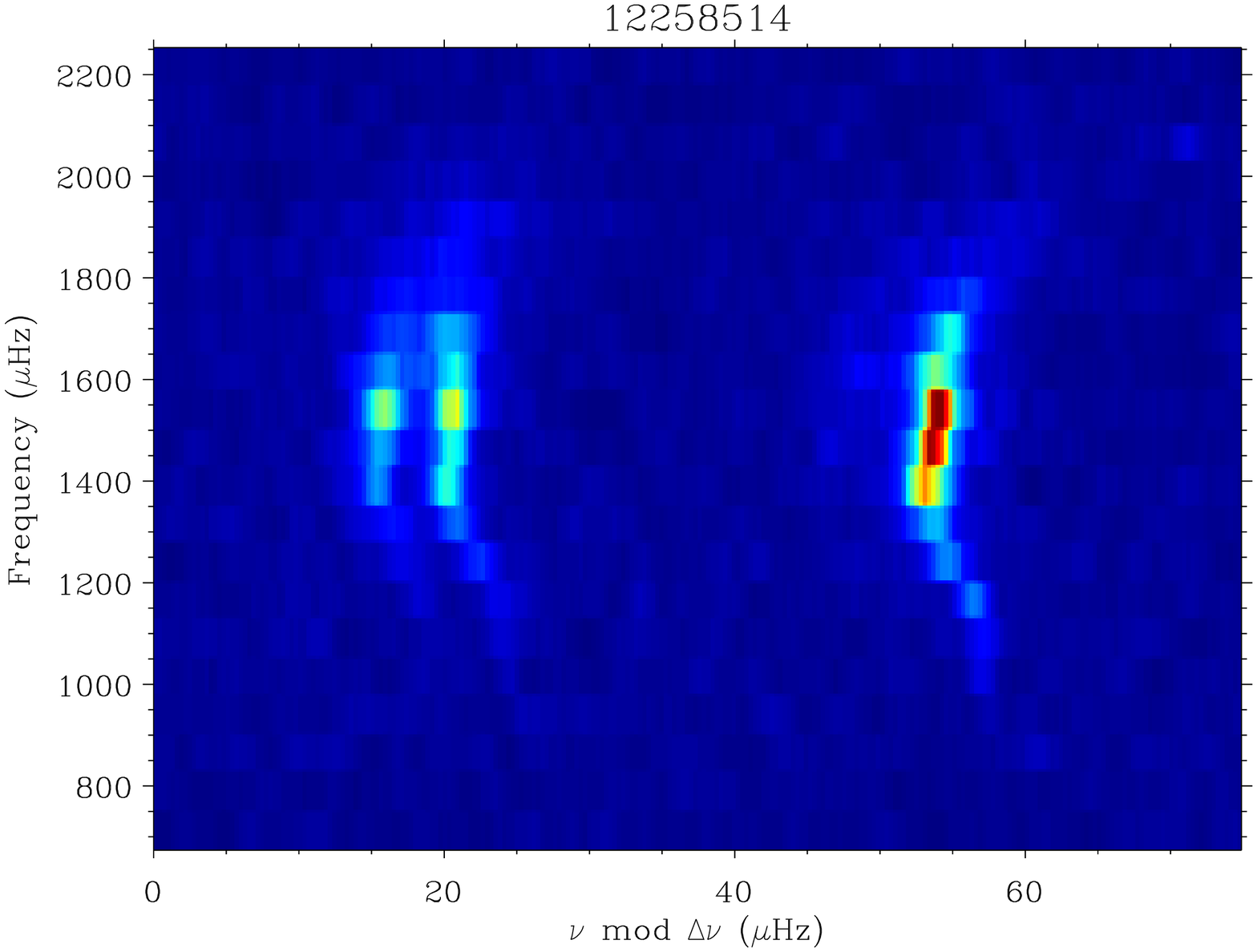}
\includegraphics[angle=90,width=9.cm]{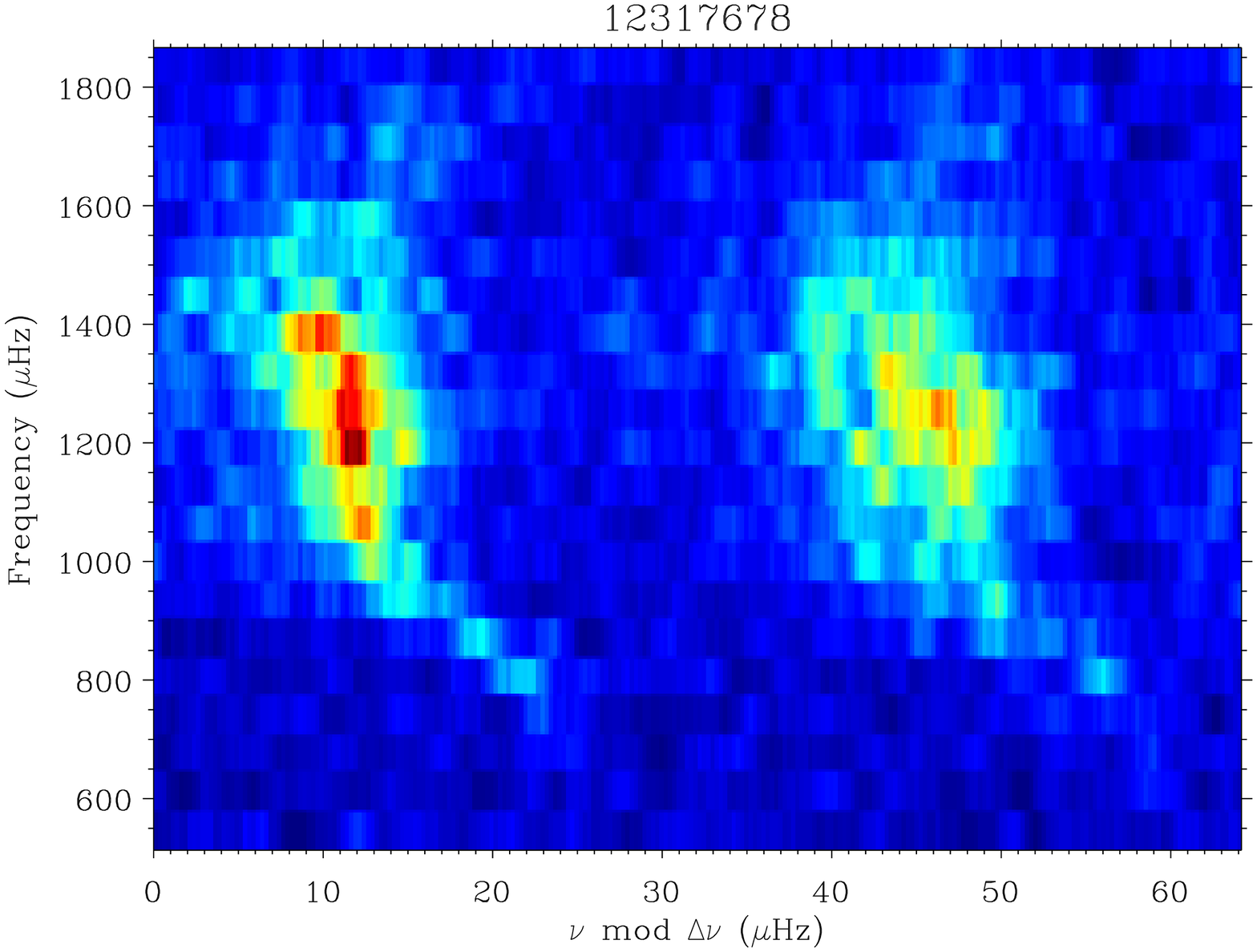}
}
\caption{Echelle diagrams of the power spectra of KIC 11717120, KIC 11771760, KIC 11772920, KIC 12009504, KIC 12258514 and KIC 12317678.  The power spectra are normalised by the background and then smoothed over 3 $\mu$Hz.}
\end{figure*}


\begin{table*}[!]
\caption{Frequencies for KIC 1435467. The first column is the degree.  The second column is the frequency.  The third column is the 68\%-credible interval quoted when the mode is fitted.  The last column provides an indication of the quality of the detection: {\it OK} indicates that the mode was correctly detected and fitted; {\it Not detected} indicates that the mode was fitted but not detected by the quality assurance test and {\it Not fitted} indicates that the mode was detected with a posterior probability provided by the quality assurance test.  When an uncertainty {\it and} a posterior probability are quoted, it means that the mode is fitted but detected using the quality assurance test with a probability lower than 90\%.}
\centering
\begin{tabular}{c c c c} 
\hline
\hline
Degree&Frequency ($\mu$Hz)& 68\% credible ($\mu$Hz) &Comment\\
\hline
\hline
       0&992.752&+0.817 / -0.635&OK\\
       0&1064.053&+0.475 / -0.561&OK\\
       0&1135.629&+0.410 / -0.450&OK\\
       0&1206.221&+0.188 / -0.226&OK\\
       0&1274.392&+0.271 / -0.316&OK\\
       0&1344.028&+0.317 / -0.357&OK\\
       0&1414.186&+0.301 / -0.338&OK\\
       0&1485.429&+0.358 / -0.318&OK\\
       0&1556.749&+0.336 / -0.378&OK\\
       0&1626.041&+0.477 / -0.477&OK\\
       0&1698.262&+0.572 / -0.531&OK\\
       0&1770.598&+0.951 / -0.951&OK\\
       0&1842.029&+0.973 / -1.022&OK\\
       0&1912.611&+0.959 / -1.012&OK\\
       0&1983.466&+1.426 / -1.473&OK\\
\hline
       1&956.270&Not fitted&0.710\\
       1&1026.739&+0.452 / -0.508&OK\\
       1&1096.903&+0.345 / -0.388&OK\\
       1&1168.054&+0.338 / -0.338&OK\\
       1&1237.866&+0.205 / -0.205&OK\\
       1&1307.203&+0.317 / -0.277&OK\\
       1&1376.074&+0.269 / -0.269&OK\\
       1&1445.906&+0.292 / -0.292&OK\\
       1&1517.838&+0.314 / -0.314&OK\\
       1&1590.530&+0.293 / -0.293&OK\\
       1&1662.385&+0.376 / -0.376&OK\\
       1&1733.286&+0.462 / -0.420&OK\\
       1&1804.266&+0.618 / -0.665&OK\\
       1&1874.800&+0.756 / -0.756&OK\\
       1&1946.806&+0.759 / -0.658&OK\\
       1&2019.615&+0.819 / -0.877&OK\\
\hline
       2&989.305&+2.308 / -2.451&OK\\
       2&1059.494&+1.300 / -1.298&OK\\
       2&1129.521&+0.746 / -0.746&OK\\
       2&1199.325&+0.449 / -0.628&OK\\
       2&1268.328&+0.642 / -0.642&OK\\
       2&1338.288&+0.709 / -0.620&OK\\
       2&1408.902&+0.547 / -0.638&OK\\
       2&1480.234&+0.648 / -0.648&OK\\
       2&1551.839&+0.698 / -0.698&OK\\
       2&1623.900&+0.685 / -0.684&OK\\
       2&1695.649&+0.898 / -0.987&OK\\
       2&1766.228&+1.285 / -1.375&OK\\
       2&1836.974&+1.317 / -1.410&OK\\
       2&1907.883&+1.362 / -1.452&OK\\
       2&1979.145&+2.073 / -2.259&OK\\
\hline
\hline
\end{tabular}
\label{1435467}
\end{table*}   %
\begin{table*}[!]
\caption{Frequencies for KIC 2837475. The first column is the degree.  The second column is the frequency.  The third column is the 68\%-credible interval quoted when the mode is fitted.  The last column provides an indication of the quality of the detection: {\it OK} indicates that the mode was correctly detected and fitted; {\it Not detected} indicates that the mode was fitted but not detected by the quality assurance test and {\it Not fitted} indicates that the mode was detected with a posterior probability provided by the quality assurance test.  When an uncertainty {\it and} a posterior probability are quoted, it means that the mode is fitted but detected using the quality assurance test with a probability lower than 90\%.}
\centering
\begin{tabular}{c c c c} 
\hline
\hline
Degree&Frequency ($\mu$Hz)& 68\% credible ($\mu$Hz) &Comment\\
\hline
\hline
       0&985.740&+0.169 / -0.225&OK\\
       0&1057.562&+0.842 / -0.841&OK\\
       0&1130.442&+0.702 / -0.654&OK\\
       0&1205.403&+0.547 / -0.638&OK\\
       0&1279.772&+0.674 / -0.674&OK\\
       0&1355.167&+0.515 / -0.566&OK\\
       0&1432.082&+0.514 / -0.608&OK\\
       0&1509.488&+0.409 / -0.368&OK\\
       0&1585.937&+0.484 / -0.537&OK\\
       0&1660.125&+0.631 / -0.631&OK\\
       0&1734.503&+0.654 / -0.654&OK\\
       0&1810.869&+0.946 / -0.946&OK\\
       0&1886.532&+1.081 / -1.029&OK\\
       0&1963.679&+1.459 / -1.400&OK\\
       0&2041.304&+2.430 / -2.246&OK\\
       0&2117.550&Not fitted&0.630\\
       0&2194.480&Not fitted&0.813\\
       0&2265.410&Not fitted&0.807\\
\hline
       1&1020.321&+0.258 / -0.309&OK\\
       1&1093.727&+0.617 / -0.566&OK\\
       1&1167.088&+0.578 / -0.525&OK\\
       1&1241.065&+0.532 / -0.577&OK\\
       1&1315.382&+0.654 / -0.653&OK\\
       1&1390.946&+0.467 / -0.467&OK\\
       1&1468.287&+0.518 / -0.518&OK\\
       1&1544.878&+0.431 / -0.431&OK\\
       1&1620.036&+0.467 / -0.513&OK\\
       1&1695.664&+0.432 / -0.432&OK\\
       1&1771.638&+0.523 / -0.522&OK\\
       1&1845.982&+0.607 / -0.560&OK\\
       1&1922.405&+0.686 / -0.686&OK\\
       1&2000.119&+0.770 / -0.770&OK\\
       1&2077.552&+1.182 / -1.128&OK\\
       1&2303.520&Not fitted&0.855\\
\hline
       2&977.959&+4.691 / -3.024&Not detected\\
       2&1051.458&+3.784 / -1.258&OK\\
       2&1125.882&+1.897 / -1.624&OK\\
       2&1199.526&+1.111 / -1.295&OK\\
       2&1272.976&+1.124 / -1.123&OK\\
       2&1347.530&+1.152 / -1.151&OK\\
       2&1423.080&+1.222 / -1.119&OK\\
       2&1499.386&+1.509 / -1.407&OK\\
       2&1575.939&+1.865 / -1.569&OK\\
       2&1652.855&+2.032 / -2.029&OK\\
       2&1729.308&+2.131 / -2.128&OK\\
       2&1805.913&+1.800 / -1.893&OK\\
       2&1882.285&+1.280 / -1.377&OK\\
       2&1959.293&+1.425 / -1.526&OK\\
       2&2036.156&+2.423 / -2.630&OK\\
       2&2265.410&Not fitted&0.807\\
\hline
\hline
\end{tabular}
\label{2837475}
\end{table*}
\begin{table*}[!]
\caption{Frequencies for KIC 3424541. The first column is the degree.  The second column is the frequency.  The third column is the 68\%-credible interval quoted when the mode is fitted.  The last column provides an indication of the quality of the detection: {\it OK} indicates that the mode was correctly detected and fitted; {\it Not detected} indicates that the mode was fitted but not detected by the quality assurance test and {\it Not fitted} indicates that the mode was detected with a posterior probability provided by the quality assurance test.  When an uncertainty {\it and} a posterior probability are quoted, it means that the mode is fitted but detected using the quality assurance test with a probability lower than 90\%.}
\centering
\begin{tabular}{c c c c} 
\hline
\hline
Degree&Frequency ($\mu$Hz)& 68\% credible ($\mu$Hz) &Comment\\
\hline
\hline
       0&515.159&+0.898 / -0.998&OK\\
       0&554.896&+0.750 / -0.847&OK\\
       0&594.851&+0.832 / -0.964&OK\\
       0&635.872&+0.854 / -1.029&OK\\
       0&677.707&+0.631 / -0.664&OK\\
       0&720.542&+0.430 / -0.460&OK\\
       0&761.406&+0.617 / -0.644&OK\\
       0&801.236&+0.460 / -0.521&OK\\
       0&842.571&+0.716 / -0.751&OK\\
       0&884.695&+0.676 / -0.675&OK\\
       0&926.904&+0.674 / -0.704&OK\\
       0&966.729&+0.432 / -0.432&OK\\
       0&1008.191&+1.030 / -1.065&OK\\
       0&1049.754&+1.868 / -2.146&Not detected\\
       0&1092.100&Not fitted&0.867\\
\hline
       1&368.423&Not fitted&0.663\\
       1&497.417&+0.413 / -0.461&0.703\\
       1&535.467&+0.548 / -0.547&OK\\
       1&574.884&+0.510 / -0.537&OK\\
       1&616.271&+0.583 / -0.583&OK\\
       1&659.448&+0.457 / -0.483&OK\\
       1&699.337&+0.372 / -0.338&OK\\
       1&740.189&+0.390 / -0.420&OK\\
       1&781.366&+0.478 / -0.478&OK\\
       1&821.304&+0.494 / -0.467&OK\\
       1&863.283&+0.564 / -0.531&OK\\
       1&906.191&+0.586 / -0.553&OK\\
       1&948.677&+0.583 / -0.582&OK\\
       1&990.946&+0.423 / -0.455&OK\\
       1&1032.211&+0.785 / -0.941&Not detected\\
       1&1070.390&Not fitted&0.825\\
       1&1115.410&Not fitted&0.870\\
\hline
       2&510.044&+1.390 / -1.142&Not detected\\
       2&549.930&+1.048 / -0.994&OK\\
       2&590.101&+1.087 / -0.930&OK\\
       2&631.710&+0.911 / -0.910&OK\\
       2&673.375&+0.930 / -0.878&OK\\
       2&714.813&+0.885 / -0.785&OK\\
       2&755.444&+0.775 / -0.774&OK\\
       2&796.314&+0.838 / -0.732&OK\\
       2&837.143&+1.043 / -1.042&OK\\
       2&878.519&+1.377 / -1.222&OK\\
       2&920.884&+1.426 / -1.319&OK\\
       2&963.606&+1.182 / -1.227&OK\\
       2&1005.239&+1.464 / -1.406&OK\\
       2&1046.639&+2.249 / -2.195&Not detected\\
\hline
\hline
\end{tabular}
\label{3424541}
\end{table*}
\begin{table*}[!]
\caption{Frequencies for KIC 3427720. The first column is the degree.  The second column is the frequency.  The third column is the 1-$\sigma$ uncertainty quoted when the mode is fitted.  The last column provides an indication of the quality of the detection: {\it OK} indicates that the mode was correctly detected and fitted; {\it Not detected} indicates that the mode was fitted but not detected by the quality assurance test and {\it Not fitted} indicates that the mode was detected with a posterior probability provided by the quality assurance test.  When an uncertainty {\it and} a posterior probability are quoted, it means that the mode is fitted but detected using the quality assurance test with a probability lower than 90\%.}
\centering
\begin{tabular}{c c c c} 
\hline
\hline
Degree&Frequency ($\mu$Hz)&1-$\sigma$ error ($\mu$Hz)&Comment\\
\hline
\hline
       0&2207.437&0.631&OK\\
       0&2325.063&0.244&OK\\
       0&2444.427&0.280&OK\\
       0&2564.300&0.218&OK\\
       0&2684.764&0.143&OK\\
       0&2804.640&0.328&OK\\
       0&2925.244&0.361&OK\\
       0&3044.276&0.396&OK\\
       0&3165.060&0.541&Not detected\\
       0&3287.661&0.293&0.866\\
\hline
       1&2024.400&Not fitted&0.706\\
       1&2144.016&0.453&OK\\
       1&2262.339&0.244&OK\\
       1&2380.617&0.285&OK\\
       1&2500.464&0.206&OK\\
       1&2621.039&0.160&OK\\
       1&2741.644&0.206&OK\\
       1&2861.405&0.186&OK\\
       1&2981.503&0.264&OK\\
       1&3101.854&0.289&OK\\
       1&3223.454&0.395&Not detected\\
\hline
       2&2195.406&0.619&OK\\
       2&2314.163&0.365&OK\\
       2&2433.841&0.259&OK\\
       2&2553.282&0.328&OK\\
       2&2674.781&0.197&OK\\
       2&2793.974&0.383&OK\\
       2&2914.192&0.419&OK\\
       2&3034.861&0.484&OK\\
       2&3156.454&0.667&0.859\\
       2&3278.512&0.748&Not detected\\
\hline
\hline
\end{tabular}
\label{3427720}
\end{table*}
\begin{table*}[!]
\caption{Frequencies for KIC 3632418. The first column is the degree.  The second column is the frequency.  The third column is the 1-$\sigma$ uncertainty quoted when the mode is fitted.  The last column provides an indication of the quality of the detection: {\it OK} indicates that the mode was correctly detected and fitted; {\it Not detected} indicates that the mode was fitted but not detected by the quality assurance test and {\it Not fitted} indicates that the mode was detected with a posterior probability provided by the quality assurance test.  When an uncertainty {\it and} a posterior probability are quoted, it means that the mode is fitted but detected using the quality assurance test with a probability lower than 90\%.}
\centering
\begin{tabular}{c c c c} 
\hline
\hline
Degree&Frequency ($\mu$Hz)&1-$\sigma$ error ($\mu$Hz)&Comment\\
\hline
\hline
       0&682.388&Not fitted&0.709\\
       0&740.494&Not fitted&0.694\\
       0&799.280&0.410&OK\\
       0&858.930&0.240&OK\\
       0&920.370&0.180&OK\\
       0&979.300&0.160&OK\\
       0&1039.310&0.190&OK\\
       0&1099.260&0.180&OK\\
       0&1159.920&0.180&OK\\
       0&1221.310&0.200&OK\\
       0&1282.550&0.200&OK\\
       0&1342.600&0.400&OK\\
       0&1405.040&0.250&OK\\
       0&1464.790&0.570&OK\\
       0&1645.870&Not fitted&0.861\\
\hline
       1&709.051&Not fitted&0.666\\
       1&765.354&Not fitted&0.658\\
       1&824.590&0.420&OK\\
       1&885.250&0.190&OK\\
       1&946.450&0.180&OK\\
       1&1006.170&0.180&OK\\
       1&1065.010&0.160&OK\\
       1&1125.860&0.140&OK\\
       1&1186.780&0.150&OK\\
       1&1248.100&0.180&OK\\
       1&1309.960&0.140&OK\\
       1&1370.890&0.200&OK\\
       1&1432.220&0.200&OK\\
       1&1493.980&0.260&OK\\
       1&1612.500&Not fitted&0.831\\
\hline
       2&682.388&Not fitted&0.709\\
       2&740.494&Not fitted&0.694\\
       2&916.940&0.430&OK\\
       2&976.340&0.290&OK\\
       2&1034.990&0.370&OK\\
       2&1094.420&0.270&OK\\
       2&1155.090&0.210&OK\\
       2&1217.370&0.350&OK\\
       2&1277.980&0.190&OK\\
       2&1342.420&0.460&OK\\
       2&1400.910&0.480&OK\\
       2&1460.440&0.760&OK\\
       2&1574.440&Not fitted&0.576\\
\hline
\hline
\end{tabular}
\label{3632418}
\end{table*}	
\begin{table*}[!]
\caption{Frequencies for KIC 3733735. The first column is the degree.  The second column is the frequency.  The third column is the 68\%-credible interval quoted when the mode is fitted.  The last column provides an indication of the quality of the detection: {\it OK} indicates that the mode was correctly detected and fitted; {\it Not detected} indicates that the mode was fitted but not detected by the quality assurance test and {\it Not fitted} indicates that the mode was detected with a posterior probability provided by the quality assurance test.  When an uncertainty {\it and} a posterior probability are quoted, it means that the mode is fitted but detected using the quality assurance test with a probability lower than 90\%.}
\centering
\begin{tabular}{c c c c} 
\hline
\hline
Degree&Frequency ($\mu$Hz)& 68\% credible ($\mu$Hz) &Comment\\
\hline
\hline
       0&1207.275&+1.933 / -1.999&Not detected\\
       0&1296.993&+0.663 / -0.783&0.701\\
       0&1386.418&+0.877 / -0.877&OK\\
       0&1475.046&+0.780 / -0.780&OK\\
       0&1564.659&+1.051 / -1.050&OK\\
       0&1655.392&+0.545 / -0.600&OK\\
       0&1746.599&+0.543 / -0.603&OK\\
       0&1840.152&+0.776 / -0.711&OK\\
       0&1934.770&+0.679 / -0.740&OK\\
       0&2027.616&+0.839 / -0.903&OK\\
       0&2118.940&+0.583 / -0.636&OK\\
       0&2210.372&+0.826 / -0.894&OK\\
       0&2301.778&+0.968 / -0.887&OK\\
       0&2394.698&+0.882 / -0.882&OK\\
       0&2487.872&+0.573 / -0.516&OK\\
       0&2580.729&+1.731 / -1.730&OK\\
       0&2761.700&Not fitted&0.867\\
       0&2947.460&Not fitted&0.852\\
\hline
       1&1158.006&+4.157 / -4.060&Not detected\\
       1&1247.271&+2.149 / -2.231&Not detected\\
       1&1336.416&+0.843 / -0.702&OK\\
       1&1425.981&+0.851 / -0.773&OK\\
       1&1515.171&+0.726 / -0.726&OK\\
       1&1605.189&+0.782 / -0.726&OK\\
       1&1695.891&+0.411 / -0.411&OK\\
       1&1788.845&+0.553 / -0.553&OK\\
       1&1882.074&+0.528 / -0.587&OK\\
       1&1974.958&+0.571 / -0.571&OK\\
       1&2068.381&+0.745 / -0.687&OK\\
       1&2161.319&+0.481 / -0.549&OK\\
       1&2254.070&+0.808 / -0.943&OK\\
       1&2346.065&+1.117 / -1.116&OK\\
       1&2437.515&+1.117 / -1.037&OK\\
       1&2527.773&+0.630 / -0.567&OK\\
       1&2618.528&+1.887 / -1.886&OK\\
\hline
       2&1193.353&+5.465 / -5.133&Not detected\\
       2&1282.766&+3.751 / -3.313&Not detected\\
       2&1372.137&+2.820 / -2.229&OK\\
       2&1462.461&+2.444 / -2.196&OK\\
       2&1553.355&+2.556 / -2.284&OK\\
       2&1645.004&+2.644 / -2.869&OK\\
       2&1737.415&+2.676 / -3.717&OK\\
       2&1830.877&+2.560 / -3.485&OK\\
       2&1924.991&+1.979 / -3.212&OK\\
       2&2018.465&+1.808 / -2.322&OK\\
       2&2110.910&+1.605 / -1.604&OK\\
       2&2203.185&+1.356 / -1.232&OK\\
       2&2295.285&+1.506 / -1.505&OK\\
       2&2388.019&+1.858 / -1.972&OK\\
       2&2480.757&+2.167 / -3.092&OK\\
       2&2573.191&+3.271 / -4.828&OK\\
       2&2932.240&Not fitted&0.863\\
\hline
\hline
\end{tabular}
\label{3733735}
\end{table*}
\begin{table*}[!]
\caption{Frequencies for KIC 3735871. The first column is the degree.  The second column is the frequency.  The third column is the 68\%-credible interval quoted when the mode is fitted.  The last column provides an indication of the quality of the detection: {\it OK} indicates that the mode was correctly detected and fitted; {\it Not detected} indicates that the mode was fitted but not detected by the quality assurance test and {\it Not fitted} indicates that the mode was detected with a posterior probability provided by the quality assurance test.  When an uncertainty {\it and} a posterior probability are quoted, it means that the mode is fitted but detected using the quality assurance test with a probability lower than 90\%.}
\centering
\begin{tabular}{c c c c} 
\hline
\hline
Degree&Frequency ($\mu$Hz)& 68\% credible ($\mu$Hz) &Comment\\
\hline
\hline
       0&2247.640&Not fitted&0.614\\
       0&2375.005&+0.760 / -0.859&OK\\
       0&2500.882&+0.293 / -0.312&OK\\
       0&2624.587&+0.190 / -0.196&OK\\
       0&2746.938&+0.289 / -0.289&OK\\
       0&2870.265&+0.527 / -0.596&OK\\
       0&2993.256&+0.331 / -0.331&OK\\
       0&3117.781&+0.370 / -0.358&OK\\
       0&3244.134&+0.695 / -0.918&Not detected\\
       0&3369.276&+1.520 / -1.814&Not detected\\
       0&3494.141&+2.348 / -3.233&Not detected\\
       0&3618.881&+3.662 / -4.980&Not detected\\
\hline
       1&1927.430&Not fitted&0.572\\
       1&2056.550&Not fitted&0.695\\
       1&2300.500&Not fitted&0.600\\
       1&2314.535&+0.538 / -0.651&OK\\
       1&2436.561&+0.283 / -0.310&OK\\
       1&2558.186&+0.251 / -0.251&OK\\
       1&2681.946&+0.156 / -0.156&OK\\
       1&2805.076&+0.338 / -0.329&OK\\
       1&2926.893&+0.447 / -0.371&OK\\
       1&3050.760&+0.470 / -0.459&OK\\
       1&3175.231&+0.545 / -0.616&OK\\
       1&3299.693&+0.827 / -0.882&0.696\\
       1&3424.191&+2.375 / -2.476&Not detected\\
       1&3548.722&+4.463 / -4.626&Not detected\\
\hline
       2&1981.510&Not fitted&0.710\\
       2&2366.009&+1.703 / -1.624&OK\\
       2&2489.364&+0.989 / -0.840&OK\\
       2&2613.201&+1.005 / -0.978&OK\\
       2&2737.463&+0.316 / -0.334&OK\\
       2&2859.516&+1.401 / -1.050&OK\\
       2&2981.966&+0.631 / -0.463&OK\\
       2&3105.303&+0.745 / -0.745&OK\\
       2&3229.273&+1.004 / -1.004&Not detected\\
       2&3354.202&+1.130 / -1.554&Not detected\\
       2&3479.350&+2.614 / -3.242&Not detected\\
       2&3604.491&+4.686 / -5.399&Not detected\\
\hline
\hline
\end{tabular}
\label{3735871}
\end{table*}
\begin{table*}[!]
\caption{Frequencies for KIC 5607242. The first column is the degree.  The second column is the frequency.  The third column is the 1-$\sigma$ uncertainty quoted when the mode is fitted.  The last column provides an indication of the quality of the detection: {\it OK} indicates that the mode was correctly detected and fitted; {\it Not detected} indicates that the mode was fitted but not detected by the quality assurance test and {\it Not fitted} indicates that the mode was detected with a posterior probability provided by the quality assurance test.  When an uncertainty {\it and} a posterior probability are quoted, it means that the mode is fitted but detected using the quality assurance test with a probability lower than 90\%.}
\centering
\begin{tabular}{c c c c} 
\hline
\hline
Degree&Frequency ($\mu$Hz)&1-$\sigma$ error ($\mu$Hz)&Comment\\
\hline
\hline
       0&504.104&0.032&OK\\
       0&543.376&0.090&OK\\
       0&583.050&0.076&OK\\
       0&623.645&0.045&OK\\
       0&664.119&0.049&OK\\
       0&704.801&0.098&OK\\
       0&745.287&0.084&OK\\
       0&786.310&0.156&OK\\
       0&827.058&0.264&OK\\
       0&868.990&0.811&OK\\
\hline
       1&371.818&Not fitted&0.622\\
       1&467.697&Not fitted&0.709\\
       1&485.242&0.343&OK\\
       1&509.017&0.031&OK\\
       1&527.565&0.034&OK\\
       1&554.995&0.129&OK\\
       1&572.791&0.089&OK\\
       1&600.775&0.061&OK\\
       1&626.905&0.055&OK\\
       1&648.623&0.059&OK\\
       1&679.921&0.037&OK\\
       1&705.359&0.076&OK\\
       1&731.227&0.075&OK\\
       1&764.822&0.104&OK\\
       1&792.989&0.145&OK\\
       1&816.677&0.150&OK\\
       1&851.846&0.286&OK\\
       1&889.136&0.730&0.868\\
       1&923.697&Not fitted&0.853\\
\hline
       2&499.255&0.013&OK\\
       2&539.702&0.132&OK\\
       2&579.354&0.086&OK\\
       2&619.687&0.047&OK\\
       2&660.523&0.071&OK\\
       2&701.137&0.062&OK\\
       2&740.434&0.133&OK\\
       2&782.653&0.145&OK\\
       2&824.289&0.365&OK\\
       2&866.243&3.016&OK\\
\hline
       3&675.487&0.157&OK\\
       3&716.263&0.138&OK\\
       3&757.426&0.458&OK\\
\hline
\hline
\end{tabular}
\label{5607242}
\end{table*}	
\begin{table*}[!]
\caption{Frequencies for KIC 5955122. The first column is the degree.  The second column is the frequency.  The third column is the 1-$\sigma$ uncertainty quoted when the mode is fitted.  The last column provides an indication of the quality of the detection: {\it OK} indicates that the mode was correctly detected and fitted; {\it Not detected} indicates that the mode was fitted but not detected by the quality assurance test and {\it Not fitted} indicates that the mode was detected with a posterior probability provided by the quality assurance test.  When an uncertainty {\it and} a posterior probability are quoted, it means that the mode is fitted but detected using the quality assurance test with a probability lower than 90\%.}
\centering
\begin{tabular}{c c c c} 
\hline
\hline
Degree&Frequency ($\mu$Hz)&1-$\sigma$ error ($\mu$Hz)&Comment\\
\hline
\hline
       0&508.907&Not fitted&0.561\\
       0&561.117&0.259&OK\\
       0&609.718&0.163&OK\\
       0&658.498&0.355&OK\\
       0&706.164&0.194&OK\\
       0&754.359&0.126&OK\\
       0&803.714&0.093&OK\\
       0&853.473&0.123&OK\\
       0&903.104&0.151&OK\\
       0&952.668&0.191&OK\\
       0&1002.451&0.263&OK\\
       0&1052.268&0.416&OK\\
       0&1104.737&0.639&OK\\
       0&1155.150&Not fitted&0.850\\
\hline
       1&508.907&Not fitted&0.561\\
       1&574.827&0.231&OK\\
       1&586.344&0.144&OK\\
       1&623.587&0.340&OK\\
       1&632.632&0.217&OK\\
       1&657.474&0.282&OK\\
       1&690.112&0.141&OK\\
       1&731.011&0.130&OK\\
       1&774.916&0.106&OK\\
       1&809.729&0.144&OK\\
       1&836.646&0.094&OK\\
       1&880.086&0.124&OK\\
       1&927.117&0.140&OK\\
       1&975.990&0.189&OK\\
       1&1025.152&0.241&OK\\
       1&1073.686&0.337&OK\\
       1&1119.196&0.381&OK\\
       1&1155.150&Not fitted&0.850\\
       1&1230.280&Not fitted&0.673\\
\hline
       2&653.068&0.350&OK\\
       2&701.490&0.344&OK\\
       2&749.720&0.253&OK\\
       2&799.480&0.148&OK\\
       2&849.042&0.183&OK\\
       2&898.478&0.250&OK\\
       2&948.528&0.323&OK\\
       2&997.886&0.396&OK\\
       2&1047.614&0.738&OK\\
       2&1097.643&0.705&OK\\
       2&1146.390&Not fitted&0.805\\
\hline
\hline
\end{tabular}
\label{5955122}
\end{table*}
\begin{table*}[!]
\caption{Frequencies for KIC 6116048. The first column is the degree.  The second column is the frequency.  The third column is the 1-$\sigma$ uncertainty quoted when the mode is fitted.  The last column provides an indication of the quality of the detection: {\it OK} indicates that the mode was correctly detected and fitted; {\it Not detected} indicates that the mode was fitted but not detected by the quality assurance test and {\it Not fitted} indicates that the mode was detected with a posterior probability provided by the quality assurance test.  When an uncertainty {\it and} a posterior probability are quoted, it means that the mode is fitted but detected using the quality assurance test with a probability lower than 90\%.}
\centering
\begin{tabular}{c c c c} 
\hline
\hline
Degree&Frequency ($\mu$Hz)&1-$\sigma$ error ($\mu$Hz)&Comment\\
\hline
\hline
       0&1550.390&0.160&OK\\
       0&1649.830&0.150&OK\\
       0&1748.080&0.140&OK\\
       0&1847.980&0.130&OK\\
       0&1948.290&0.110&OK\\
       0&2049.460&0.130&OK\\
       0&2149.990&0.110&OK\\
       0&2250.410&0.150&OK\\
       0&2352.280&0.260&OK\\
       0&2452.990&0.240&OK\\
       0&2554.950&0.610&OK\\
       0&2658.100&Not fitted&0.834\\
       0&2753.350&Not fitted&0.819\\
       0&2856.990&Not fitted&0.822\\
\hline
       1&1393.700&Not fitted&0.694\\
       1&1494.980&0.260&OK\\
       1&1595.300&0.200&OK\\
       1&1694.420&0.150&OK\\
       1&1793.340&0.150&OK\\
       1&1893.950&0.110&OK\\
       1&1995.150&0.100&OK\\
       1&2096.030&0.110&OK\\
       1&2196.690&0.110&OK\\
       1&2298.040&0.150&OK\\
       1&2399.210&0.190&OK\\
       1&2501.850&0.300&OK\\
       1&2604.150&0.490&OK\\
       1&2703.760&Not fitted&0.863\\
\hline
       2&1342.680&Not fitted&0.699\\
       2&1543.300&0.590&OK\\
       2&1642.530&0.530&OK\\
       2&1741.440&0.340&OK\\
       2&1841.140&0.150&OK\\
       2&1941.650&0.140&OK\\
       2&2043.370&0.160&OK\\
       2&2143.880&0.150&OK\\
       2&2245.240&0.240&OK\\
       2&2346.010&0.410&OK\\
       2&2448.300&0.780&OK\\
       2&2551.400&1.740&OK\\
       2&2753.350&Not fitted&0.819\\
\hline
\hline
\end{tabular}
\label{6116048}
\end{table*}
\begin{table*}[!]
\caption{Frequencies for KIC 6508366. The first column is the degree.  The second column is the frequency.  The third column is the 68\%-credible interval quoted when the mode is fitted.  The last column provides an indication of the quality of the detection: {\it OK} indicates that the mode was correctly detected and fitted; {\it Not detected} indicates that the mode was fitted but not detected by the quality assurance test and {\it Not fitted} indicates that the mode was detected with a posterior probability provided by the quality assurance test.  When an uncertainty {\it and} a posterior probability are quoted, it means that the mode is fitted but detected using the quality assurance test with a probability lower than 90\%.}
\centering
\begin{tabular}{c c c c} 
\hline
\hline
Degree&Frequency ($\mu$Hz)& 68\% credible ($\mu$Hz) &Comment\\
\hline
\hline
       0&622.782&+0.586 / -0.586&0.704\\
       0&672.204&+0.629 / -0.629&OK\\
       0&722.846&+0.445 / -0.445&OK\\
       0&775.109&+0.294 / -0.324&OK\\
       0&826.341&+0.341 / -0.341&OK\\
       0&877.627&+0.341 / -0.341&OK\\
       0&928.955&+0.392 / -0.392&OK\\
       0&979.470&+0.429 / -0.400&OK\\
       0&1030.870&+0.286 / -0.314&OK\\
       0&1083.772&+0.429 / -0.400&OK\\
       0&1137.175&+0.401 / -0.400&OK\\
       0&1189.993&+0.381 / -0.413&OK\\
       0&1241.728&+0.608 / -0.608&OK\\
       0&1293.248&+0.802 / -0.768&OK\\
       0&1345.443&+1.359 / -1.402&OK\\
       0&1394.910&Not fitted&0.852\\
\hline
       1&596.156&+1.169 / -1.133&Not detected\\
       1&645.356&+0.600 / -0.567&OK\\
       1&694.004&+0.391 / -0.423&OK\\
       1&744.731&+0.382 / -0.411&OK\\
       1&796.168&+0.326 / -0.293&OK\\
       1&848.680&+0.351 / -0.324&OK\\
       1&901.024&+0.314 / -0.314&OK\\
       1&951.267&+0.278 / -0.250&OK\\
       1&1002.882&+0.271 / -0.301&OK\\
       1&1054.058&+0.264 / -0.234&OK\\
       1&1107.459&+0.323 / -0.294&OK\\
       1&1159.366&+0.452 / -0.422&OK\\
       1&1212.665&+0.476 / -0.476&OK\\
       1&1265.253&+0.485 / -0.447&OK\\
       1&1317.549&+0.818 / -0.818&OK\\
       1&1371.172&+1.547 / -1.596&OK\\
       1&1422.500&Not fitted&0.871\\
       1&1471.300&Not fitted&0.634\\
\hline
       2&618.025&+2.549 / -2.739&Not detected\\
       2&668.970&+1.476 / -1.807&OK\\
       2&719.958&+1.153 / -1.079&OK\\
       2&771.570&+0.660 / -0.733&OK\\
       2&823.269&+0.579 / -0.506&OK\\
       2&874.571&+0.573 / -0.573&OK\\
       2&925.154&+0.770 / -0.770&OK\\
       2&976.915&+0.698 / -0.837&OK\\
       2&1028.317&+0.543 / -0.543&OK\\
       2&1080.846&+0.589 / -0.588&OK\\
       2&1133.837&+0.642 / -0.641&OK\\
       2&1186.441&+0.734 / -0.660&OK\\
       2&1237.987&+0.826 / -0.756&OK\\
       2&1289.505&+0.993 / -1.054&OK\\
       2&1341.134&+1.950 / -2.103&OK\\
       2&1489.910&Not fitted&0.579\\
\hline
\hline
\end{tabular}
\label{6508366}
\end{table*}
\begin{table*}[!]
\caption{Frequencies for KIC 6603624. The first column is the degree.  The second column is the frequency.  The third column is the 1-$\sigma$ uncertainty quoted when the mode is fitted.  The last column provides an indication of the quality of the detection: {\it OK} indicates that the mode was correctly detected and fitted; {\it Not detected} indicates that the mode was fitted but not detected by the quality assurance test and {\it Not fitted} indicates that the mode was detected with a posterior probability provided by the quality assurance test.  When an uncertainty {\it and} a posterior probability are quoted, it means that the mode is fitted but detected using the quality assurance test with a probability lower than 90\%.}
\centering
\begin{tabular}{c c c c} 
\hline
\hline
Degree&Frequency ($\mu$Hz)&1-$\sigma$ error ($\mu$Hz)&Comment\\
\hline
\hline
       0&1819.570&0.120&OK\\
       0&1928.480&0.140&OK\\
       0&2036.940&0.070&OK\\
       0&2146.780&0.070&OK\\
       0&2256.960&0.090&OK\\
       0&2367.050&0.060&OK\\
       0&2477.100&0.070&OK\\
       0&2587.510&0.080&OK\\
       0&2698.360&0.170&OK\\
       0&2806.710&0.950&OK\\
       0&3028.490&Not fitted&0.861\\
\hline
       1&1760.530&0.110&OK\\
       1&1869.750&0.110&OK\\
       1&1978.720&0.140&OK\\
       1&2088.330&0.050&OK\\
       1&2198.460&0.070&OK\\
       1&2309.010&0.050&OK\\
       1&2419.470&0.060&OK\\
       1&2529.730&0.070&OK\\
       1&2640.460&0.090&OK\\
       1&2751.160&0.080&OK\\
       1&2862.450&0.620&OK\\
       1&2974.460&Not fitted&0.871\\
\hline
       2&1812.780&0.110&OK\\
       2&1921.720&0.180&OK\\
       2&2030.900&0.130&OK\\
       2&2141.040&0.090&OK\\
       2&2251.610&0.070&OK\\
       2&2362.080&0.050&OK\\
       2&2472.460&0.100&OK\\
       2&2583.100&0.080&OK\\
       2&2693.980&0.100&OK\\
       2&2806.710&1.890&OK\\
       2&3028.490&Not fitted&0.861\\
\hline
\hline
\end{tabular}
\label{6603624}
\end{table*}

\begin{table*}[!]
\caption{Frequencies for KIC 6679371. The first column is the degree.  The second column is the frequency.  The third column is the 68\%-credible interval quoted when the mode is fitted.  The last column provides an indication of the quality of the detection: {\it OK} indicates that the mode was correctly detected and fitted; {\it Not detected} indicates that the mode was fitted but not detected by the quality assurance test and {\it Not fitted} indicates that the mode was detected with a posterior probability provided by the quality assurance test.  When an uncertainty {\it and} a posterior probability are quoted, it means that the mode is fitted but detected using the quality assurance test with a probability lower than 90\%.}
\centering
\begin{tabular}{c c c c} 
\hline
\hline
Degree&Frequency ($\mu$Hz)& 68\% credible ($\mu$Hz) &Comment\\
\hline
\hline
       0&298.997&Not fitted&0.569\\
       0&555.363&+1.167 / -1.209&Not detected\\
       0&606.509&+0.301 / -0.335&0.697\\
       0&656.927&+0.387 / -0.457&OK\\
       0&704.583&+0.492 / -0.527&OK\\
       0&752.698&+0.420 / -0.455&OK\\
       0&803.166&+0.301 / -0.334&OK\\
       0&854.637&+0.273 / -0.273&OK\\
       0&904.941&+0.228 / -0.228&OK\\
       0&956.589&+0.294 / -0.326&OK\\
       0&1007.842&+0.301 / -0.301&OK\\
       0&1057.374&+0.458 / -0.425&OK\\
       0&1106.851&+0.447 / -0.479&OK\\
       0&1157.878&+0.622 / -0.621&OK\\
       0&1209.282&+0.446 / -0.480&OK\\
       0&1262.380&+0.537 / -0.536&OK\\
       0&1313.877&+0.834 / -0.794&OK\\
       0&1365.205&+0.938 / -0.871&OK\\
       0&1417.344&+1.363 / -1.322&OK\\
       0&1465.930&Not fitted&0.861\\
       0&1527.010&Not fitted&0.843\\
\hline
       1&577.997&+0.337 / -0.412&0.693\\
       1&628.633&+0.319 / -0.319&OK\\
       1&677.564&+0.343 / -0.342&OK\\
       1&725.750&+0.286 / -0.250&OK\\
       1&774.772&+0.335 / -0.335&OK\\
       1&825.029&+0.243 / -0.243&OK\\
       1&876.900&+0.309 / -0.278&OK\\
       1&928.205&+0.253 / -0.253&OK\\
       1&978.385&+0.255 / -0.255&OK\\
       1&1028.600&+0.276 / -0.241&OK\\
       1&1079.790&+0.326 / -0.326&OK\\
       1&1129.762&+0.392 / -0.360&OK\\
       1&1181.655&+0.408 / -0.408&OK\\
       1&1234.010&+0.476 / -0.515&OK\\
       1&1284.774&+0.501 / -0.468&OK\\
       1&1336.540&+0.694 / -0.693&OK\\
       1&1389.323&+0.631 / -0.586&OK\\
       1&1442.837&+0.833 / -0.885&OK\\
       1&1492.930&Not fitted&0.856\\
       1&1653.830&Not fitted&0.865\\
\hline
       2&550.249&+0.548 / -1.149&Not detected\\
       2&599.852&+0.781 / -0.851&Not detected\\
       2&649.285&+0.712 / -0.711&OK\\
       2&698.002&+0.749 / -0.686&OK\\
       2&747.779&+0.969 / -0.774&OK\\
       2&799.343&+0.773 / -0.772&OK\\
       2&851.598&+0.640 / -0.639&OK\\
       2&902.103&+0.563 / -0.633&OK\\
       2&951.708&+0.640 / -0.703&OK\\
       2&1002.033&+0.694 / -0.624&OK\\
       2&1053.508&+0.589 / -0.655&OK\\
       2&1104.216&+0.702 / -0.766&OK\\
       2&1155.486&+0.744 / -0.811&OK\\
       2&1207.218&+0.770 / -0.770&OK\\
       2&1259.383&+0.918 / -0.983&OK\\
       2&1311.287&+1.016 / -1.015&OK\\
       2&1362.461&+1.138 / -1.137&OK\\
       2&1413.792&+1.846 / -1.914&OK\\
       2&1465.930&Not fitted&0.861\\
\hline
\hline
\end{tabular}
\label{6679371}
\end{table*}	
\begin{table*}[!]
\caption{Frequencies for KIC 6933899. The first column is the degree.  The second column is the frequency.  The third column is the 1-$\sigma$ uncertainty quoted when the mode is fitted.  The last column provides an indication of the quality of the detection: {\it OK} indicates that the mode was correctly detected and fitted; {\it Not detected} indicates that the mode was fitted but not detected by the quality assurance test and {\it Not fitted} indicates that the mode was detected with a posterior probability provided by the quality assurance test.  When an uncertainty {\it and} a posterior probability are quoted, it means that the mode is fitted but detected using the quality assurance test with a probability lower than 90\%.}
\centering
\begin{tabular}{c c c c} 
\hline
\hline
Degree&Frequency ($\mu$Hz)&1-$\sigma$ error ($\mu$Hz)&Comment\\
\hline
\hline
       0&965.310&0.200&OK\\
       0&1036.720&0.160&OK\\
       0&1107.750&0.170&OK\\
       0&1177.640&0.070&OK\\
       0&1249.050&0.100&OK\\
       0&1321.660&0.110&OK\\
       0&1393.940&0.090&OK\\
       0&1466.220&0.110&OK\\
       0&1538.170&0.100&OK\\
       0&1610.740&0.310&OK\\
       0&1683.510&0.650&OK\\
       0&1756.380&Not fitted&0.853\\
       0&1830.710&Not fitted&0.814\\
\hline
       1&853.349&Not fitted&0.705\\
       1&996.190&0.130&OK\\
       1&1067.540&0.080&OK\\
       1&1137.320&0.050&OK\\
       1&1208.220&0.090&OK\\
       1&1279.930&0.100&OK\\
       1&1352.410&0.080&OK\\
       1&1424.560&0.110&OK\\
       1&1496.640&0.120&OK\\
       1&1569.120&0.070&OK\\
       1&1641.110&0.190&OK\\
       1&1714.790&0.500&OK\\
       1&1787.650&Not fitted&0.870\\
       1&1856.360&Not fitted&0.732\\
\hline
       2&961.410&0.250&OK\\
       2&1032.190&0.350&OK\\
       2&1101.520&0.370&OK\\
       2&1172.450&0.170&OK\\
       2&1243.480&0.080&OK\\
       2&1316.700&0.120&OK\\
       2&1389.070&0.140&OK\\
       2&1460.900&0.100&OK\\
       2&1533.630&0.160&OK\\
       2&1606.220&0.310&OK\\
       2&1679.560&0.960&OK\\
       2&1756.380&Not fitted&0.853\\
       2&1830.710&Not fitted&0.814\\
\hline
\hline
\end{tabular}
\label{6933899}
\end{table*}
\begin{table*}[!]
\caption{Frequencies for KIC 7103006. The first column is the degree.  The second column is the frequency.  The third column is the 68\%-credible interval quoted when the mode is fitted.  The last column provides an indication of the quality of the detection: {\it OK} indicates that the mode was correctly detected and fitted; {\it Not detected} indicates that the mode was fitted but not detected by the quality assurance test and {\it Not fitted} indicates that the mode was detected with a posterior probability provided by the quality assurance test.  When an uncertainty {\it and} a posterior probability are quoted, it means that the mode is fitted but detected using the quality assurance test with a probability lower than 90\%.}
\centering
\begin{tabular}{c c c c} 
\hline
\hline
Degree&Frequency ($\mu$Hz)& 68\% credible ($\mu$Hz) &Comment\\
\hline
\hline
       0&654.507&Not fitted&0.703\\
       0&722.230&+0.332 / -0.295&OK\\
       0&779.406&+0.210 / -0.210&OK\\
       0&835.967&+0.250 / -0.250&OK\\
       0&894.308&+0.206 / -0.206&OK\\
       0&954.831&+0.334 / -0.334&OK\\
       0&1014.741&+0.293 / -0.293&OK\\
       0&1073.733&+0.257 / -0.257&OK\\
       0&1132.806&+0.288 / -0.320&OK\\
       0&1191.865&+0.367 / -0.330&OK\\
       0&1251.640&+0.372 / -0.372&OK\\
       0&1311.689&+0.496 / -0.523&OK\\
       0&1372.332&+0.716 / -0.754&OK\\
       0&1432.079&+0.647 / -0.588&OK\\
       0&1493.964&+0.758 / -0.757&OK\\
       0&1556.446&+0.824 / -0.782&OK\\
       0&1620.231&+1.333 / -1.378&OK\\
       0&1674.000&Not fitted&0.867\\
\hline
       1&633.713&Not fitted&0.698\\
       1&693.045&+1.431 / -1.389&OK\\
       1&748.872&+0.232 / -0.283&OK\\
       1&804.302&+0.228 / -0.199&OK\\
       1&862.204&+0.235 / -0.235&OK\\
       1&921.588&+0.213 / -0.213&OK\\
       1&980.934&+0.266 / -0.240&OK\\
       1&1041.136&+0.323 / -0.323&OK\\
       1&1100.874&+0.240 / -0.266&OK\\
       1&1159.107&+0.240 / -0.266&OK\\
       1&1218.379&+0.321 / -0.320&OK\\
       1&1279.409&+0.385 / -0.360&OK\\
       1&1339.755&+0.353 / -0.382&OK\\
       1&1400.431&+0.645 / -0.611&OK\\
       1&1459.991&+0.606 / -0.646&OK\\
       1&1521.031&+0.726 / -0.686&OK\\
       1&1584.351&+0.770 / -0.924&OK\\
       1&1710.390&Not fitted&0.854\\
\hline
       2&654.507&Not fitted&0.703\\
       2&714.286&+3.623 / -2.797&OK\\
       2&773.211&+2.247 / -1.160&OK\\
       2&832.655&+0.692 / -0.846&OK\\
       2&891.692&+0.958 / -0.877&OK\\
       2&950.164&+0.674 / -0.589&OK\\
       2&1009.253&+0.665 / -0.665&OK\\
       2&1067.324&+0.756 / -0.604&OK\\
       2&1126.088&+0.632 / -0.553&OK\\
       2&1186.034&+0.700 / -0.700&OK\\
       2&1247.803&+0.986 / -0.985&OK\\
       2&1308.653&+1.001 / -0.924&OK\\
       2&1369.219&+0.828 / -0.827&OK\\
       2&1429.926&+0.972 / -1.046&OK\\
       2&1490.720&+1.204 / -1.203&OK\\
       2&1551.843&+1.456 / -1.697&OK\\
       2&1613.238&+2.455 / -2.789&OK\\
       2&1674.000&Not fitted&0.867\\
\hline
\hline
\end{tabular}
\label{7103006}
\end{table*}
\include{table_7206837}
\include{table_7341231}	
\include{table_7747078}
\include{table_7799349}	
\begin{table*}[!]
\caption{Frequencies for KIC 7871531. The first column is the degree.  The second column is the frequency.  The third column is the 1-$\sigma$ uncertainty quoted when the mode is fitted.  The last column provides an indication of the quality of the detection: {\it OK} indicates that the mode was correctly detected and fitted; {\it Not detected} indicates that the mode was fitted but not detected by the quality assurance test and {\it Not fitted} indicates that the mode was detected with a posterior probability provided by the quality assurance test.  When an uncertainty {\it and} a posterior probability are quoted, it means that the mode is fitted but detected using the quality assurance test with a probability lower than 90\%.}
\centering
\begin{tabular}{c c c c} 
\hline
\hline
Degree&Frequency ($\mu$Hz)&1-$\sigma$ error ($\mu$Hz)&Comment\\
\hline
\hline
       0&2658.353&0.998&OK\\
       0&2952.707&0.260&OK\\
       0&3103.788&0.181&OK\\
       0&3254.641&0.189&OK\\
       0&3405.707&0.170&OK\\
       0&3556.559&0.153&OK\\
       0&3708.340&0.368&OK\\
       0&3860.650&0.781&OK\\
\hline
       1&1967.590&Not fitted&0.572\\
       1&2572.881&0.393&OK\\
       1&2724.260&0.587&OK\\
       1&3025.279&0.267&OK\\
       1&3176.721&0.138&OK\\
       1&3327.866&0.151&OK\\
       1&3479.092&0.181&OK\\
       1&3630.458&0.166&OK\\
       1&3782.359&0.338&OK\\
       1&3934.169&0.260&OK\\
\hline
       2&2643.529&0.806&OK\\
       2&2943.651&1.112&OK\\
       2&3094.944&0.379&OK\\
       2&3247.483&0.355&OK\\
       2&3394.363&0.331&OK\\
       2&3549.802&0.324&OK\\
       2&3695.585&0.734&OK\\
       2&3850.309&0.570&OK\\
\hline
\hline
\end{tabular}
\label{7871531}
\end{table*}

\begin{table*}[!]
\caption{Frequencies for KIC 8006161. The first column is the degree.  The second column is the frequency.  The third column is the 1-$\sigma$ uncertainty quoted when the mode is fitted.  The last column provides an indication of the quality of the detection: {\it OK} indicates that the mode was correctly detected and fitted; {\it Not detected} indicates that the mode was fitted but not detected by the quality assurance test and {\it Not fitted} indicates that the mode was detected with a posterior probability provided by the quality assurance test.  When an uncertainty {\it and} a posterior probability are quoted, it means that the mode is fitted but detected using the quality assurance test with a probability lower than 90\%.}
\centering
\begin{tabular}{c c c c} 
\hline
\hline
Degree&Frequency ($\mu$Hz)&1-$\sigma$ error ($\mu$Hz)&Comment\\
\hline
\hline
       0&2774.670&0.080&OK\\
       0&2922.700&0.050&OK\\
       0&3070.940&0.080&OK\\
       0&3220.010&0.070&OK\\
       0&3369.420&0.070&OK\\
       0&3518.270&0.060&OK\\
       0&3667.650&0.070&OK\\
       0&3817.470&0.110&OK\\
       0&3966.600&0.110&OK\\
       0&4117.390&0.100&OK\\
\hline
       1&2546.720&Not fitted&0.698\\
       1&2695.970&Not fitted&0.707\\
       1&2844.880&0.110&OK\\
       1&2992.960&0.050&OK\\
       1&3142.020&0.100&OK\\
       1&3291.260&0.080&OK\\
       1&3440.660&0.080&OK\\
       1&3590.130&0.080&OK\\
       1&3739.220&0.090&OK\\
       1&3888.960&0.120&OK\\
       1&4039.520&0.060&OK\\
       1&4190.060&0.170&OK\\
       1&4338.690&Not fitted&0.857\\
\hline
       2&2613.460&Not fitted&0.699\\
       2&2911.810&0.100&OK\\
       2&3060.230&0.150&OK\\
       2&3209.920&0.140&OK\\
       2&3359.190&0.130&OK\\
       2&3508.760&0.110&OK\\
       2&3658.160&0.130&OK\\
\hline
\hline
\end{tabular}
\label{8006161}
\end{table*}
\include{table_8026226}
\include{table_8228742}	
\include{table_8379927}	
\begin{table*}[!]
\caption{Frequencies for KIC 8394589. The first column is the degree.  The second column is the frequency.  The third column is the 1-$\sigma$ uncertainty quoted when the mode is fitted.  The last column provides an indication of the quality of the detection: {\it OK} indicates that the mode was correctly detected and fitted; {\it Not detected} indicates that the mode was fitted but not detected by the quality assurance test and {\it Not fitted} indicates that the mode was detected with a posterior probability provided by the quality assurance test.  When an uncertainty {\it and} a posterior probability are quoted, it means that the mode is fitted but detected using the quality assurance test with a probability lower than 90\%.}
\centering
\begin{tabular}{c c c c} 
\hline
\hline
Degree&Frequency ($\mu$Hz)&1-$\sigma$ error ($\mu$Hz)&Comment\\
\hline
\hline
       0&1569.350&Not fitted&0.643\\
       0&1787.581&0.751&OK\\
       0&1889.847&6.558&OK\\
       0&2001.614&0.305&OK\\
       0&2109.726&0.289&OK\\
       0&2219.077&0.150&OK\\
       0&2328.624&0.294&OK\\
       0&2438.007&0.265&OK\\
       0&2546.914&0.499&OK\\
       0&2657.331&0.498&OK\\
       0&2764.523&0.926&OK\\
       0&2875.150&3.064&0.868\\
\hline
       1&1508.980&Not fitted&0.707\\
       1&1838.458&1.913&OK\\
       1&1944.536&0.335&OK\\
       1&2051.116&0.418&OK\\
       1&2161.014&0.291&OK\\
       1&2270.413&0.242&OK\\
       1&2380.092&0.234&OK\\
       1&2489.487&0.347&OK\\
       1&2597.882&0.550&OK\\
       1&2708.954&0.615&OK\\
       1&2819.976&1.305&OK\\
       1&3043.308&1.129&Not detected\\
\hline
       2&1778.667&0.986&OK\\
       2&1879.080&5.824&OK\\
       2&1993.845&0.530&OK\\
       2&2099.815&1.025&OK\\
       2&2209.195&0.476&OK\\
       2&2319.629&0.366&OK\\
       2&2429.626&0.569&OK\\
       2&2538.401&0.746&OK\\
       2&2648.172&0.660&OK\\
       2&2757.504&1.659&OK\\
       2&2871.844&4.567&Not detected\\
\hline
\hline
\end{tabular}
\label{8394589}
\end{table*}

\begin{table*}[!]
\caption{Frequencies for KIC 8524425. The first column is the degree.  The second column is the frequency.  The third column is the 1-$\sigma$ uncertainty quoted when the mode is fitted.  The last column provides an indication of the quality of the detection: {\it OK} indicates that the mode was correctly detected and fitted; {\it Not detected} indicates that the mode was fitted but not detected by the quality assurance test and {\it Not fitted} indicates that the mode was detected with a posterior probability provided by the quality assurance test.  When an uncertainty {\it and} a posterior probability are quoted, it means that the mode is fitted but detected using the quality assurance test with a probability lower than 90\%.}
\centering
\begin{tabular}{c c c c} 
\hline
\hline
Degree&Frequency ($\mu$Hz)&1-$\sigma$ error ($\mu$Hz)&Comment\\
\hline
\hline
       0&743.501&0.095&OK\\
       0&803.332&0.217&OK\\
       0&861.330&0.127&OK\\
       0&919.528&0.085&OK\\
       0&979.033&0.084&OK\\
       0&1038.730&0.063&OK\\
       0&1098.265&0.066&OK\\
       0&1157.937&0.133&OK\\
       0&1217.826&0.206&OK\\
       0&1278.207&0.303&OK\\
       0&1400.570&Not fitted&0.617\\
\hline
       1&727.466&0.052&0.691\\
       1&782.275&0.043&OK\\
       1&832.894&0.241&OK\\
       1&888.712&0.091&OK\\
       1&945.280&0.098&OK\\
       1&1001.843&0.083&OK\\
       1&1044.914&0.081&OK\\
       1&1073.963&0.061&OK\\
       1&1127.282&0.073&OK\\
       1&1185.414&0.128&OK\\
       1&1245.099&0.168&OK\\
       1&1305.489&0.284&OK\\
       1&1364.969&0.445&OK\\
\hline
       2&736.722&0.070&Not detected\\
       2&797.203&0.446&OK\\
       2&856.538&0.154&OK\\
       2&914.522&0.123&OK\\
       2&973.600&0.138&OK\\
       2&1032.870&0.093&OK\\
       2&1093.268&0.081&OK\\
       2&1153.257&0.122&OK\\
       2&1212.986&0.205&OK\\
       2&1274.186&0.393&OK\\
\hline
       3&1055.426&0.256&OK\\
       3&1177.944&0.577&OK\\
       3&1235.932&0.638&OK\\
\hline
\hline
\end{tabular}
\label{8524425}
\end{table*}
\include{table_8694723}
\include{table_8702606}	
\begin{table*}[!]
\caption{Frequencies for KIC 8760414. The first column is the degree.  The second column is the frequency.  The third column is the 1-$\sigma$ uncertainty quoted when the mode is fitted.  The last column provides an indication of the quality of the detection: {\it OK} indicates that the mode was correctly detected and fitted; {\it Not detected} indicates that the mode was fitted but not detected by the quality assurance test and {\it Not fitted} indicates that the mode was detected with a posterior probability provided by the quality assurance test.  When an uncertainty {\it and} a posterior probability are quoted, it means that the mode is fitted but detected using the quality assurance test with a probability lower than 90\%.}
\centering
\begin{tabular}{c c c c} 
\hline
\hline
Degree&Frequency ($\mu$Hz)&1-$\sigma$ error ($\mu$Hz)&Comment\\
\hline
\hline
       0&1578.160&Not fitted&0.697\\
       0&1813.760&0.550&OK\\
       0&1925.990&0.130&OK\\
       0&2041.160&0.100&OK\\
       0&2158.340&0.130&OK\\
       0&2274.790&0.090&OK\\
       0&2391.480&0.100&OK\\
       0&2508.650&0.130&OK\\
       0&2626.050&0.120&OK\\
       0&2744.180&0.150&OK\\
       0&2862.970&0.360&OK\\
       0&2980.080&0.240&OK\\
       0&3099.130&Not fitted&0.681\\
\hline
       1&1861.020&0.090&OK\\
       1&1976.870&0.130&OK\\
       1&2093.400&0.130&OK\\
       1&2210.920&0.100&OK\\
       1&2328.110&0.090&OK\\
       1&2445.410&0.090&OK\\
       1&2563.390&0.120&OK\\
       1&2681.430&0.120&OK\\
       1&2800.490&0.110&OK\\
       1&2918.240&0.370&OK\\
       1&3041.170&0.200&OK\\
\hline
       2&1802.590&0.110&OK\\
       2&1918.100&0.740&OK\\
       2&2034.310&0.100&OK\\
       2&2151.110&0.120&OK\\
       2&2268.920&0.110&OK\\
       2&2386.090&0.130&OK\\
       2&2503.680&0.130&OK\\
       2&2622.050&0.170&OK\\
       2&2740.530&0.210&OK\\
       2&2857.750&0.290&OK\\
       2&2975.240&0.470&OK\\
\hline
\hline
\end{tabular}
\label{8760414}
\end{table*}
\begin{table*}[!]
\caption{Frequencies for KIC 9025370. The first column is the degree.  The second column is the frequency.  The third column is the 1-$\sigma$ uncertainty quoted when the mode is fitted.  The last column provides an indication of the quality of the detection: {\it OK} indicates that the mode was correctly detected and fitted; {\it Not detected} indicates that the mode was fitted but not detected by the quality assurance test and {\it Not fitted} indicates that the mode was detected with a posterior probability provided by the quality assurance test.  When an uncertainty {\it and} a posterior probability are quoted, it means that the mode is fitted but detected using the quality assurance test with a probability lower than 90\%.}
\centering
\begin{tabular}{c c c c} 
\hline
\hline
Degree&Frequency ($\mu$Hz)&1-$\sigma$ error ($\mu$Hz)&Comment\\
\hline
\hline
       0&2446.782&0.391&OK\\
       0&2583.443&0.325&OK\\
       0&2715.683&0.106&OK\\
       0&2848.290&0.104&OK\\
       0&2980.918&0.168&OK\\
       0&3113.779&0.183&OK\\
       0&3246.101&0.200&OK\\
       0&3379.536&0.229&OK\\
\hline
       1&2513.462&0.186&OK\\
       1&2645.981&0.250&OK\\
       1&2778.757&0.150&OK\\
       1&2911.677&0.144&OK\\
       1&3044.525&0.236&OK\\
       1&3176.898&0.290&OK\\
       1&3310.026&0.265&OK\\
       1&3441.600&Not fitted&0.870\\
\hline
       2&2439.798&0.514&Not detected\\
       2&2576.073&1.253&OK\\
       2&2706.664&0.356&OK\\
       2&2839.358&0.236&OK\\
       2&2972.265&0.537&OK\\
       2&3104.379&0.357&OK\\
       2&3237.619&0.968&OK\\
       2&3368.469&0.683&OK\\
\hline
\hline
\end{tabular}
\label{9025370}
\end{table*}
\begin{table*}[!]
\caption{Frequencies for KIC 9098294. The first column is the degree.  The second column is the frequency.  The third column is the 1-$\sigma$ uncertainty quoted when the mode is fitted.  The last column provides an indication of the quality of the detection: {\it OK} indicates that the mode was correctly detected and fitted; {\it Not detected} indicates that the mode was fitted but not detected by the quality assurance test and {\it Not fitted} indicates that the mode was detected with a posterior probability provided by the quality assurance test.  When an uncertainty {\it and} a posterior probability are quoted, it means that the mode is fitted but detected using the quality assurance test with a probability lower than 90\%.}
\centering
\begin{tabular}{c c c c} 
\hline
\hline
Degree&Frequency ($\mu$Hz)&1-$\sigma$ error ($\mu$Hz)&Comment\\
\hline
\hline
       0&1900.398&0.244&OK\\
       0&2008.168&0.179&OK\\
       0&2117.053&0.111&OK\\
       0&2225.792&0.166&OK\\
       0&2334.965&0.160&OK\\
       0&2443.801&0.311&OK\\
       0&2552.619&0.460&OK\\
       0&2661.744&0.489&OK\\
       0&2774.688&0.498&OK\\
\hline
       1&1842.447&0.191&OK\\
       1&1949.654&0.243&OK\\
       1&2058.885&0.129&OK\\
       1&2167.927&0.141&OK\\
       1&2276.743&0.147&OK\\
       1&2386.202&0.174&OK\\
       1&2495.467&0.300&OK\\
       1&2605.407&0.188&OK\\
       1&2716.053&0.557&OK\\
\hline
       2&1893.690&0.521&OK\\
       2&2001.824&0.537&OK\\
       2&2110.041&0.246&OK\\
       2&2219.983&0.224&OK\\
       2&2329.685&0.279&OK\\
       2&2438.017&0.333&OK\\
       2&2547.606&0.649&OK\\
       2&2656.647&0.771&OK\\
       2&2768.198&4.452&OK\\
       2&3198.360&Not fitted&0.608\\
\hline
\hline
\end{tabular}
\label{9098294}
\end{table*}

\include{table_9206432}	
\begin{table*}[!]
\caption{Frequencies for KIC 9410862. The first column is the degree.  The second column is the frequency.  The third column is the 1-$\sigma$ uncertainty quoted when the mode is fitted.  The last column provides an indication of the quality of the detection: {\it OK} indicates that the mode was correctly detected and fitted; {\it Not detected} indicates that the mode was fitted but not detected by the quality assurance test and {\it Not fitted} indicates that the mode was detected with a posterior probability provided by the quality assurance test.  When an uncertainty {\it and} a posterior probability are quoted, it means that the mode is fitted but detected using the quality assurance test with a probability lower than 90\%.}
\centering
\begin{tabular}{c c c c} 
\hline
\hline
Degree&Frequency ($\mu$Hz)&1-$\sigma$ error ($\mu$Hz)&Comment\\
\hline
\hline
       0&1864.701&1.081&OK\\
       0&1969.975&0.140&OK\\
       0&2077.738&0.188&OK\\
       0&2184.876&0.313&OK\\
       0&2291.988&0.531&OK\\
       0&2399.497&0.526&OK\\
       0&2506.797&0.503&OK\\
       0&2729.886&0.381&OK\\
\hline
       1&1487.060&Not fitted&0.710\\
       1&1805.510&Not fitted&0.705\\
       1&1912.076&0.782&OK\\
       1&2019.115&0.165&OK\\
       1&2127.074&0.149&OK\\
       1&2235.010&0.255&OK\\
       1&2342.143&0.457&OK\\
       1&2448.996&0.402&OK\\
       1&2559.003&0.544&OK\\
       1&2776.668&0.549&OK\\
\hline
       2&1853.731&0.767&OK\\
       2&1962.460&0.153&OK\\
       2&2069.540&0.257&OK\\
       2&2176.861&0.517&OK\\
       2&2285.465&0.697&OK\\
       2&2392.977&0.518&OK\\
       2&2498.973&0.864&OK\\
       2&2723.195&1.828&OK\\
\hline
\hline
\end{tabular}
\label{9410862}
\end{table*}
\include{table_9574283}	
\include{table_9812850}
\begin{table*}[!]
\caption{Frequencies for KIC 9955598. The first column is the degree.  The second column is the frequency.  The third column is the 1-$\sigma$ uncertainty quoted when the mode is fitted.  The last column provides an indication of the quality of the detection: {\it OK} indicates that the mode was correctly detected and fitted; {\it Not detected} indicates that the mode was fitted but not detected by the quality assurance test and {\it Not fitted} indicates that the mode was detected with a posterior probability provided by the quality assurance test.  When an uncertainty {\it and} a posterior probability are quoted, it means that the mode is fitted but detected using the quality assurance test with a probability lower than 90\%.}
\centering
\begin{tabular}{c c c c} 
\hline
\hline
Degree&Frequency ($\mu$Hz)&1-$\sigma$ error ($\mu$Hz)&Comment\\
\hline
\hline
       0&2842.968&0.493&0.694\\
       0&2995.260&0.316&OK\\
       0&3147.413&0.053&OK\\
       0&3300.495&0.115&OK\\
       0&3453.438&0.048&OK\\
       0&3606.396&0.102&OK\\
       0&3759.417&0.298&OK\\
       0&3913.358&0.217&OK\\
       0&4067.178&0.368&OK\\
\hline
       1&2763.317&0.610&Not detected\\
       1&2914.890&0.236&Not detected\\
       1&3067.401&0.166&OK\\
       1&3220.773&0.090&OK\\
       1&3373.738&0.050&OK\\
       1&3526.992&0.156&OK\\
       1&3680.497&0.206&OK\\
       1&3833.821&0.204&OK\\
       1&3987.506&0.258&OK\\
       1&4142.053&0.554&OK\\
       1&4295.320&Not fitted&0.856\\
\hline
       2&2830.646&0.583&Not detected\\
       2&2984.733&0.252&OK\\
       2&3136.979&0.297&OK\\
       2&3291.075&0.159&OK\\
       2&3444.565&0.153&OK\\
       2&3597.662&0.372&OK\\
       2&3750.276&0.522&OK\\
       2&3899.979&4.820&OK\\
       2&4062.023&0.692&OK\\
       2&4666.180&Not fitted&0.699\\
\hline
\hline
\end{tabular}
\label{9955598}
\end{table*}

\include{table_10018963} 
\include{table_10355856}
\include{table_10454113}
\begin{table*}[!]
\caption{Frequencies for KIC 10644253. The first column is the degree.  The second column is the frequency.  The third column is the 1-$\sigma$ uncertainty quoted when the mode is fitted.  The last column provides an indication of the quality of the detection: {\it OK} indicates that the mode was correctly detected and fitted; {\it Not detected} indicates that the mode was fitted but not detected by the quality assurance test and {\it Not fitted} indicates that the mode was detected with a posterior probability provided by the quality assurance test.  When an uncertainty {\it and} a posterior probability are quoted, it means that the mode is fitted but detected using the quality assurance test with a probability lower than 90\%.}
\centering
\begin{tabular}{c c c c} 
\hline
\hline
Degree&Frequency ($\mu$Hz)&1-$\sigma$ error ($\mu$Hz)&Comment\\
\hline
\hline
       0&2379.902&1.134&OK\\
       0&2500.879&0.271&OK\\
       0&2623.063&0.343&OK\\
       0&2746.737&0.336&OK\\
       0&2870.518&0.359&OK\\
       0&2993.077&0.314&OK\\
       0&3116.035&0.797&0.864\\
       0&3235.641&4.097&Not detected\\
       0&3360.194&1.883&0.865\\
       0&3486.588&1.448&Not detected\\
\hline
       1&2315.207&0.346&0.695\\
       1&2436.154&0.218&OK\\
       1&2558.471&0.208&OK\\
       1&2680.718&0.252&OK\\
       1&2804.066&0.200&OK\\
       1&2927.156&0.190&OK\\
       1&3049.676&0.505&OK\\
       1&3172.635&1.985&0.868\\
       1&3297.161&1.770&0.732\\
       1&3421.872&1.698&Not detected\\
\hline
       2&2371.960&0.778&OK\\
       2&2491.793&1.798&OK\\
       2&2611.786&0.394&OK\\
       2&2734.842&0.303&OK\\
       2&2858.804&0.434&OK\\
       2&2981.418&0.751&OK\\
       2&3104.574&0.753&OK\\
       2&3229.069&5.457&Not detected\\
       2&3350.597&3.399&Not detected\\
       2&3481.222&1.657&Not detected\\
\hline
\hline
\end{tabular}
\label{10644253}
\end{table*}
\include{table_10909629}
\include{table_10963065}
\include{table_11026764} 
\include{table_11081729}
\include{table_11193681}
\include{table_11244118} 
\include{table_11253226}
\begin{table*}[!]
\caption{Frequencies for KIC 11395018. The first column is the degree.  The second column is the frequency.  The third column is the 1-$\sigma$ uncertainty quoted when the mode is fitted.  The last column provides an indication of the quality of the detection: {\it OK} indicates that the mode was correctly detected and fitted; {\it Not detected} indicates that the mode was fitted but not detected by the quality assurance test and {\it Not fitted} indicates that the mode was detected with a posterior probability provided by the quality assurance test.  When an uncertainty {\it and} a posterior probability are quoted, it means that the mode is fitted but detected using the quality assurance test with a probability lower than 90\%.}
\centering
\begin{tabular}{c c c c} 
\hline
\hline
Degree&Frequency ($\mu$Hz)&1-$\sigma$ error ($\mu$Hz)&Comment\\
\hline
\hline
       0&638.990&0.090&OK\\
       0&685.400&0.130&OK\\
       0&732.020&0.150&OK\\
       0&779.710&0.070&OK\\
       0&827.520&0.070&OK\\
       0&875.340&0.110&OK\\
       0&923.430&0.100&OK\\
       0&970.250&0.270&OK\\
       0&1067.700&Not fitted&0.717\\
\hline
       1&571.850&0.160&OK\\
       1&603.110&0.140&OK\\
       1&631.860&0.160&OK\\
       1&667.490&0.120&OK\\
       1&707.940&0.140&OK\\
       1&740.230&0.130&OK\\
       1&763.800&0.100&OK\\
       1&805.800&0.070&OK\\
       1&851.480&0.090&OK\\
       1&897.560&0.090&OK\\
       1&940.810&0.110&OK\\
       1&997.800&0.290&OK\\
       1&1015.050&Not fitted&0.849\\
       1&1043.930&Not fitted&0.872\\
       1&1061.040&Not fitted&0.872\\
\hline
       2&681.510&0.190&OK\\
       2&728.120&0.310&OK\\
       2&775.570&0.150&OK\\
       2&823.510&0.070&OK\\
       2&871.390&0.110&OK\\
       2&918.500&0.140&OK\\
       2&965.880&0.170&OK\\
       2&1015.050&Not fitted&0.849\\
       2&1061.040&Not fitted&0.872\\
\hline
\hline
\end{tabular}
\label{11395018}
\end{table*}

\include{table_11414712}
\include{table_11717120} 
\include{table_11771760} 
\include{table_11772920} 
\include{table_12009504}
\include{table_12258514}
\include{table_12317678} 
\include{table_12508433} 

\end{document}